\documentclass[12pt]{article}
\usepackage{amsmath,lscape}
\usepackage{graphicx,psfrag,epsf}
\usepackage{enumerate}
\usepackage{natbib, bm}
\usepackage{url} % not crucial - just used below for the URL 
\usepackage{xcolor}
\newcommand{\blind}{0}

\usepackage{comment}
\usepackage{amsfonts, amssymb}
\usepackage{booktabs}
\usepackage{subcaption}
\usepackage{graphicx,array}
\usepackage{placeins}
\usepackage{makecell} 

\newtheorem{lemma}{Lemma}
\newtheorem{theorem}{Theorem}

\newtheorem{proposition}{Proposition}
\numberwithin{proposition}{section}
\numberwithin{theorem}{section}
\usepackage{float}
\newcommand{\bbeta}{\boldsymbol{\beta}}
\newcommand{\bA}{\mathbf{A}}
\newcommand{\bS}{\mathbf{S}}
\newcommand{\bB}{\mathbf{B}}
\newcommand{\bJ}{\mathbf{J}}

\newcommand{\bU}{\mathbf{U}}
\newcommand{\Var}{\textrm{Var}}
\newcommand{\Cov}{\textrm{Cov}}

\newcommand{\bV}{\mathbf{V}}
\newcommand{\bW}{\mathbf{W}}
\newcommand{\bM}{\mathbf{M}}

\newcommand{\bZ}{\mathbf{Z}}
\newcommand{\bQ}{\mathbf{Q}}
\newcommand{\bXstar}{\mathbf{X}^{*}}
\newcommand{\bZstar}{\mathbf{Z}^{*}}
\newcommand{\bSigma}{\bm\Sigma}
\newcommand{\bgamma}{\bm\gamma}
\newtheorem{condition}{Condition}
\newcommand{\rev}[1]{\begingroup\color{black}#1\endgroup}

\newcommand{\by}{\mathbf{y}}
\newcommand{\bv}{\mathbf{v}}

\newcommand{\T}{\textsc{T}}
\newcommand{\bI}{\mathbf{I}}

\newcommand{\bmu}{\mbox{\boldmath$\mu$}}

\newcommand{\bb}{\mbox{\boldmath$b$}}

\newcommand{\bX}{\mathbf{X}}

\newcommand{\bC}{\mathbf{C}}
\usepackage{authblk} %Allows author to have block and multiple affiliations
\newcommand{\boldone}{\mathbf{1}}
\newcolumntype{H}{>{\setbox0=\hbox\bgroup}c<{\egroup}@{}}

\begin{document}

\def\spacingset#1{\renewcommand{\baselinestretch}%
{#1}\small\normalsize} \spacingset{1}

%%%%%%%%%%%%%%%%%%%%%%%%%%%%%%%%%%%%%%%%%%%%%%%%%%%%%%%%%%%%%%%%%%%%%%%%%%%%%%

\if0\blind
{
 \title{\rev{Logistic} Regression Model for Differentially-Private Matrix Masking Data} 
 \author[1]{Linh H. Nghiem\thanks{Corresponding author: linh.nghiem@sydney.edu.au}} 

\author[2]{Aidong Adam Ding}
\author[3]{Samuel S. Wu}
\affil[1]{School of Mathematics and Statistics, University of Sydney, Sydney, NSW, Australia}
\affil[2]{Department of Mathematics, Northeastern University, Boston, MA, United States}
\affil[3]{Health Informatics Institute
University of South Florida, Tampa, FL, United States}
\date{}
  \maketitle
} \fi

\if1\blind
{
  \bigskip
  \bigskip
  \bigskip
  \begin{center}
    {\LARGE\bf \rev{Logistic} Regression Model for Differentially-Private Matrix Masked Data}
\end{center}
  \medskip
} \fi

\bigskip
\begin{abstract}
A recently proposed scheme utilizing local noise addition and matrix masking enables data collection while protecting individual privacy from all parties, including the central data manager. Statistical analysis of such privacy-preserved data is particularly challenging for nonlinear models like logistic regression. By leveraging a relationship between logistic regression and linear regression estimators, we propose the first valid statistical analysis method for logistic regression under this setting. Theoretical analysis of the proposed estimator confirmed its validity under an asymptotic framework with large noise variances %, which is uncommon in traditional measurement error model analysis, 
to account for strict privacy requirements. Simulations and real data analyses demonstrate the superiority of the proposed estimators over naive logistic regression methods on privacy-preserved datasets.
\end{abstract}

\noindent%
\vfill

\newpage
\spacingset{1.45} % DON'T change the spacing!
\section{Introduction}
\label{sec:intro}

In the age of data-driven innovation, privacy-protected data collection methods have become essential for balancing the interests of organizations and individuals. These methods ensure that personal and sensitive information is collected, stored, and processed in ways that protect user privacy and comply with regulatory frameworks like GDPR, HIPAA, and CCPA. Protecting privacy is not only necessary for legal compliance but also critical to building and maintaining consumer trust. To ensure that data analysis yields valid statistical inferences from privacy-protected data, new statistical methods are required. Here, we propose a statistical inference procedure for logistic regression on data collected using matrix masking combined with noise addition.

Over the years, many privacy-protected data collection methods have been developed. Common approaches include anonymization (removing personal identifiers) and pseudonymization (replacing sensitive information with fictitious identifiers). While widely used, these methods do not offer formal mathematical privacy guarantees. Differential privacy~\citep{Dwork:2006} addresses this gap by adding statistical noise to datasets, providing a theoretical, quantitative guarantee of privacy. It has since become one of the most popular privacy-preserving methods, with noise addition to summary statistics enabling privacy-preserving logistic regression studies, such as that of~\citet{zhang2012functional}. However, noise addition at the summary level requires a trustworthy central data manager, which does not protect user privacy from data collectors themselves. To achieve user privacy at the individual level, local differential privacy adds noise to data before it is sent to the data collector. 
Local differential privacy schemes have been successfully implemented by companies like Apple~\citep{apple} and Google~\citep{google} to protect individual-level data.

Another privacy protection scheme, matrix masking uses a random orthogonal matrix to scramble raw data and prevent identification of individual records~\citep{Ting:2008}. 
\cite{ding2020privacy} proved that a triple matrix-masking (TM$^2$) data collection procedure~\citep{samuel2017new}, using multi-party computing concepts, can collect a matrix masked data set while preserving individual-level privacy. 
This method, combined with local differential privacy’s noise addition, achieves the same quantitative privacy guarantees with less noise than local differential privacy only scheme ~\citep{ding2023preprint}. The combined scheme is referred to here as the TM²+Noise scheme.

The TM$^2$+Noise data presents unique challenges for logistic regression. Although simple noise addition to the covariates results in data that follows a measurement error model, which has been studied extensively, the existing statistical methods for logistic regression models with covariate measurement errors, such as those in \citep[Chapter 6]{buonaccorsi2010measurement}, are not directly applicable to this setting for the following reason. First, under this TM$^2$+Noise scheme, the observed response variable is no longer binary, but a real-valued continuous variable. Secondly, the matrix masking only preserves the first and second sample moments of the data, the presence of a non-linear link function in the logistic regression model (and a non-linear model in general) typically requires other information to conduct statistical inference for the model parameters. This other information, however, is not preserved under matrix masking.  Hence, incorporating matrix masking with noise addition requires new statistical methods for non-linear models like logistic regression.  

In this paper, we propose the first valid statistical estimator and inference for the logistic regression model under the TM$^2$+Noise setting. Building upon the seminal work of  \cite{haggstrom1983logistic} who relate the logistic regression to the linear regression through a Gaussian mixture model, we generalize it to a conditional Gaussian mixture model that allows confounding covariates, and obtain the corresponding form of the logistic regression model. Then, we establish both a point and an interval estimator of this model using only the first and second sample moments of the raw data, whose expectations are preserved under the TM$^2$+Noise scheme. We study the properties of this estimator under an asymptotic framework that does \textit{not} assume the variance of added noise to be fixed, demonstrating the tradeoff between statistical accuracy and privacy. Simulation studies and a data application on the relationship between hypertension prevalence and various covariates in a medical database,  illustrate the desirable performance of the proposed statistical approach. Our paper also contributes to a growing statistics literature on making inferences from summary statistics \citep{bhowmik2015generalized, whitaker2021logistic}.

\section{Notations and model setup of logistic regression data collected from TM$^2$+Noise scheme}

Data is collected from $n$ individuals, where for each individual $i=1,...,n$, the data includes a binary outcome variable $y_i^* \in \{0, 1\}$ and and a set of $p$ covariates represented by a $p$-dimensional row vector $\bX^*_i \in \mathbb{R}^{1 \times p}$. 

The logistic regression model defines the relationship between the binary response variable and the covariates as follows:
\begin{equation}
P(y^*_i = 1 \mid \bX_i^*) = \dfrac{\exp(\beta_0 + \bX_i^*\bm\beta_1) }{1 + \exp(\beta_0 + \bX_i^*\bm\beta_1)} = \dfrac{1}{1 + \exp\left\{ - \left(\bm\beta_0 + \bX_i^* \bm\beta_1 \right) \right\}},
\label{eq:logistic}
\end{equation}
where $\beta_0 \in \mathbb{R}$ is the intercept and $\bm\beta_1 \in \mathbb{R}^{p \times 1}$ is the  coefficient vector. The primary goal is to make statistical inferences on the parameters $\bm\beta_1$ based on observed data.

If collected directly, the raw data from the $n$ individuals can be represented as the response vector  $\by^* = (y_1^*, \ldots, y_n^*)^\top$  and the $n \times p$ covariate matrix $\bX^*$ with $\bX^*_i$ as its $i$-th row. However, for privacy protection, the data is collected using the TM$^2$+Noise scheme, so the analyst only has access to
\begin{equation}\label{eq:TM2.noise}
\bX = \bM \bX^* + \bU, ~ \by = \bM \by^* + \bv,  
\end{equation}
where $\bM$  is an unknown $n\times n$ orthogonal matrix that preserves row sums, i.e $\bM^\T \bM = \bI_n$ and $\bM^\top \mathbf{1}_n = \mathbf{1}_n$, and $\bU$ and $\bv$ are noise terms with dimensions $n\times p$ and $n\times 1$, respectively. \rev{The elements of these noise terms $\bU$ and $\bv$ are drawn from the normal distribution $N(0, \sigma_x^2)$ and $N(0, \sigma_y^2)$ respectively, with $\sigma_x$ and $\sigma_y$ known to the analyst.} Due to the masking matrix $\bM$ and added noise, all elements in the masked data $\bX$ and $\by$ are real-valued. The focus is on estimating and making inferences for $\beta_0$ and $\bm\beta_1$ based on the masked data $\bX$  and $\by$.  

If the raw data $\bX^*$ and $\by^*$ were directly available, the \textit{conditional maximum likelihood estimator} (cMLE) would typically be used to estimate $\beta_0$ and $\bm\beta_1$. 
This involves minimizing the conditional negative log-likelihood function $\ell^*(\beta_0, \bbeta_1 \mid \by^*, \bX^*) = -\sum_{i=1}^{n}\log P(y_i^*=1\mid \bX_i^*)$. Equivalently, this estimator solves the score equation
\begin{equation}
\bS(\beta_0, \bm\beta_1; \by^*, \bX^*) = \dfrac{1}{n} \sum_{i=1}^{n} \left(\by^*_i\widetilde{\bX}_i^{*}  -  \dfrac{\widetilde{\bX}_i^*}{1 + \exp\left\{ -  \widetilde{\bX}_i^{*} \bm\beta  \right\}} \right) = \bm{0}.
\label{eq:scoreMLEraw}
\end{equation}
Here and in the following, for any matrix $\mathbf{B}$, we let $\widetilde{\mathbf{B}} = (\boldone, \mathbf{B})$ denote the matrix formed by stacking a column of one to the left of $\mathbf{B}$, and $\bm\beta = (\beta_0, \bbeta_1^T)^T \in \mathbb{R}^{(p+1) \times 1}$ denote the combined coefficients vector. 

While this cMLE approach works well for raw data, adapting it to masked data is challenging due to the non-linearity in terms of $\bX_i^*$ within this score equation. 
To address this, we first review an alternative estimation method based on the unconditional maximum likelihood for the logistic model \eqref{eq:logistic} and then adapt this approach to the masked data setting.

\section{Methodology}
\subsection{{Logistic} model from mixture normal}
A key ingredient of our proposed methodology is to relate the logistic regression model to a normal mixture model, as done in \citet{haggstrom1983logistic}. Particularly, for the raw observation $(\bX_i^*, y^*_i)$ for $i = 1,\ldots, n$ with $y^*_i \in \{0, 1\}$, \citet{haggstrom1983logistic} models $\bX_i^* \mid (y^*_i = j) \sim N(\bmu_j, \bm{\Sigma})$ and hence the marginal distribution of $\bX_i^*$ is a normal mixture
\begin{equation}
\bX_i^* \sim p_1 N(\bmu_1, \bm{\Sigma}) + p_0 N(\bmu_0, \bm{\Sigma}),
\label{eq: conddistribution}
\end{equation}
and $p_j = P(y^*_i = j)$ for $j=0,1$, with $\bmu_j \in \mathbb{R}^{1\times p}$ being mean row vectors, $\bm\Sigma \in \mathbb{R}^{p \times p}$ being a positive definite variance matrix, and $p_0 + p_1 = 1$. To see the connection between the model \eqref{eq: conddistribution} and the logistic model \eqref{eq:logistic}, the model \eqref{eq: conddistribution} implied
\begin{equation}
\begin{aligned}
P(y^*_i = 1 | \bX^*_i) 
= & \frac{P(y^*_i = 1, \bX^*_i)}{P(y^*_i = 1, \bX^*_i) + P(y^*_i = 0, \bX^*_i)} \\
= & \frac{ p_1\exp\{-\frac{1}{2} (\bX_i^* - \bmu_1) \bm\Sigma^{-1} (\bX_i^* - \bmu_1)^T\}}{p_1 \exp\{-\frac{1}{2} (\bX_i^* - \bmu_1) \bm\Sigma^{-1} (\bX_i^* - \bmu_1)^T\} + p_0  \exp\{-\frac{1}{2} (\bX_i^* - \bmu_0) \bm\Sigma^{-1} (\bX_i^* - \bmu_0)^T\} } \\
%= & \frac{ 1}{1 + \frac{p_0}{p_1}  \exp\{\frac{1}{2} [(\bX_i^* - \bmu_1) \bm\Sigma^{-1} (\bX_i^* - \bmu_1)^T - (\bX_i^* - \bmu_0) \bm\Sigma^{-1} (\bX_i^* - \bmu_0)^T]\}} \\
%= & \frac{ 1}{1 + \frac{p_0}{p_1}  \exp\{\frac{1}{2} [ \bmu_1 \bm\Sigma^{-1} \bmu_1^T - \bmu_0 \bm\Sigma^{-1} \bmu_0^T - 2 \bX_i^* \  \bm\Sigma^{-1} \bmu_1^T  + 2 \bX_i^* \ \bm\Sigma^{-1} \bmu_0^T]\}} \\
%= & \frac{ 1}{1 + \frac{p_0}{p_1}  \exp\{\frac{1}{2} [(\bmu_1 + \bmu_0) \bm\Sigma^{-1} (\bmu_1 - \bmu_0)^T  + 2 \bX_i^* \ \bm\Sigma^{-1} (\bmu_0 - \bmu_1)^T]\}} \\
= & \frac{1}{1 + \frac{p_0}{p_1}  \exp\{ \frac{\bmu_1 + \bmu_0}{2} \bm\Sigma^{-1} (\bmu_1 - \bmu_0)^T  + \bX_i^* \ \bm\Sigma^{-1} (\bmu_0 - \bmu_1)^T\}} \\
= & \frac{ 1}{1 +   \exp\{ [\log(\frac{p_0}{p_1}) + \frac{\bmu_1 + \bmu_0}{2} \bm\Sigma^{-1} (\bmu_1 - \bmu_0)^T]  + \bX_i^* [\bm\Sigma^{-1} (\bmu_0 - \bmu_1)^T]\}} \\
= & \frac{1}{1 +   \exp\{ - (\beta_0  + \bX_i^* \bm\beta_1) \}},
\end{aligned}
\label{eq:logisticmodel1}
\end{equation}
where $\bbeta_1=\bm\Sigma^{-1}(\bmu_1-\bmu_0)^T$ and $\beta_0= \log(p_1/p_0) - (1/2)(\bmu_1+\bmu_0)\bbeta_1$. It is straightforward to see that the last expression has the same form as the logistic model \eqref{eq:logistic}. 

Since the (conditional) logistic model can be viewed as an implication of a normal mixture model, we can estimate the coefficients vector $\bm\beta = (\beta_0, \bbeta_1^T)^T$ of the logistic model via estimating parameters of the normal mixture model \eqref{eq: conddistribution}. Particularly, the maximum likelihood estimators for $p_1$, $\bmu_0$, $\bmu_1$ and $\bm\Sigma$ are respectively
\begin{equation}
\label{eq:MLE.quantities}
\begin{aligned}
\hat p_1 = \frac{1}{n} \sum_{i=1}^{n} y_i^*, \qquad \hat \bmu_0 = \frac{\sum_{i=1}^{n} (1-y_i^*) \bX_i^*}{\sum_{i=1}^{n} (1- y_i^*)}, \qquad \hat \bmu_1 =\frac{\sum_{i=1}^{n} y_i^* \bX_i^*}{\sum_{i=1}^{n} y_i^*}, \\[1em]
\hat{\bm\Sigma} =  \frac{1}{n} \sum_{i=1}^{n} [y_i^* (\bX_i^* - \hat \bmu_1)^T (\bX_i^* - \hat \bmu_1) + (1 - y_i^*) (\bX_i^* - \hat \bmu_0)^T (\bX_i^* - \hat \bmu_0) ].
\end{aligned}
\end{equation}
Hence the maximum likelihood estimator for ${\bm\beta}_1$ is 
\begin{equation}\label{eq:MLE.beta1}
\hat{\bm\beta}_1 = \hat{\bm\Sigma}^{-1}(\hat\bmu_1 - \hat\bmu_0)^\T
\end{equation}
with corresponding quantities from \eqref{eq:MLE.quantities}. 

\citet{haggstrom1983logistic} related the estimator in~\eqref{eq:MLE.beta1} to a \textit{linear regression} of $\by^*$ on $\bX^*$:
\begin{equation*} \label{eq:lr}
    y_i^* = b_0 + \bX_i^*\bm{b}_1,
\end{equation*}
for $i=1,\ldots,n$. A special property, as shown by \citet{haggstrom1983logistic}, is that that $\hat{\bbeta}_1$ in ~\eqref{eq:MLE.beta1} is \textit{exactly} the same as the maximum likelihood estimator of the mixture model \eqref{eq: conddistribution} for the parameter $\bm{b}_1/\tau^2$ where 
\begin{align}\label{eq:para.b.tao}
\bm{b}_1 = [\Var(\bX_i^*)]^{-1}\Cov(\bX_i^*, y^*_i), 
\ \ \tau^2 = \Var(y_i^* - b_0 -\bX_i^* \bm{b}_1) 
\ \ \mbox{and } b_0 = E(y^*_i) -  E(\bX_i^*)\bm{b}_1.
\end{align}
Here and in the following, for any random row vectors $\bV_1$ and $\bV_2$, we have
$
\Cov(\bV_1,\bV_2) = E[\left\{\bV_1 - E (\bV_1)\right\}^\T\left\{\bV_2 - E(\bV_2)\right\}]$, and $\Var(\bV_1) = \Cov(\bV_1,\bV_1).
$
The quantities $\bm{b} = (b_0, {\bm{b}_1^T})^T$ and $\tau$ in \eqref{eq:para.b.tao} can be estimated by the ordinary least squaress (OLS) estimators:
\begin{equation}
\hat{\bm{b}} = (\hat b_0, \hat{\bm{b}}_1^T)^T = (\widetilde{\bX}^{*\T} \widetilde{\bX}^*)^{-1} \widetilde{\bX}^{*\T}\by^*, 
\qquad \hat\tau^2 = \dfrac{1}{n} \left\{ \by^{*\T}\by^* - \by^{*\T} \widetilde{\bX}^* (\widetilde{\bX}^{*\top} \widetilde{\bX}^{*})^{-1} \widetilde{\bX}^{*\T}\by^* \right\}.
\label{eq:rawestimators}
\end{equation}

\noindent Hence, the estimator $\hat{\bm\beta}_1$ defined in \eqref{eq:MLE.beta1} can be written as   $\hat{\beta}_1 = \hat{\bm{b}}_1^T/\hat{\tau}^2$, which is the sub-vector consisting of the last $p$ components of $\hat{\bm\theta} = \hat{\bm{b}}/\hat{\tau}^2$. Notice that the OLS estimators in~\eqref{eq:rawestimators} is from linear regression  which can be adapted to matrix masked data easily. 
This key property will be used in section 3.3 to develop a consistent estimator and valid confidence interval for $\bm\beta_1$ when the data are masked. Before doing it, we generalize the mixture model \eqref{eq: conddistribution} to allow a more flexible distribution of the raw covariate $\bXstar$.

\subsection{Conditional mixture model}
The mixture model \eqref{eq: conddistribution} is restrictive because it assumes the covariate $\bXstar$ corresponding to each class of $y$ is multivariate Gaussian, and that the mixture weight is constant. In this section, we extend it to a  conditional mixture model 
\begin{align}
& \bXstar_i \mid (y_i^* = j, \bZstar_i) \sim N(\bmu_j+  \bZstar_i \bC, \bSigma),  \label{conditionmixturemodel} \\
& p_1(\bZstar_i) = P(y_i^* = 1 \mid \bZstar_i) = \dfrac{\exp(\gamma_0 + \bZstar_i\bm\gamma_1)}{1 + \exp(\gamma_0 + \bZstar_i\bm\gamma_1)}
\label{eq:logisticZ},
\end{align}
for $j=0,1$ and $i=1,\ldots, n$, where $\bZstar_i$ is a $1\times q$ vector and $\bC \in \mathbb{R}^{q \times p}$, $\gamma_0 \in \mathbb{R}$ and $\bm\gamma_1 \in \mathbb{R}^{q\times 1}$ are the matrix of coefficients.

In other words,  Model \eqref{conditionmixturemodel}-\eqref{eq:logisticZ} extends model \eqref{eq: conddistribution} by assuming the conditional distribution, rather than the unconditional distribution, of $\bXstar_i$ given $\bZstar_i$ within each class of $y_i$ is multivariate Gaussian. Furthermore, the probability belonging to each class can potentially vary as a function of $\bZstar$, rather than being a constant. It is straightforward to see that the mixture model \eqref{eq: conddistribution} is a special case of model  \eqref{conditionmixturemodel}-\eqref{eq:logisticZ} with $\bZ^*=1$ always. For this more flexible model, only part of covariates, $\bXstar$, follow the mixture model. The covariates $\bZ$ can be considered as a confounding variable, so accounting for it is essential in estimating the effects of the covariates $\bX$ on the outcome $\by$. %We show that inferences for its coefficients can be similarly based on OLS. 

By a similar argument to the equation \eqref{eq:logisticmodel1}, the conditional mixture model \eqref{conditionmixturemodel}-\eqref{eq:logisticZ} implies a logistic model of $y_i^*$ on $\bXstar_i$ and $\bZstar_i$, i.e 
\begin{equation}
P(y_i^* = 1 \mid \bZstar_i, \bXstar_i) = \dfrac{1}{1 + \exp(-\beta_0 - \bXstar_i\bbeta_1  - \bZstar_i\bbeta_2)},
\label{eq:log_relationship}
\end{equation}
where
\begin{equation}\label{eq:beta}
\beta_0 = \gamma_0 - (1/2) (\bmu_1 \bSigma^{-1}\bmu_1^\T - \bmu_0\bSigma^{-1}\bmu_0^\T ), \ \ \bm\beta_1 = \bm\Sigma^{-1}(\bmu_1 - \bmu_0)^\T \mbox{ and } \bm\beta_2 = \bgamma_1 - \bC \bbeta_1.
\end{equation}
The detailed derivations are provided in Appendix~\ref{sec:log_relationship}.
With this model, we are interested in the estimation and coefficient for $\bm\beta_1$. 

Furthermore, if the raw data $(y_i^*, \bXstar_i, \bZstar_i)$ are i.i.d from the model \eqref{conditionmixturemodel}-\eqref{eq:logisticZ},  for $i=1,\ldots, n$, the coefficient $\bbeta_1$ can also be estimated using ordinary least squares from a linear regression of $y^*$ on the set of $p+q$ covariates $\bW^* = (\bXstar, \bZstar)$. Indeed, let $\bar{\bm{b}} = \left[\Var(\bW_1^*)\right]^{-1} \Cov(\bW_1^*, y_1^*) = (\bar{\bm{b}}_1, \bar{\bm{b}}_2)$, where $\bar{\bm{b}}_1\in \mathbb{R}^{1\times p}$ and $\bar{\bm{b}}_2 \in \mathbb{R}^{1\times q}$ are the coefficients associated with $\bX$ and $\bZ$, respectively, and $\varepsilon_1 = y^*_1 - \bW_1^*\bar{\bm{b}}$.   The following lemma establishes the relationship between $\bar{\bm{b}}_1$ and $\bbeta_1$, whose proof is provided in Appendix~\ref{sec:lemma1}. 
\begin{lemma}\label{est.b1.cond.mixturemodel}
Under the conditional mixture model \eqref{conditionmixturemodel}-\eqref{eq:logisticZ}, we have $\bm\beta_1 = \left[\Var(\varepsilon_1)\right]^{-1} \bar{\bm{b}}_1$.
\end{lemma}

Note that in Lemma \ref{est.b1.cond.mixturemodel}, there is no relationship between $\bm\beta_2$ and $\bar{\bm{b}}_2$; generally, estimation of $\bm\beta_2$ is not possible via ordinary least squares unless further assumption on the distribution of $\bZstar \mid y^*$ is imposed. Nevertheless, Lemma \ref{est.b1.cond.mixturemodel} implies that for estimating $\bm\beta_1$, we can still use the OLS estimators for the linear regression
$$
y_i^* = b_0 + \bX_i^*\bar{\bm{b}}_1 + \bZ_i^*\bar{\bm{b}}_2.
$$
The OLS estimators for the regression coefficients $\bm{b}=(b_0,\bar{\bm{b}}_1^T,,\bar{\bm{b}}_2^T)^T$ and error variance $\tau^2 = \Var(\varepsilon_1)$ are
\begin{equation}
\hat{\bm{b}} = (\hat b_0,\hat{\bm{b}}_1^\T,\hat{\bm{b}}_2^\T)^\T =  (\widetilde{\bW}^{*\T} \widetilde{\bW}^*)^{-1} \widetilde{\bW}^{*\T}\by^*, \ \
\hat\tau^2 = \dfrac{1}{n} \left\{ \by^{*\T}\by^* - \by^{*\T} \widetilde{\bW}^* (\widetilde{\bW}^{*\top} \widetilde{\bW}^{*})^{-1} \widetilde{\bW}^{*\T}\by^* \right\}.
\label{eq:rawestimatorsCONDITIONAL}
\end{equation}
Then an estimator $\hat{\bbeta}_1$ is obtained as $\hat{\bm{b}}_1^T/\hat{\tau}^2$, a sub-vector consisting of $p$-components, from the second to the $(p+1)$-th component, of $\hat{\bm\theta} = \hat{\bm{b}}/\tau^2$. 

To facilitate the derivation of an estimator from the masked data in the next subsection, we justify this estimator through an estimating equation perspective. To this end, let $b_0 = E(y_1^*) - E(\bW_1^*)\bar{\bm{b}}$,  $\bm\theta =  (b_0, \bar{\bm{b}})^\T/\tau^2$ and $\phi = 1/\tau^2$, and $\widetilde{\bW}_i^* = (1, \bXstar_i, \bZstar_i)$ for $i=1,\ldots,n$. Consider the two following estimating equations 
$
m_{\bm\theta}^*(\bm\theta, \phi) = \sum_{i=1}^{n} m_{i, \bm\theta}^*(\bm\theta, \phi) = \bm{0}$ and $ m_{\phi}^*(\bm\theta, \phi) = \sum_{i=1}^{n} m_{i, \bm\phi}^*(\bm\theta, \phi) = 0, 
$
where 
\begin{equation}    
m_{i,\bm\theta}^*(\bm\theta, \phi) = \widetilde{\bW}_i^{*}y_i^* - \dfrac{1}{\phi} \widetilde{\mathbf{W}}_{i}^{*\T} \widetilde{\bW}_{i}^{*}\bm\theta, \quad m_{i,\bm\phi}^*(\bm\theta, \phi) = \dfrac{1}{2\phi} - \dfrac{1}{2} y_i^{*2} + \dfrac{1}{2\phi^2}\bm\theta^\top \widetilde{\mathbf{W}}_{i}^{*\T}\widetilde{\mathbf{W}}_{i}^{*}\bm\theta 
\label{eq:scorefunctionsraw}
\end{equation}
We will derive in the next subsection that,  at the true parameter $\bm\theta$ and  $\phi$, the two estimating equations $m_{i,\bm\theta}^*(\bm\theta, \phi)$ and $m_{i,\phi}^*(\bm\theta, \phi)$ are unbiased. These unbiased estimating equations imply that under regularity conditions, the $\hat{\bm\theta}$ (and hence $\hat{\bm\beta}_1)$ is asymptotically unbiased and normally distributed for ${\bm\theta}$ (and for $\bm\beta_1$). This property will be used to derive a valid confidence interval for $\bm\beta_1$, which we will make precise in the next subsection.

\subsection{Estimation and inference from masked data}
\label{section: maskeddata}
We first consider the effect of matrix masking but \textit{without} noise addition, i.e $\by = \bM \by^*$, $\bW = \bM \bW^*$ where $\bM$ is an $n\times n$ orthogonal matrix that preserves row sums, satisfying $\bM^\T \bM = \bI_n$ and $\bM^\top \mathbf{1}_n = \mathbf{1}_n$. In this case, it is straightforward to see that the first and second sample moments of the raw data are preserved, i.e $\widetilde{\bW}^\T \widetilde{\bW} = \widetilde{\bW}^{*\T} \widetilde{\bW}^*$, $\widetilde{\bW}^\T \by = \widetilde{\bW}^{*\T} \by^*$, and $\by^\T \by = \by^{*\T}\by^*$. Consequently, the estimators $\hat{\bm\theta}$ and $\hat\tau^2$ defined in \eqref{eq:rawestimators} are invariant to $\bM$, hence so is the coefficient $\hat{\bm\beta}_1$. 

Thus, any estimator, such as the OLS estimator $\hat{\bm\beta}_1$, that relies on these invariant quantities will yield the same results on data with or without matrix masking. This aligns with the original intention of the matrix masking scheme’s designers, in that linear regression analysis using OLS estimators requires no adjustment on the matrix-masked dataset.

However, it is important to note that the regular conditional maximum likelihood estimator, obtained by solving~\eqref{eq:scoreMLEraw}, does not possess this invariance property, making it challenging to adapt for masked data. Here, we focus on OLS-based estimators adapted for logistic regression to leverage this invariance.

We now propose a corrected OLS-based estimator $\hat{\bm\beta}_1$ for TM$^2$+Noise data, utilizing estimating equations that depend solely on the invariant quantities described above. The theoretical analysis of this estimator is significantly simplified by its invariance property, as the estimator yields the same result as if there were no masking by the matrix $\bM$. This allows us to perform the analysis of the estimator within an equivalent data model under the assumption of no matrix masking, i.e., setting $\bM=\bI_{n}$ in~\eqref{eq:TM2.noise}:
\begin{equation}
\bW= \bW^* + \rev{\sigma_x} \bU, \qquad \by = \by^* + \rev{\sigma_y} \bv,
\end{equation}  
where $\bU$ and $\bv$ are $n \times p$ and $n\times 1$ matrices with each entry generated from $N(0, 1)$ distribution, with known \rev{$\sigma_x^2$ and $\sigma_y^2$}. In this case, we propose an estimator $(\hat{\bm\theta}^{(c)}, \hat{\phi}^{(c)})$ for $(\bm\theta, \phi)$ to be the solution of the corrected estimating equations
\begin{equation}
\begin{aligned}
  \sum_{i=1}^{n}\bm{m}_{i, \bm\theta}(\bm\theta, \phi) & 
  = \sum_{i=1}^{n} \left\{\widetilde{\bW}_i^\T y_i - \dfrac{1}{\phi} (\widetilde{\bW}_i\widetilde{\bW}_i^\T - \rev{\sigma_x^2}\bJ_{p+q}) \bm\theta \right\} = \boldsymbol{0},
\\ 
\sum_{i=1}^{n}m_{i,\phi}(\bm\theta, \phi) 
& = \sum_{i=1}^{n} \left\{\dfrac{1}{2\phi} - \dfrac{1}{2} \left(y_i^2 - \rev{\sigma_y^2} \right) +  \dfrac{1}{2\phi^2}\bm\theta^\top \left(\widetilde{\bW}_i\widetilde{\bW}_i^\T - \rev{\sigma_x^2} \bJ_{p+q} \right)\bm\theta     \right\} = 0,
\end{aligned}
\label{eq:esteqmasked}
\end{equation}
where $\bJ_{p+q} = \text{diag}(\boldsymbol{0}, \bI_{p+q})$.  
The explicit formulas for the resulting estimator are
\begin{equation}
\begin{aligned}
\hat\phi^{(c)} &= \dfrac{n}{\Vert \bm{y}\Vert_2^2 - n\rev{\sigma_y^2} - \bm{y}^\T \widetilde{\bW}(\widetilde{\bW}^{\T}\widetilde{\bW} - n\rev{\sigma_x^2} \bJ_{p+q})^{-1} \widetilde{\bW}^\T \bm{y}}, \\ \hat{\bm\theta}^{(c)} &= \hat\phi^{(c)} (\widetilde{\bW}^{\T}\widetilde{\bW} - n\rev{\sigma_x^2} \bJ_{p+q})^{-1} \widetilde{\bW}^\T \bm{y},
\end{aligned}
\label{eq:correctedestimator}
\end{equation}
and the final estimator $\hat{\bm\beta}_1^{(c)}$ for $\bm\beta_1$ is the subvector of $\hat{\bm\theta}^{(c)}$ consisting of $p$ elements, from the second to the $(p+1)$-th element.  With this estimating equation approach, we can leverage the theory of estimating equations, see for example \citet{carroll1998local}, to establish the asymptotic properties of the proposed corrected estimator. 
Since the variances of the added noises, \rev{$\sigma_x^2$ and $\sigma_y^2$}, determine the level of privacy protection, they can become large under strict privacy requirements. Consequently, we derive the theoretical results under an asymptotic framework where both the sample size, $n$, and the noise variance, \rev{$\sigma^2 = \max(\sigma_x^2, \sigma_y^2)$}, diverge to infinity.
 
For these derivations, we assume that the raw data $(\bXstar_i, \bZstar_i, y^*_i)$ are independently and identically distributed according to the mixture model specified in \eqref{conditionmixturemodel}-\eqref{eq:logisticZ}, subject to the following technical conditions:
\begin{condition}
The marginal expectation and variance-covariance matrix $E(\bZ^*_i)$ and $\Var(\bZ^*_i)$ are finite. 
\label{cond:marginalmeanandvarianceZ}
\end{condition}
\begin{condition}
The parameter space for $\bm{\Xi} = \{\bmu_0, \bmu_1, \bC, \bSigma, \gamma_0, \bgamma_1\}$ is compact and the marginal variance-covariance matrix of $\Var(\bW^*_i)$ is not degenerate .
\label{cond:parameterspace}
\end{condition}
These two mild conditions imply that the marginal covariance matrix of $\bW_i^* = (\bXstar_i, \bZstar_i)$ are finite, and condition \ref{cond:parameterspace} also implies the parameter space for $(\bm\theta, \phi)$ is compact. To establish the asymptotic properties of the corrected estimator $\bm\beta_1^{(c)}$, we first show that the corresponding equations are unbiased in the following proposition whose proof is in the Supplementary Materials S1. 

\begin{proposition}
At the true parameter $\bm\theta$ and $\phi$, $E\left\{\bm{m}_{i,\bm\theta}(\bm\theta, \bm\phi)\right\} =E\left\{m_{i,\phi}(\bm\theta, \phi)\right\} = \bm{0}$.
\label{prop:unbiasedscore}
\end{proposition}
We note that when $\sigma = 0$, the estimating equations $\bm{m}_{i,\bm\theta}(\bm\theta, \bm\phi)$ and $m_{i,\phi}(\bm\theta, \phi)$ become $\bm{m}_i^*(\bm\theta, \bm\phi)$ and $m^*_{i,\phi}(\bm\theta, \phi) $ defined in \eqref{eq:scorefunctionsraw}, respectively, so Proposition \ref{prop:unbiasedscore} also establishes the unbiasedness of these estimating equations when the raw data are available.

\begin{theorem}
Under conditions \ref{cond:marginalmeanandvarianceZ}-\ref{cond:parameterspace}, if $\sigma^4/n \to 0$ as $n \to \infty$ where $\sigma = \max\{\sigma_x, \sigma_y \}$, then the estimator $\hat{\bm\beta}_1^{(c)}$ is a consistent estimator for $\bm\beta_1$, and
$$
\sqrt{n}(\hat{\bm\beta}_1^{(c)} - \bm\beta_1) \sim N\left(\boldsymbol{0},  \bV\right),
$$
where $\bV$ is the middle $p \times p$ sub-matrix of the matrix $\mathbf{A}^{-1} \mathbf{B} \mathbf{A}^{-1}$, with $\mathbf{A} = E_{\bm\theta,\phi} \left[\nabla \bm{m}_1(\bm\theta, \phi)\right]$, and $\mathbf{B} = E_{\bm\theta,\phi} \left[ \bm{m}_1(\bm\theta, \phi) \bm{m}_1(\bm\theta, \phi)^\T\right]$.
\label{thm: asymptotic.normality.correctedestimator}
\end{theorem}
Proof of the theorem is provided in the Supplementary Materials S1.

Based on Theorem \ref{thm: asymptotic.normality.correctedestimator}, we can  compute the standard error of $\hat{\bm\beta}_1$ from estimating $\bV$.  Specifically, the matrix $\bA$ and $\bB$ in the formula for $\bV$ can be estimated by 
$$
\widehat{\mathbf{A}} = \begin{bmatrix}
\widehat{\bA}_{11} & \widehat{\bA}_{12} \\
\widehat{\bA}_{21} & \widehat{\bA}_{22}
\end{bmatrix}, 
$$
with
\begin{alignat*}{2}
& \widehat{\bA}_{11} = \frac{1}{n}\sum_{i=1}^{n} \dfrac{\partial \bm{m}_{i,\bm\theta}} {\partial \bm\theta} \bigg|_{(\bm\theta,\phi) = (\hat{\bm\theta},\hat\phi)} && = 
\dfrac{1}{n\hat{\phi}} (\widetilde{\bW}^{\T}\widetilde{\bW} - n \rev{\sigma_x^2}\bJ_{p+q}), \\
& \widehat{\bA}_{12} = \widehat{\bA}_{21}^\top = \frac{1}{n} \sum_{i=1}^{n} \dfrac{\partial \bm{m}_{i,\bm\theta}} {\partial \phi} \bigg|_{(\bm\theta,\phi) = (\hat{\bm\theta},\hat\phi)} && =  \dfrac{1}{n\hat{\phi}^2}(\widetilde{\bW}^{\T}\widetilde{\bW} - n \rev{\sigma_x^2} \bJ_{p+q}) \hat{\bm\theta}, \\ 
& \widehat{\bA}_{22} = \frac{1}{n} \sum_{i=1}^{n} \dfrac{\partial \bm{m}_{i,\phi}} {\partial \phi} \bigg|_{(\bm\theta,\phi) = (\hat{\bm\theta},\hat\phi)} && = \dfrac{1}{2\hat{\phi}^2} + \dfrac{1}{n\hat{\phi}^3} \hat{\bm\theta}^\top (\widetilde{\bW}^\T\widetilde{\bW} - n \rev{\sigma_x^2} \bJ_{p+q}) \hat{\bm\theta},
\end{alignat*}
and 
$$
\widehat{\mathbf{B}} = \frac{1}{n} \sum_{i=1}^{n} \bm{m}_i(\hat{\bm\theta}, \hat{\phi})\bm{m}_i^\T(\hat{\bm\theta}, \hat{\phi}).
$$
Then, a $100(1-\alpha)\%$ asymptotic confidence interval for each component $\beta_{1j}$ of $\bm\beta_1$ is
$$
\hat\beta^{(c)}_{1j} \pm z_{\alpha/2} \sqrt{v_{j+1, j+1}}.
$$
where $z_{\alpha/2}$ denotes the $1-\alpha$ quantile of the standard normal distribution, for $j=1,\ldots, p$. 

\section{Simulation}\label{sec:simulation}

We conducted simulation studies to demonstrate the performance of the proposed estimation procedure with varying variances \rev{$\sigma_x^2$ and $\sigma_y^2$ for random noises added to the covariates and responses, respectively.} We set $p=3$ and generated the raw data $(\bX_i^*, y_i^*)$ from the mixture normal model \eqref{eq: conddistribution} and $(\bW_i^*, y_i^*)$ from the conditional mixture model \eqref{conditionmixturemodel}-\eqref{eq:logisticZ} for $i=1,\ldots,n$. For the mixture model \eqref{eq: conddistribution}, we vary the probability $p_1 \in \{0.1, 0.5, 0.9\}$, set the true vector $\bm\mu_1 = (1, 1, 1)$, and the covariance matrix $\bm\Sigma$ to have an AR(1) structure with autocorrelation $0.5$. We set $\bm\beta_1 = (1, -1, 0)^\top$ and computed $\bm\mu_0 = \bm\mu_1 - \bm\Sigma \bbeta$. For the conditional mixture model \eqref{conditionmixturemodel}-\eqref{eq:logisticZ}, we set $q=2$ and generated the true $\bZ_i^*$  from a bivariate uniform distribution on $[-1,1]^{q}$, and set the matrix $\bC$ to be a random $q \times p$ matrix with each element from a $\text{uniform}(1,2)$ distribution. We set $\gamma_0 = 0$ and $\gamma_1 = (1.5, 1)^\top$.

After generating the raw data, the masked covariates $\bX$ or $\bW = [\bX, \bZ]$ and the masked response $\by$ were generated by adding random noises to the raw data matrices and masking with a random orthogonal matrix. That is, we generated $\bW$ and $\by$ according to model~\eqref{eq:TM2.noise}: \rev{$\bW = \bM \bW^* + \sigma_x \bU$ and $\by = \bM \by^* + \sigma_y \bv$}, where $\bU$ and $\bv$ are $n \times p$ matrix and $n \times 1$ vector of independent standard normal random variables. \rev{We varied the sample size $n \in \{10^4, 10^5,  10^6\}$, and selected the values for $\sigma_x^2$ and $\sigma_y^2$ to satisfy three $(\varepsilon,0.01)$ levels of differential privacy (DP)  with $\varepsilon \in \{5, 1, 0.5\}$. We provided a brief review of DP and the DP guarantees of the matrix masking and noise addition mechanism in Appendix~\ref{sec:priv_level_sigma}. Since the outcome $y^*$ has a different range from the covariates $\bX^*$ or $\bW^*$, different values of $\sigma_x$ and $\sigma_y$ were needed to achieve the same $(\varepsilon,0.01)$-DP at the same sample size. The $\sigma_x$ and $\sigma_y$ corresponding to the $(\varepsilon,0.01)$-DP were reported in the second and the third columns of Tables \ref{tab:sim_results_mixturemodel} and \ref{tab:sim_results_conditionalmixturemodel}.} 

On each simulated dataset, we computed several point and interval estimators. First, the naive conditional MLE estimator is defined to be the solution of the estimating equation $\bS(\beta_0, \bm\beta_1; \bX, \by) = \mathbf{0}$,  obtained by naively substituting the masked data ($\widetilde{\bX}$, $\by$)  in place of the raw data ($\widetilde{\bX}^*$, $\by^*$)  in equation \eqref{eq:scoreMLEraw}. The corresponding confidence intervals for this naive conditional MLE were computed using the asymptotic Wald method.
Next, we computed the naive least squaress estimator (LS) by similarly substitution of masked data in \eqref{eq:rawestimators}.  Finally, we computed our corrected least squaress (cLS) estimator in \eqref{eq:correctedestimator}, which accounts for the effects of noise addition and masking.

We evaluated the performance of the point estimators by examining their bias and median squared errors (MSE) and evaluate the confidence intervals based on their coverage probability across $1000$ datasets per simulation settings.  \rev{When $\sigma_x = \sigma_y=0$, corresponding to $\varepsilon = \infty$}, the raw data and masked data are identical, making the naive and corrected least squaress estimators equivalent. This setup provides a baseline for comparing the performance of the conditional MLE and the LS estimator on raw data. When noise is added ($\varepsilon<\infty$) for privacy protection, data utility decreases, resulting in a degradation of estimator performance from this baseline. We reported the results for the unconditional mixture model normal distribution \eqref{eq: conddistribution} with $p_1 = 0.5$  in Table \ref{tab:sim_results_mixturemodel} and the conditional mixture model \eqref{conditionmixturemodel}-\eqref{eq:logisticZ} in Table \ref{tab:sim_results_conditionalmixturemodel}. The results for $p_1 \in \{0.1, 0.9\}$ show similar conclusions and are presented in the Supplementary Materials S2.  

\begin{table}[ht]
\centering
\caption{\rev{Simulation results when the raw data were generated from the mixture normal distribution~\eqref{eq: conddistribution}. Reported are bias, median squared error (MSE) for point estimators (MLE: naive conditional MLE, LS: naive least squares, and cLS: corrected least squares), and coverage probability for the confidence intervals corresponding to each coefficient of the slope vector.}}
\label{tab:sim_results_mixturemodel}

\resizebox{!}{.42\textheight}{\begin{tabular}{llrrlrrrrrrrrr}
\toprule
$\epsilon$ & \rev{$\sigma_x$} &\rev{ $\sigma_y$} & $n$ & Coef.  & \multicolumn{3}{c}{Bias } & \multicolumn{3}{c}{MSE } & \multicolumn{3}{c}{Coverage} \\
\cmidrule(lr){6-8} \cmidrule(lr){9-11} \cmidrule(lr){12-14}
& & & & &  MLE & LS & cLS & MLE & LS & cLS & MLE & LS & cLS \\
\midrule

$\infty$ & 0 & 0 & $10^4$ & $\beta_1$ & 0.00 & 0.00 & 0.00 & 0.00 & 0.00 & 0.00 & 0.93 & 0.94 & 0.94\\
 &  &  &  & $\beta_2$ & 0.00 & 0.00 & 0.00 & 0.00 & 0.00 & 0.00 & 0.95 & 0.96 & 0.96\\
 &  &  &  & $\beta_3$ & 0.00 & 0.00 & 0.00 & 0.00 & 0.00 & 0.00 & 0.96 & 0.95 & 0.95\\
\addlinespace[0.5pt]
 & 0 & 0 & $10^5$ & $\beta_1$ & 0.00 & 0.00 & 0.00 & 0.00 & 0.00 & 0.00 & 0.96 & 0.96 & 0.96\\
 &  &  &  & $\beta_2$ & 0.00 & 0.00 & 0.00 & 0.00 & 0.00 & 0.00 & 0.97 & 0.97 & 0.97\\
 &  &  &  & $\beta_3$ & 0.00 & 0.00 & 0.00 & 0.00 & 0.00 & 0.00 & 0.95 & 0.95 & 0.95\\
\addlinespace[0.5pt]
 & 0 & 0 & $10^6$ & $\beta_1$ & 0.00 & 0.00 & 0.00 & 0.00 & 0.00 & 0.00 & 0.96 & 0.96 & 0.96\\
 &  &  &  & $\beta_2$ & 0.00 & 0.00 & 0.00 & 0.00 & 0.00 & 0.00 & 0.97 & 0.97 & 0.97\\
 &  &  &  & $\beta_3$ & 0.00 & 0.00 & 0.00 & 0.00 & 0.00 & 0.00 & 0.96 & 0.97 & 0.97\\
\addlinespace[0.5pt]
5 & 4.20 & 0.60 & $10^4$ & $\beta_1$ & 0.97 & 0.99 & 0.01 & 0.94 & 0.98 & 0.20 & 0.00 & 0.00 & 0.95\\
 &  &  &  & $\beta_2$ & -0.97 & -0.99 & -0.05 & 0.95 & 0.98 & 0.29 & 0.00 & 0.00 & 0.93\\
 &  &  &  & $\beta_3$ & 0.01 & 0.01 & 0.12 & 0.00 & 0.00 & 0.19 & 0.29 & 0.52 & 1.00\\
\addlinespace[0.5pt]
 & 4.18 & 0.60 & $10^5$ & $\beta_1$ & 0.97 & 0.99 & -0.02 & 0.95 & 0.98 & 0.02 & 0.00 & 0.00 & 0.96\\
 &  &  &  & $\beta_2$ & -0.97 & -0.99 & 0.02 & 0.95 & 0.98 & 0.03 & 0.00 & 0.00 & 0.96\\
 &  &  &  & $\beta_3$ & 0.01 & 0.01 & -0.02 & 0.00 & 0.00 & 0.01 & 0.00 & 0.00 & 0.98\\
\addlinespace[0.5pt]
 & 4.14 & 0.59 & $10^6$ & $\beta_1$ & 0.97 & 0.99 & -0.01 & 0.95 & 0.98 & 0.00 & 0.00 & 0.00 & 0.96\\
 &  &  &  & $\beta_2$ & -0.97 & -0.99 & 0.00 & 0.95 & 0.98 & 0.00 & 0.00 & 0.00 & 0.96\\
 &  &  &  & $\beta_3$ & 0.01 & 0.01 & 0.00 & 0.00 & 0.00 & 0.00 & 0.00 & 0.00 & 0.95\\
\addlinespace[0.5pt]
1 & 9.51 & 1.36 & $10^4$ & $\beta_1$ & 0.99 & 1.00 & 1.00 & 0.99 & 1.00 & 1.32 & 0.00 & 0.00 & 0.80\\
 &  &  &  & $\beta_2$ & -0.99 & -1.00 & -0.98 & 0.99 & 1.00 & 1.22 & 0.00 & 0.00 & 0.82\\
 &  &  &  & $\beta_3$ & 0.00 & 0.00 & 0.08 & 0.00 & 0.00 & 0.31 & 0.45 & 0.93 & 1.00\\
\addlinespace[0.5pt]
 & 9.34 &1.33  & $10^5$ & $\beta_1$ & 0.99 & 1.00 & 0.23 & 0.99 & 1.00 & 0.46 & 0.00 & 0.00 & 0.94\\
 &  &  &  & $\beta_2$ & -0.99 & -1.00 & -0.27 & 0.99 & 1.00 & 0.69 & 0.00 & 0.00 & 0.92\\
 &  &  &  & $\beta_3$ & 0.00 & 0.00 & 0.26 & 0.00 & 0.00 & 0.49 & 0.21 & 0.68 & 1.00\\
\addlinespace[0.5pt]
 & 9.27 & 1.32 & $10^6$ & $\beta_1$ & 0.99 & 1.00 & -0.05 & 0.99 & 1.00 & 0.04 & 0.00 & 0.00 & 0.95\\
 &  &  &  & $\beta_2$ & -0.99 & -1.00 & 0.02 & 0.99 & 1.00 & 0.05 & 0.00 & 0.00 & 0.95\\
 &  &  &  & $\beta_3$ & 0.00 & 0.00 & 0.02 & 0.00 & 0.00 & 0.02 & 0.00 & 0.01 & 0.98\\
\addlinespace[0.5pt]
0.5 & 13.4 & 1.92 & $10^4$ & $\beta_1$ & 1.00 & 1.00 & 1.03 & 0.99 & 1.00 & 1.19 & 0.00 & 0.00 & 0.80\\
 &  &  &  & $\beta_2$ & -1.00 & -1.00 & -1.08 & 0.99 & 1.00 & 1.36 & 0.00 & 0.00 & 0.78\\
 &  &  &  & $\beta_3$ & 0.00 & 0.00 & 0.06 & 0.00 & 0.00 & 0.18 & 0.38 & 0.94 & 1.00\\
\addlinespace[0.5pt]
 & 13.2 & 1.89 & $10^5$ & $\beta_1$ & 1.00 & 1.00 & 0.94 & 0.99 & 1.00 & 1.38 & 0.00 & 0.00 & 0.84\\
 &  &  &  & $\beta_2$ & -1.00 & -1.00 & -0.91 & 0.99 & 1.00 & 1.23 & 0.00 & 0.00 & 0.85\\
 &  &  &  & $\beta_3$ & 0.00 & 0.00 & 0.27 & 0.00 & 0.00 & 0.48 & 0.28 & 0.89 & 1.00\\
\addlinespace[0.5pt]
 & 13.1 & 1.87 & $10^6$ & $\beta_1$ & 1.00 & 1.00 & -0.01 & 0.99 & 1.00 & 0.14 & 0.00 & 0.00 & 0.94\\
 &  &  &  & $\beta_2$ & -1.00 & -1.00 & -0.10 & 0.99 & 1.00 & 0.20 & 0.00 & 0.00 & 0.93\\
 &  &  &  & $\beta_3$ & 0.00 & 0.00 & 0.08 & 0.00 & 0.00 & 0.12 & 0.02 & 0.32 & 0.98\\
 \bottomrule
\end{tabular}}
\end{table}

% latex table generated in R 4.3.0 by xtable 1.8-4 package
% Thu Apr  4 17:30:03 2024
\begin{table}[ht]
\centering
\caption{\rev{Simulation results when the raw data were generated from the (conditional) mixture model~\eqref{conditionmixturemodel}-\eqref{eq:logisticZ}. Reported are bias, median squared error (MSE) for point estimators (MLE: naive conditional MLE, LS: naive least squares, and cLS: corrected least squares), and coverage probability for the confidence interval corresponding to each coefficient of the slope vector.}}
\label{tab:sim_results_conditionalmixturemodel}
\resizebox{!}{.42\textheight}{\begin{tabular}{llrrlrrrrrrrrr}
\toprule
$\epsilon$ & \rev{$\sigma_x$} &\rev{ $\sigma_y$} & $n$ & Coef.  & \multicolumn{3}{c}{Bias } & \multicolumn{3}{c}{MSE } & \multicolumn{3}{c}{Coverage} \\
\cmidrule(lr){6-8} \cmidrule(lr){9-11} \cmidrule(lr){12-14}
& & & & &  MLE & LS & cLS & MLE & LS & cLS & MLE & LS & cLS \\
\midrule
$\infty$ & 0 & 0 & $10^4$ & $\beta_1$ & 0.00 & 0.00 & 0.00 & 0.00 & 0.00 & 0.00 & 0.94 & 0.94 & 0.94\\
 &  &  &  & $\beta_2$ & 0.00 & 0.00 & 0.00 & 0.00 & 0.00 & 0.00 & 0.96 & 0.95 & 0.95\\
 &  &  &  & $\beta_3$ & 0.00 & 0.00 & 0.00 & 0.00 & 0.00 & 0.00 & 0.97 & 0.97 & 0.97\\
\addlinespace[0.5pt]
 &  0& 0  & $10^5$ & $\beta_1$ & 0.00 & 0.00 & 0.00 & 0.00 & 0.00 & 0.00 & 0.93 & 0.93 & 0.93\\
 &  &  &  & $\beta_2$ & 0.00 & 0.00 & 0.00 & 0.00 & 0.00 & 0.00 & 0.95 & 0.96 & 0.96\\
 &  &  &  & $\beta_3$ & 0.00 & 0.00 & 0.00 & 0.00 & 0.00 & 0.00 & 0.96 & 0.96 & 0.96\\
\addlinespace[0.5pt]
 &  0&0  & $10^6$ & $\beta_1$ & 0.00 & 0.00 & 0.00 & 0.00 & 0.00 & 0.00 & 0.96 & 0.96 & 0.96\\
 &  &  &  & $\beta_2$ & 0.00 & 0.00 & 0.00 & 0.00 & 0.00 & 0.00 & 0.95 & 0.95 & 0.95\\
 &  &  &  & $\beta_3$ & 0.00 & 0.00 & 0.00 & 0.00 & 0.00 & 0.00 & 0.95 & 0.94 & 0.94\\
\addlinespace[0.5pt]
5 & 4.55 & 0.65 & $10^4$ & $\beta_1$ & -0.97 & -0.99 & -0.29 & 0.94 & 0.98 & 0.63 & 0.00 & 0.00 & 0.97\\
 &  &  &  & $\beta_2$ & 0.99 & 1.00 & 0.51 & 0.97 & 0.99 & 1.13 & 0.00 & 0.00 & 0.97\\
 &  &  &  & $\beta_3$ & 0.00 & 0.00 & -0.16 & 0.00 & 0.00 & 0.42 & 0.57 & 0.87 & 1.00\\
\addlinespace[0.5pt]
 & 4.53 & 0.65 & $10^5$ & $\beta_1$ & -0.97 & -0.99 & 0.02 & 0.94 & 0.98 & 0.14 & 0.00 & 0.00 & 0.99\\
 &  &  &  & $\beta_2$ & 0.98 & 1.00 & 0.00 & 0.97 & 0.99 & 0.19 & 0.00 & 0.00 & 0.99\\
 &  &  &  & $\beta_3$ & 0.00 & 0.00 & 0.03 & 0.00 & 0.00 & 0.14 & 0.28 & 0.46 & 1.00\\
\addlinespace[0.5pt]
 & 4.52 & 0.64 & $10^6$ & $\beta_1$ & -0.97 & -0.99 & 0.01 & 0.94 & 0.98 & 0.01 & 0.00 & 0.00 & 0.97\\
 &  &  &  & $\beta_2$ & 0.98 & 1.00 & -0.03 & 0.97 & 0.99 & 0.01 & 0.00 & 0.00 & 0.96\\
 &  &  &  & $\beta_3$ & 0.00 & 0.00 & 0.00 & 0.00 & 0.00 & 0.01 & 0.11 & 0.17 & 0.98\\
\addlinespace[0.5pt]
1 & 10.17 & 1.45 & $10^4$ & $\beta_1$ & -0.99 & -1.00 & -1.02 & 0.99 & 1.00 & 1.25 & 0.00 & 0.00 & 0.84\\
 &  &  & & $\beta_2$ & 1.00 & 1.00 & 1.05 & 1.00 & 1.00 & 1.27 & 0.00 & 0.00 & 0.83\\
 &  &  &  & $\beta_3$ & 0.00 & 0.00 & 0.08 & 0.00 & 0.00 & 0.22 & 0.47 & 0.93 & 1.00\\
\addlinespace[0.5pt]
 &  10.13 &  1.45 & $10^5$ & $\beta_1$ & -0.99 & -1.00 & -0.65 & 0.99 & 1.00 & 1.03 & 0.00 & 0.00 & 0.94\\
 &  &  &  & $\beta_2$ & 1.00 & 1.00 & 0.79 & 1.00 & 1.00 & 1.38 & 0.00 & 0.00 & 0.94\\
 &  &  &  & $\beta_3$ & 0.00 & 0.00 & -0.15 & 0.00 & 0.00 & 0.54 & 0.37 & 0.89 & 1.00\\
\addlinespace[0.5pt]
 &  10.11 &    1.44 & $10^6$ & $\beta_1$ & -0.99 & -1.00 & -0.04 & 0.99 & 1.00 & 0.30 & 0.00 & 0.00 & 0.99\\
 &  &  &  & $\beta_2$ & 1.00 & 1.00 & 0.14 & 1.00 & 1.00 & 0.48 & 0.00 & 0.00 & 0.98\\
 &  &  &  & $\beta_3$ & 0.00 & 0.00 & 0.00 & 0.00 & 0.00 & 0.26 & 0.18 & 0.55 & 1.00\\
\addlinespace[0.5pt]
0.5 & 14.38 & 2.05 & $10^4$ & $\beta_1$ & -1.00 & -1.00 & -1.00 & 0.99 & 1.00 & 1.19 & 0.00 & 0.00 & 0.81\\
 &  &  &  & $\beta_2$ & 1.00 & 1.00 & 1.02 & 1.00 & 1.00 & 1.26 & 0.00 & 0.00 & 0.85\\
 &  &  &  & $\beta_3$ & 0.00 & 0.00 & 0.04 & 0.00 & 0.00 & 0.19 & 0.32 & 0.95 & 1.00\\
\addlinespace[0.5pt]
 & 14.32 & 2.05  & $10^5$ & $\beta_1$ & -1.00 & -1.00 & -1.01 & 0.99 & 1.00 & 1.15 & 0.00 & 0.00 & 0.86\\
 &  &  &  & $\beta_2$ & 1.00 & 1.00 & 1.05 & 1.00 & 1.00 & 1.35 & 0.00 & 0.00 & 0.83\\
 &  &  &  & $\beta_3$ & 0.00 & 0.00 & -0.06 & 0.00 & 0.00 & 0.26 & 0.31 & 0.95 & 1.00\\
\addlinespace[0.5pt]
 & 14.30 & 2.04  & $10^6$ & $\beta_1$ & -1.00 & -1.00 & -0.23 & 0.99 & 1.00 & 0.54 & 0.00 & 0.00 & 0.99\\
 &  &  &  & $\beta_2$ & 1.00 & 1.00 & 0.48 & 1.00 & 1.00 & 0.99 & 0.00 & 0.00 & 0.98\\
 &  &  &  & $\beta_3$ & 0.00 & 0.00 & -0.16 & 0.00 & 0.00 & 0.59 & 0.22 & 0.77 & 1.00\\
\bottomrule
\end{tabular}}
\end{table}

\begin{table}[ht]
\centering
\caption{\rev{\emph{Sensitivity analysis}: Simulation results when the raw data were generated from the logistic regression model ~\eqref{eq:logistic} when $\bX^* \sim N_3 (0, \bm\Sigma)$. Reported are bias, MSE for point estimators (MLE: naive conditional MLE, LS: naive least squares, and cLS: corrected least squares), and coverage probability for the confidence interval corresponding to each coefficient of the slope vector.}}
\label{tab:sim_results_sensitivity}

\resizebox{!}{.42\textheight}{\begin{tabular}{llrrlrrrrrrrrr}
\toprule
$\epsilon$ & \rev{$\sigma_x$} &\rev{ $\sigma_y$} & $n$ & Coef.  & \multicolumn{3}{c}{Bias} & \multicolumn{3}{c}{MSE} & \multicolumn{3}{c}{Coverage} \\
\cmidrule(lr){6-8} \cmidrule(lr){9-11} \cmidrule(lr){12-14}
& & & & &  MLE & LS & cLS & MLE & LS & cLS & MLE & LS & cLS \\
\midrule
$\infty$ & 0 & 0 & $10^4$ & $\beta_1$ & 0.00 & 0.00 & 0.00 & 0.00 & 0.00 & 0.00 & 0.95 & 0.95 & 0.95\\
 &  &  &  & $\beta_2$ & 0.00 & 0.00 & 0.00 & 0.00 & 0.00 & 0.00 & 0.95 & 0.95 & 0.95\\
 &  &  &  & $\beta_3$ & 0.00 & 0.00 & 0.00 & 0.00 & 0.00 & 0.00 & 0.96 & 0.96 & 0.96\\
\addlinespace[0.5pt]
 &0  & 0 & $10^5$ & $\beta_1$ & 0.00 & 0.00 & 0.00 & 0.00 & 0.00 & 0.00 & 0.95 & 0.92 & 0.92\\
 &  &  &  & $\beta_2$ & 0.00 & 0.00 & 0.00 & 0.00 & 0.00 & 0.00 & 0.96 & 0.92 & 0.92\\
 &  &  &  & $\beta_3$ & 0.00 & 0.00 & 0.00 & 0.00 & 0.00 & 0.00 & 0.96 & 0.96 & 0.96\\
\addlinespace[0.5pt]
 & 0 & 0 & $10^6$ & $\beta_1$ & 0.00 & 0.00 & 0.00 & 0.00 & 0.00 & 0.00 & 0.94 & 0.82 & 0.82\\
 &  &  &  & $\beta_2$ & 0.00 & 0.00 & 0.00 & 0.00 & 0.00 & 0.00 & 0.97 & 0.85 & 0.85\\
 &  &  &  & $\beta_3$ & 0.00 & 0.00 & 0.00 & 0.00 & 0.00 & 0.00 & 0.95 & 0.94 & 0.94\\
\addlinespace[0.5pt]
5 & 4.20 & 0.60 & $10^4$ & $\beta_1$ & 0.98 & 0.99 & 0.11 & 0.95 & 0.98 & 0.26 & 0.00 & 0.00 & 0.92\\
 &  &  &  & $\beta_2$ & -0.98 & -0.99 & -0.12 & 0.95 & 0.98 & 0.32 & 0.00 & 0.00 & 0.93\\
 &  &  &  & $\beta_3$ & 0.01 & 0.00 & 0.11 & 0.00 & 0.00 & 0.15 & 0.40 & 0.68 & 1.00\\
\addlinespace[0.5pt]
 & 4.18 & 0.60 & $10^5$ & $\beta_1$ & 0.98 & 0.99 & -0.03 & 0.95 & 0.98 & 0.02 & 0.00 & 0.00 & 0.94\\
 &  &  &  & $\beta_2$ & -0.98 & -0.99 & 0.02 & 0.95 & 0.98 & 0.03 & 0.00 & 0.00 & 0.94\\
 &  &  &  & $\beta_3$ & 0.01 & 0.00 & -0.01 & 0.00 & 0.00 & 0.01 & 0.00 & 0.01 & 0.96\\
\addlinespace[0.5pt]
 & 4.14 & 0.59 & $10^6$ & $\beta_1$ & 0.98 & 0.99 & 0.01 & 0.95 & 0.98 & 0.00 & 0.00 & 0.00 & 0.94\\
 &  &  &  & $\beta_2$ & -0.98 & -0.99 & -0.01 & 0.96 & 0.98 & 0.00 & 0.00 & 0.00 & 0.94\\
 &  &  &  & $\beta_3$ & 0.01 & 0.00 & 0.00 & 0.00 & 0.00 & 0.00 & 0.00 & 0.00 & 0.94\\
\addlinespace[0.5pt]
1 & 9.51 & 1.36 & $10^4$ & $\beta_1$ & 1.00 & 1.00 & 0.98 & 0.99 & 1.00 & 1.15 & 0.00 & 0.00 & 0.81\\
 &  &  &  & $\beta_2$ & -1.00 & -1.00 & -1.00 & 0.99 & 1.00 & 1.29 & 0.00 & 0.00 & 0.81\\
 &  &  &  & $\beta_3$ & 0.00 & 0.00 & 0.06 & 0.00 & 0.00 & 0.29 & 0.50 & 0.95 & 1.00\\
\addlinespace[0.5pt]
 & 9.34 &1.33  & $10^5$ & $\beta_1$ & 1.00 & 1.00 & 0.32 & 0.99 & 1.00 & 0.54 & 0.00 & 0.00 & 0.91\\
 &  &  &  & $\beta_2$ & -1.00 & -1.00 & -0.39 & 0.99 & 1.00 & 0.74 & 0.00 & 0.00 & 0.90\\
 &  &  &  & $\beta_3$ & 0.00 & 0.00 & 0.24 & 0.00 & 0.00 & 0.48 & 0.26 & 0.75 & 0.99\\
\addlinespace[0.5pt]
 & 9.27 & 1.32 & $10^6$ & $\beta_1$ & 1.00 & 1.00 & -0.03 & 0.99 & 1.00 & 0.04 & 0.00 & 0.00 & 0.95\\
 &  &  &  & $\beta_2$ & -1.00 & -1.00 & 0.05 & 0.99 & 1.00 & 0.05 & 0.00 & 0.00 & 0.93\\
 &  &  &  & $\beta_3$ & 0.00 & 0.00 & -0.01 & 0.00 & 0.00 & 0.02 & 0.00 & 0.02 & 0.98\\
\addlinespace[0.5pt]
0.5 & 13.4 & 1.92 & $10^4$ & $\beta_1$ & 1.00 & 1.00 & 1.05 & 1.00 & 1.00 & 1.28 & 0.00 & 0.00 & 0.82\\
 &  &  &  & $\beta_2$ & -1.00 & -1.00 & -1.03 & 1.00 & 1.00 & 1.28 & 0.00 & 0.00 & 0.80\\
 &  &  &  & $\beta_3$ & 0.00 & 0.00 & 0.01 & 0.00 & 0.00 & 0.21 & 0.34 & 0.94 & 1.00\\
\addlinespace[0.5pt]
 & 13.2 & 1.89 & $10^5$ & $\beta_1$ & 1.00 & 1.00 & 0.91 & 0.99 & 1.00 & 1.22 & 0.00 & 0.00 & 0.79\\
 &  &  &  & $\beta_2$ & -1.00 & -1.00 & -1.02 & 1.00 & 1.00 & 1.35 & 0.00 & 0.00 & 0.79\\
 &  &  &  & $\beta_3$ & 0.00 & 0.00 & 0.18 & 0.00 & 0.00 & 0.35 & 0.33 & 0.89 & 1.00\\
\addlinespace[0.5pt]
 & 13.1 & 1.87 & $10^6$ & $\beta_1$ & 1.00 & 1.00 & 0.04 & 1.00 & 1.00 & 0.24 & 0.00 & 0.00 & 0.91\\
 &  &  &  & $\beta_2$ & -1.00 & -1.00 & -0.03 & 1.00 & 1.00 & 0.31 & 0.00 & 0.00 & 0.92\\
 &  &  &  & $\beta_3$ & 0.00 & 0.00 & 0.10 & 0.00 & 0.00 & 0.10 & 0.05 & 0.49 & 0.99\\
\bottomrule
\end{tabular}}
\end{table}

Both tables demonstrate that, for raw data \rev{($\sigma_x = \sigma_y = 0$)}, the conditional MLE and LS estimators exhibit comparable performance for both point and interval estimation across all sample sizes considered. \rev{At any considered level of privacy $\varepsilon \in \{5, 1, 0.5\}$, both the MLE and the naive least squares estimators set estimates to zero, evidenced by bias and MSE being approximately equal to the true value and its square of each coefficient, respectively. As a result, these estimators only have the smallest MSE for a true zero coefficient, such as $\beta_3$, while the corresponding coverage is 1.  

For non-zero coefficients $\beta_1$ and $\beta_2$, the proposed corrected LS has the smallest bias and MSE when $\varepsilon \in \{5, 1\}$, and it is the only estimator with both bias and MSE decreasing when the sample size increases. Turning to the confidence intervals, the MLE and naive LS estimator have zero coverage when computed from the masked data. On the other hand, the coverage probability of the corrected LS estimator approaches the nominal level when the sample size increases, validating the theoretical results in Section \ref{section: maskeddata}. 
Additionally, the cLS estimator has a slightly worse performance when the data were generated from the conditional mixture model \eqref{conditionmixturemodel}-\eqref{eq:logisticZ}, as shown in Table~\ref{tab:sim_results_conditionalmixturemodel}, compared to when data are generated from the simpler mixture model\eqref{eq: conddistribution}, as shown in Table~\ref{tab:sim_results_mixturemodel}, which is expected both because the former has a higher number of covariates and requires higher magnitudes for $\sigma_x$ and $\sigma_y$ to achieve the same level of privacy.}

\rev{Finally, we conducted a sensitivity analysis of the results when the raw data were not generated from the previous mixture models. Instead, we generated the raw data from the logistic regression model \eqref{eq:logistic} where the covariate $\bX^*_i$ was marginally simulated from a multivariate Gaussian distribution with zero mean and covariance $\bm{\Sigma}$, and the slope vector $\bbeta_1 = (1,-1,0)^\top$ as before. Table \ref{tab:sim_results_sensitivity} demonstrated that in the case where the raw data were available (i.e no privacy protection with $\varepsilon = \infty$), both the conditional MLE and LS estimators give equivalent performance in the sample size $n \in \{10^4, 10^5\}$. Nevertheless, at a large sample size such as when $n=10^6$, although the two point estimators have similar bias and MSE, the coverage probability of the confidence intervals corresponding to the LS estimators \emph{decreases}, showing that a violation of mixture model assumption may hurt the performance of the cLS confidence interval in this case. This result confirms a well-known practice that if the raw data were available, the conditional MLE estimator should be a default method to fit a logistic regression model. 
Nevertheless, when noises were added to the raw data to protect privacy, similar to what we observed in the previous simulation, the proposed cLS estimator is the only one that still maintains estimation consistency and has meaningful coverage probability for the non-zero coefficients $\beta_1$ and $\beta_2$ in all three considered levels of privacy.     }

%However, when the data are masked with a moderate level of noise, such as $\sigma \in \{0.3, 1\}$, the cLS estimator outperforms the naive conditional MLE and naive LS estimators by exhibiting significantly lower bias and mean square error (MSE).
%Moreover, the cLS estimator is the only one among the three whose MSE decreases as the sample size increases. For interval estimators, the confidence interval based on the cLS estimator consistently achieves correct coverage probability. In contrast, the confidence intervals corresponding to the conditional MLE and naive LS estimators have zero or very low coverage for non-zero coefficients $\beta_1$ and $\beta_2$.

%For higher noise levels, such as $\sigma = 1$ and $\sigma = 3$, the cLS estimator can become unstable when the sample size is small, as reflected in its inflated MSE and confidence intervals with suboptimal coverage. Nonetheless, as the sample size increases, the cLS estimator improves its performance in both MSE and coverage probability, unlike the other two estimators. Notably, an increasing $\sigma^2$ shrinks naive LS estimator toward zero, leading to its apparent low bias and MSE for true zero coefficients like $\beta_3 = 0$. However, this comes at the cost of deteriorating coverage probabilities for its confidence intervals as sample size increases.

In summary, for TM$^2$+Noise data, the cLS estimator is crucial for valid statistical inference. When noise is added to protect privacy, both the conditional MLE and naive LS estimators suffer from bias and low-coverage confidence intervals. In contrast, the cLS estimator generally achieves proper coverage, %except in smaller sample sizes ($n=1000$) when noise levels are high ($\sigma=3$). 
and generally,  the more stringent level of privacy (i.e. smaller $\epsilon$), a larger sample size is needed for the cLS estimator to yield statistically significant inferences.

\FloatBarrier
\section{Real data analysis}\label{sec:realdata}

We illustrate our estimation methods to study the relationship between an individual's health outcome and three personal characteristics. The raw data was obtained from the {\it All of US} Registered Tier Dataset v7 database (version R2022Q4R9, released on 04/20/2023). This database contains numerous private information of individuals and cannot be released directly to the public; our access to this database requires a strict license. Hence, the focus of this data analysis is to compare the results of a logistic regression obtained on the raw data versus those obtained on the masked data that may potentially be released to the public. 

For this analysis, we were interested in modeling the probability of hypertension as a function of three key factors, including gender, race, and age \citep{chandler2021hypertension}.  The raw dataset includes $169,772$ individuals with complete observations for all four variables. Among these individuals, $60,706$ had been diagnosed with hypertension, corresponding to a crude prevalence rate of  $35.8\%$. There is an imbalance in gender distribution, with $106,897$ females, representing $63.0\%$ of the population. Additionally, $32,914$ individuals ($19.4\%$) identify themselves as Black or African American. The mean age in the dataset is $51.9$ years, with a standard deviation of $16.6$ years. 

For the logistic regression, the response variable $y$ represents an individual’s hypertension status (Yes/No). Gender and race are binary covariates, where we set gender = 1 to indicate female and race = 1 to indicate Black or African American. To ensure comparability between the continuous and binary covariates, the age variable was scaled to a range between $0$ and $1$. Given that the actual ages in the dataset range from 18 to 88 years, we applied the transformation $x_{age}=(age-18)/70$ and used it as a covariate in the logistic model. 

We first performed logistic regression analysis on the raw dataset  \rev{($\varepsilon=\infty$, i.e., $\sigma_x = \sigma_y=0$)}, and to establish a baseline for scenarios without privacy protection, we compared the least squares (LS) estimates to standard maximum likelihood estimates (MLE) on this raw data. \rev{Next, we generated pseudo-datasets by adding noise at three levels after applying random matrix masking following \eqref{eq:TM2.noise}. Since  we scaled the age variable resulting in the same range of [0,1] for all the covariates and response for raw data set, we set $\sigma_x = \sigma_y = \sigma$, and chose the value of $\sigma$ to achieve $(\varepsilon,0.01)$-DP as described in Section~\ref{sec:priv_level_sigma}. For $\varepsilon \in\{5, 1, 0.5\}$, the corresponding value for $\sigma$ is $\{0.59, 1.33, 1.87\}$}.  %whose values are listed in Table~\ref{tab:realdata} of Appendix~\ref{sec:priv_level_sigma}.}

Logistic regression was then performed on these pseudo-datasets using our corrected least squaress (cLS) estimator. Comparing the baseline results from the raw data to the cLS results on pseudo-datasets demonstrates the potential impact of privacy-protection mechanisms on analysis outcomes. It is important to note that standard MLE estimates for logistic regression cannot be applied to the pseudo-datasets because the binary response variable $y$ becomes real-valued due to the noise addition and matrix masking procedures.

\begin{table}[t]
\centering
\caption{\rev{Coefficient estimates from multiple logistic regression with three covariates on {\it All of US} raw data and \emph{one} pseudo-dataset obtained from TM$^2$+Noise scheme for each noise level to achieve $(\varepsilon,0.01)$-DP of varying $\varepsilon$ values.}} 
\begin{tabular}{ccccc}
  \toprule[1.5pt]
$x$ &  $\varepsilon$ & method  &  $\hat \bbeta_1$   & $95\%$ CI\\
  \hline
  \addlinespace
Gender & $\infty$ & MLE & -0.178 & (-0.200, -0.155) \\
 &  & LS & -0.187 & (-0.211, -0.164) \\
 & 5 & cLS &  -0.169 & (-0.242, -0.095) \\
  & 1 & cLS & -0.152 &  (-0.430, 0.127)\\
  & 0.5 & cLS & -0.144 &  (-0.679, 0.391) \\ [0.5em]
Race & $\infty$ & MLE & 0.772 & (0.745, 0.799) \\
 &  &  LS & 0.768 & (0.739, 0.797) \\
 & 5 & cLS &  0.755 & (0.650, 0.861) \\
  & 1 & cLS &  0.790 &  (0.342, 1.238) \\
  & 0.5 & cLS & 0.850 &  (-0.106, 1.806) \\
[0.5em] 
Age  & $\infty$ & MLE & 4.168 & (4.114, 4.223) \\ 
 &  & LS & 4.137 & (4.085, 4.189) \\
 &  5 & cLS & 4.208 &  (3.843, 4.573)\\
  &  1 & cLS & 4.472 &  (2.768, 6.177) \\
  &  0.5 & cLS &  4.800 & (1.249, 8.352)\\  \bottomrule[1.5pt]
\end{tabular}
\label{tab:multi.logreg.aou}
\end{table}

The results of the multiple logistic regression of hypertension status on the three variables (gender, race, and age) on the raw data and on \emph{one} pseudo-dataset corresponding to each noise level are summarized in Table~\ref{tab:multi.logreg.aou}. 
In the raw data, the LS and MLE estimates are closely aligned, with largely overlapping confidence intervals, demonstrating consistency between the two methods. Particularly, all three variables -- gender, race, and age -- are statistically significantly associated with hypertension prevalence in this raw data. Specifically, both LS and MLE estimates for gender yield $\hat \beta_{gender}^{\text{raw}} \approx -0.18$, indicating that the log-odds of hypertension are lower for females by $0.18$. Similarly, for race and age, the log-odds of hypertension are higher by approximately $0.77$ for African Americans and increase by about $0.06$ per year of age (calculated as $4/70$ based on the age scaling). Although gender and race are binary variables and do not strictly follow the conditional mixture model~\eqref{conditionmixturemodel}-\eqref{eq:logisticZ}, their MLE and LS estimators show strong agreement in this raw dataset.
%\textcolor{magenta}{We note that although Gender and Race cannot be assumed to follow the conditional mixture model \eqref{conditionmixturemodel}-\eqref{eq:logisticZ}, the results in this data analysis can show that the LS estimator may be robust against deviation from the underlying mixture model. Any further possible explanation?}

On the pseudo-datasets, the corrected LS  (cLS) estimates align closely with the estimates from the raw data, as shown by the overlapping confidence intervals. However, the confidence intervals on the pseudo-datasets are substantially wider, reflecting the increased uncertainty in the cLS estimates caused by the noise added for privacy protection. 
\rev{This uncertainty grows as the privacy requirement becomes stricter (i.e., smaller $\varepsilon$), since larger noise can obscure statistical significance. For instance, while all three covariates are statistically significant at $\varepsilon = 5$, the gender effect becomes insignificant at $\varepsilon = 1$. At $\varepsilon = 0.5$, both gender and race effects lose significance, with only the age effect remaining statistically significant.
} 
%This uncertainty increases with the magnitude of the added noise, $\sigma$, which can obscure statistical significance. For example, when $\sigma=1$, the gender effect is no longer statistically significant, though race and age effects remain significant. 
%When $\sigma=2$, race effect also becomes insignificant, with only the effect of age retains statistical significance. When $\sigma=3$, none of the three covariate effects is significant. 

To demonstrate the stability of the results, we further generated $100$ pseudo-datasets for each level of noise variance $\sigma$. Since the raw data has a relatively large sample size, we treated the LS point  estimates $\hat{\bbeta}_{1}^{\text{raw}}$ and the confidence intervals on the raw data as the ``true'' values and confidence intervals. On each pseudo-dataset, we computed our cLS estimates and the corresponding confidence interval. We then calculated biases, standard errors, coverage probability and the proportion of confidence intervals that do not contain zero and imply significant effects across 100 samples.  The results are reported in 
Table~\ref{tab:multi.logreg.aou.100}.   

\begin{table}[htb]
\centering
\caption{\rev{Results from computing the corrected OLS point and interval estimators on 100 pseudo-datasets generated by applying the TM$^2+$Noise scheme to the {\it All of US} data for each value of $\varepsilon$ satisfying $(\varepsilon,\delta)$-DP.}} 
\begin{tabular}{cccHccHc}
  \toprule[1.5pt]
$x$ & $\varepsilon$ & Bias & \makecell{ median \\ bias} & \makecell{Standard \\ error}  & \makecell{Proportion of \\ CI containing \\  $\hat{\bbeta}_{1}^{\text{raw}}$}  & \makecell{Proportion of \\  CI containing \\  raw CI} & \makecell{Proportion of \\ significant \\ effect}\\
  \hline
  \addlinespace
Gender & 5 &  0.001 & 0.009 &  0.038 & 96\% & 95\% & 100\% \\
  & 1 &  0.014 & 0.018 & 0.147  & 98\% & 99\% &  24\%\\
  & 0.5 &  0.042 & 0.045 &  0.300 & 99\% & 95\% & 7\%\\ [0.5em]
Race & 5 &   0.008 & 0.016 & 0.054 & 99\% & 95\% & 100\% \\
& 1 &   0.059 & 0.107 & 0.227 & 98\% & 99\% &  98\% \\
& 0.5 &   0.173 & 0.157  & 0.477 & 99\% & 99\% &  44\%\\
[0.5em] 
Age  & 5 &  0.021 & 0.055 & 0.185 & 98\% & 98\% & 100\% \\
&  1 &   0.292 & 0.548 & 0.846 & 98\% & 97\% &  100\%\\
&  0.5 &  1.053 & -0.136 &  1.808 & 97\% & 92\% & 81\%\\
 \bottomrule[1.5pt]
\end{tabular}
\label{tab:multi.logreg.aou.100}
\end{table}

\rev{Across 100 runs of the TM$^2$+Noise scheme, the cLS estimates exhibit only small bias. As privacy protection strengthens (i.e., smaller $\varepsilon$ and hence larger noise), the standard errors of the cLS estimates increase, reflecting greater uncertainty. The 95\% confidence intervals of the cLS estimates almost always contain those from the raw dataset. However, with higher noise levels, covariate effects become harder to detect due to this added uncertainty.

Specifically, as $\varepsilon$ decreases from 5 to 1 and then to 0.5, the detection probability for the gender effect drops from 100\% to 24\% to 7\%. For the race effect, detection falls from 100\% to 98\% to 44\%. The age effect remains consistently detectable at $\varepsilon = 5$ and $\varepsilon = 1$, but is detected only 81\% of the time at $\varepsilon = 0.5$.

Overall, stronger privacy protection (i.e., higher noise levels) leads to greater attenuation of statistical effects, thereby requiring larger sample sizes to reliably detect covariate associations.
}
%From the $100$ runs of the TM$^2$+Noise scheme, the cLS estimates have small bias. The standard error of cLS increases with the noise level, reflecting higher level of uncertainty in estimates. The $95\%$ confidence interval of cLS estimates almost always contains the confidence interval from raw data set. When the noise level increases, the covariate effects are less likely to be detected due to the increased uncertainty in estimates. When $\sigma=3$, none of the three covariate effects are ever detected as significant. In general, datasets with stronger privacy protection (i.e., higher noise levels) experience greater obscuration of statistical effects, requiring larger sample sizes to reliably detect covariate effects.

\FloatBarrier

\section{Conclusion}

As privacy concerns escalate in the era of big data, privacy-preserving statistical analysis has emerged as a critical area of research. While most existing methods focus on differential privacy schemes that perturb sufficient statistics, these approaches generally require a trusted central data manager. In contrast, local noise addition combined with matrix masking facilitates privacy-preserving data collection without relying on a central authority. This paper presents the first valid statistical analysis procedure for logistic regression — a nonlinear statistical model — applied to data collected via noise addition and matrix masking. The method exploits an extended relationship with linear regression-based estimators to ensure valid analysis. Our proposed estimator is proven to be valid in large samples, even under stringent privacy requirements (i.e., as noise magnitude $\sigma \to \infty$). Simulation studies and real data analyses confirm the validity of the proposed procedure and demonstrate its substantial improvement over naive logistic regression methods, which fail under these privacy-preserving settings.

%\section{\rev{Extension to multi-class logistic regression.}}

\rev{An intermediate future work would be to extend the framework to multinomial regression model where the raw response is categorical with $J>2$ categories, i.e $y_i^* \in \{1,2,...,J\}$ for $j=1,\ldots, n$. Since $\sum_{j=1}^{J} P(y_i^* = j \mid \bXstar_i) = 1$,  a common approach is to select a baseline category, say $J$ without loss of generality, and then model  
%The regression follows 
%\begin{equation}
%P(y^*_i = j \mid \bX_i^*) = \dfrac{\exp(\gamma_j + \bX_i^*\bm\delta_j) }{\sum_{k=1}^J\exp(\gamma_k + \bX_i^*\bm\delta_k) }, \qquad j=1,...,J.
%\end{equation}
%Defining $\beta_{0,k}=\gamma_k - \gamma_J$ and $\beta_{1,k}=\delta_k - \delta_J$, this can be equivalently written as
\begin{equation}
P(y^*_i = j \mid \bX_i^*) = \dfrac{\exp(\beta_{0,j} + \bX_i^*\bm\beta_{1,j}) }{1 + \sum_{k=1}^{J} \exp(\beta_{0,k} + \bX_i^*\bm\beta_{1,k})},
\label{eq:logisticMulti}
\end{equation}
for $j=1,...,J-1$ and $P(y_i^* = J \mid \bXstar_i) = 1 - \sum_{j=1}^{J-1}P(y^*_i = j \mid \bX_i^*)$. \cite{haggstrom1983logistic} demonstrated that logistic regression~\eqref{eq:logisticMulti} can also be derived from a mixture of $J$ Gaussian distributions over the covariates, thereby linking the regression coefficients to least squares estimators. Therefore, our approach can be extended to multinomial logistic regression under matrix masking and noise addition. However, instead of a single $J$-class response variable $y^*$, the raw data must include $J-1$ binary indicator variables for classes $k = 1, \dots, J-1$. This requirement arises because, after applying matrix masking and adding noise, the class information contained in the resulting single variable $y$ cannot be separately utilized in the $J-1$ logistic regressions of~\eqref{eq:logisticMulti}. The resulting formulas is more complex, as each logistic regression in~\eqref{eq:logisticMulti} involves comparisons among $J-1$ classes. A detailed treatment of these derivations is left for future work.}

\bigskip
\begin{center}
{\large\bf SUPPLEMENTARY MATERIAL}
\end{center}
The Supplementary Material contains the proofs for Proposition 3.1 and Theorem 3.1, as well as additional simulation results.
\appendix
\section{Appendix}

\rev{
\subsection{Differential privacy guarantee for the TM$^2$+Noise scheme}\label{sec:priv_level_sigma}
In this section, we briefly review the concept of differential privacy and how the TM$^2$+Noise scheme guarantees different levels of DP. More details can be found in \citet{ding2020privacy} and \citet{ding2023preprint}.

Broadly speaking, differential privacy (DP) quantifies privacy protection by measuring the closeness of the distributions of two pseudo-datasets generated from raw datasets that differ in only one individual~\citep{DP_0}. Two raw datasets, $\mathbf{D}^{\text{raw}}$ and $\mathbf{D}^{\text{raw}\prime}$, are called {\it neighbors} if they differ in at most one row (i.e., one individual’s data). Formally, a perturbation mechanism $\mathbf{D}^{\textrm{ps}}(\cdot)$ satisfies  $(\varepsilon, \delta)$-DP if, with probability at least $1-\delta$, for any measurable set $\mathcal{S}$ and any pair of neighbors $\mathbf{D}^{\text{raw}}$ and $\mathbf{D}^{\text{raw}\prime}$, 
\begin{equation}\label{eq:DP}
e^{-\varepsilon} \le \frac{P[\mathbf{D}^{\textrm{ps}}(\mathbf{D}^{\text{raw}}) \in \mathcal{S}]}{P[\mathbf{D}^{\textrm{ps}}(\mathbf{D}^{\text{raw}\prime}) \in \mathcal{S}]} \le e^{\varepsilon}.
\end{equation}

Most of the differential privacy literature considers privacy guarantees only for a statistic $T(\mathbf{D}^{\textrm{ps}})$ derived from the pseudo-dataset, such as summary statistics, which protects the released statistic but not against data collectors who have access to the full pseudo-data. To protect against the latter, DP must be established at the level of the entire pseudo-dataset $\mathbf{D}^{\textrm{ps}}$, as in Definition~\eqref{eq:DP}.

\citet{ding2023preprint} analyzed such full-dataset DP guarantees under the neighboring definition where datasets differ in exactly one row by at most $1$ in $\ell_2$-norm. Within that framework, for TM$^2$+Noise scheme with independent $N(0,\sigma^2)$ noise for all entries, ensuring $(\varepsilon, \delta)$-DP requires
\begin{equation}
\label{eq:sigma.bound.new}
\sigma \ge \text{the upper root of } g(\sigma) = \frac{(2 \sqrt{d}+1)}{2(n-d)} \gamma_{\delta; 2(n-d), d/\sigma^2} + \sqrt{d} - \sigma^2 \varepsilon,
\end{equation}
where $d$ is the number of columns in $\mathbf{D}^{\text{raw}}$ and $\gamma_{\delta; df, a}$ is the upper $\delta$-quantile of a non-central $\chi^2$ distribution with $df$ degrees of freedom and non-centrality parameter $a$. For a better interpretation, the root of \eqref{eq:sigma.bound.new} implies a more relaxed bound:
$$\sigma^2 \ge \frac{2n-d+\log(\frac{1}{\delta})}{2(n-d)} \frac{9 \sqrt{d}}{\varepsilon}.$$ 
When the sample size $n > \log(1/\delta)$, this requirement scales as $O(\varepsilon^{-1})$ for strict privacy ($\varepsilon<1$). In contrast, a pure noise-addition scheme (without matrix masking) requires $$\sigma^2 \ge \frac{1.7 \log(\delta^{-1})}{\varepsilon^2},$$ which scales as $O(\varepsilon^{-2})$. 

We used \eqref{eq:sigma.bound.new} to compute the $\sigma$ values to achieve $(\varepsilon, 0.01)$-DP in both the real-data analysis and simulated settings. 
Notice that \eqref{eq:sigma.bound.new} is derived for a dataset $\mathbf{D}^{\text{raw}}$ containing both the $n\times 1$  raw response vector  $\by^*$ and the $n \times p$ covariate matrix $\bX^*$, i.e. $d=p+1$. %For example, in Section~\ref{sec:realdata} logistic regression of $y$ on three covariates $x$, so  $d=4$. 
Furthermore, \eqref{eq:sigma.bound.new} was derived under the assumption that all entries of $\mathbf{D}^{\text{raw}}$ lie in $[-1,1]$, which needs to be converted to $\sigma_y$ and $\sigma_x$ according to the observed ranges of $\by^*$ and $\bX^*$ in the data, respectively. Particularly, for the simulation studies in Section~\ref{sec:simulation}, the raw response variable $y^*$ and covariates $\bX^*$ have different ranges, resulting in $\sigma_y \ne \sigma_x$ in Table~\ref{tab:sim_results_mixturemodel} and Table~\ref{tab:sim_results_conditionalmixturemodel}.
For the real data analysis in Section~\ref{sec:realdata}, all the variables were scaled to the range $[0,1]$, thus $\sigma_y=\sigma_x$ is half of the $\sigma$ value from the bound \eqref{eq:sigma.bound.new}.  }

\subsection{Derivation of equation~\eqref{eq:log_relationship}}\label{sec:log_relationship}
\begin{equation}
\footnotesize
\begin{aligned}
& P(y_i^* = 1 \mid \bZstar_i, \bXstar_i) \\
= & \frac{P(y^*_i = 1, \bZstar_i, \bX^*_i)}{P(y^*_i = 1, \bZstar_i, \bX^*_i) + P(y^*_i = 0, \bZstar_i, \bX^*_i)} \\
= & \frac{ p_1(\bZstar_i)\exp\{-\frac{1}{2} (\bX_i^* - \bmu_1 - \bZstar_i \bC) \bm\Sigma^{-1} (\bX_i^* - \bmu_1- \bZstar_i \bC)^T\}}{p_1(\bZstar_i) \exp\{-\frac{1}{2} (\bX_i^* - \bmu_1 - \bZstar_i \bC) \bm\Sigma^{-1} (\bX_i^* - \bmu_1 - \bZstar_i \bC)^T\} + p_0(\bZstar_i)  \exp\{-\frac{1}{2} (\bX_i^* - \bmu_0 - \bZstar_i \bC) \bm\Sigma^{-1} (\bX_i^* - \bmu_0 - \bZstar_i \bC)^T\} } \\
= & \frac{1}{1 + \frac{p_0(\bZstar_i)}{p_1(\bZstar_i)}  \exp\{ \frac{\bmu_1 + \bmu_0 + 2 \bZstar_i \bC}{2} \bm\Sigma^{-1} (\bmu_1 - \bmu_0)^T  + \bX_i^* \ \bm\Sigma^{-1} (\bmu_0 - \bmu_1)^T\}} \\
= & \frac{1}{1 + \frac{1}{\exp(\gamma_0 + \bZstar_i\bm\gamma_1)}  \exp\{ \frac{\bmu_1 + \bmu_0 + 2 \bZstar_i \bC}{2} \bm\Sigma^{-1} (\bmu_1 - \bmu_0)^T  + \bX_i^* \ \bm\Sigma^{-1} (\bmu_0 - \bmu_1)^T\}} \\
= & \frac{ 1}{1 +   \exp\{ - \gamma_0 - \bZstar_i\bm\gamma_1  + \frac{\bmu_1 + \bmu_0 }{2} \bm\Sigma^{-1} (\bmu_1 - \bmu_0)^T + \bZstar_i \bC \bm\Sigma^{-1} (\bmu_1 - \bmu_0)^T - \bX_i^* \bm\Sigma^{-1} (\bmu_1 - \bmu_0)^T\}} \\
= & \frac{ 1}{1 +   \exp\{ - [\gamma_0 -\frac{\bmu_1 + \bmu_0 }{2} \bm\Sigma^{-1} (\bmu_1 - \bmu_0)^T]  - \bX_i^* [\bm\Sigma^{-1} (\bmu_1 - \bmu_0)^T] - \bZstar_i\bm [\bgamma_1 - \bC \bm\Sigma^{-1} (\bmu_1 - \bmu_0)^T]
\}}.
\end{aligned}
\end{equation}
The last expression reduces to 
$\dfrac{1}{1 + \exp(-\beta_0 - \bXstar_i\bbeta_1  - \bZstar_i\bbeta_2)}$ as in \eqref{eq:log_relationship} when we denote
$$
\beta_0 = \gamma_0 -\frac{\bmu_1 + \bmu_0 }{2} \bm\Sigma^{-1} (\bmu_1 - \bmu_0)^T = \gamma_0 - \frac{1}{2} (\bmu_1 \bSigma^{-1}\bmu_1^\T - \bmu_0\bSigma^{-1}\bmu_0^\T ),
$$ 
$\bm\beta_1 = \bm\Sigma^{-1}(\bmu_1 - \bmu_0)^\T$ and $\bm\beta_2 = \bgamma_1 - \bC \bm\Sigma^{-1} (\bmu_1 - \bmu_0)^T = \bgamma_1 - \bC \bbeta_1$.

\subsection{Proof of Lemma~\ref{est.b1.cond.mixturemodel}}\label{sec:lemma1}
To simplify the notation, we let $\bSigma_{xz}$ denote the covariance between $\mathbf{X}_1^*$ and $\mathbf{Z}_1^*$, where $\mathbf{X}_1^*$ and $\bZ_1^*$ are the variables in the first row of the raw data, i.e $\bSigma_{xz} = E(\bX_1^{*\T} \bZ_1^*) - \left\{E(\bX_1^*\right\}^\T E(\bZ_1^*)$. Similar definitions hold for $\bSigma_{xy}$, $\bSigma_{zy}$, $\bSigma_{xx}$ and $\bSigma_{zz}$.

Hence, we can write 
$$
\bar{\bm{b}} = \begin{bmatrix}
\bSigma_{xx} & \bSigma_{xz} \\
\bSigma_{zx} & \bSigma_{zz} \\
\end{bmatrix}^{-1} \begin{bmatrix}
\bSigma_{xy}\\
\bSigma_{zy}
\end{bmatrix} = \begin{bmatrix}
\bar{\bm{b}}_1 \\
\bar{\bm{b}}_2 \\
\end{bmatrix}
$$
Using the Schur's matrix inversion formula, we have
\begin{align*}
\footnotesize
& \begin{bmatrix}
\bSigma_{xx} & \bSigma_{xz} \\
\bSigma_{zx} & \bSigma_{zz} \\
\end{bmatrix}^{-1} 
\\ = & \begin{bmatrix}
(\bSigma_{xx} - \bSigma_{xz}\bSigma_{zz}^{-1} \bSigma_{zx})^{-1}& - \left( \bSigma_{xx} - \bSigma_{xz}\bSigma_{zz}^{-1} \bSigma_{zx}\right)^{-1}\bSigma_{xz}\bSigma_{zz}^{-1}\\
-\bSigma_{zz}^{-1}\bSigma_{zx}  \left( \bSigma_{xx} - \bSigma_{xz}\bSigma_{zz}^{-1} \bSigma_{zx}\right)^{-1} & \bSigma_{zz}^{-1} + \bSigma_{zz}^{-1} \bSigma_{zx} (\bSigma_{xx} - \bSigma_{xz}\bSigma_{zz}^{-1} \bSigma_{zx})^{-1} \bSigma_{xz} \bSigma_{zz}^{-1}
\end{bmatrix}
\end{align*}
so we have
\begin{align}
&\bar\bb_1 = 
(\bSigma_{xx} - \bSigma_{xz}\bSigma_{zz}^{-1} \bSigma_{zx})^{-1}(\bSigma_{xy} - \bSigma_{xz}\bSigma_{zz}^{-1}\bSigma_{zy}),
\label{eq:barb1}
\end{align} and 
\begin{equation}
\begin{array}{cl}
     \bar\bb_2 & =  -\bSigma_{zz}^{-1}\bSigma_{zx}  \left( \bSigma_{xx} - \bSigma_{xz}\bSigma_{zz}^{-1} \bSigma_{zx}\right)^{-1} \bSigma_{xy} + \bSigma_{zz}^{-1}\bSigma_{zy} \\
 & \qquad + \bSigma_{zz}^{-1} \bSigma_{zx} (\bSigma_{xx} - \bSigma_{xz}\bSigma_{zz}^{-1} \bSigma_{zx})^{-1} \bSigma_{xz} \bSigma_{zz}^{-1}\bSigma_{zy} \\
 & = \bSigma_{zz}^{-1}\bSigma_{zy} - \bSigma_{zz}^{-1} \bSigma_{zx} (\bSigma_{xx} - \bSigma_{xz}\bSigma_{zz}^{-1} \bSigma_{zx})^{-1}(\bSigma_{xy} - \bSigma_{xz}\bSigma_{zz}^{-1}\bSigma_{zy}) \\
 & = \bSigma_{zz}^{-1}\bSigma_{zy} - \bSigma_{zz}^{-1} \bSigma_{zx} \bar\bb_1.
\end{array}
\label{eq:barb2}
\end{equation}
Now we will express $\bar\bb_1$ in terms of parameters of the mixture model \eqref{conditionmixturemodel}-\eqref{eq:logisticZ}, through finding the expressions for the two factors in \eqref{eq:barb1}: $\bSigma_{xx} - \bSigma_{xz}\bSigma_{zz}^{-1} \bSigma_{zx}$ and $\bSigma_{xy} - \bSigma_{xz}\bSigma_{zz}^{-1}\bSigma_{zy}$. First, we have
\begin{equation}
\begin{array}{cl}
& \ \ \ \bSigma_{xx}  = \Var(\bX_1^*)  \\
& = E\left[\Var(\bX_1^*\mid y_1^*, \bZ_1^* ) \right] + \Var\left[E\left(\bX_1^*\mid y_1^*, \bZ_1^* \right) \right] \\
& = E\left[\bm\Sigma \right]  + \Var\left[(\bmu_1 - \bmu_0) y_1^* + \bmu_0 + \bZ_1^* \bC \right]\\
& = \bm\Sigma + \Var\left[(\bmu_1 - \bmu_0) y_1^* + \bZ_1^* \bC \right]\\
& = \bm\Sigma + \Var[(\bmu_1 - \bmu_0) y_1^*] + \Var[\bZ_1^* \bC ] 
+ \Cov[(\bmu_1 - \bmu_0) y_1^*, \bZ_1^* \bC] \\
& \qquad + \Cov[\bZ_1^* \bC, (\bmu_1 - \bmu_0) y_1^*]\\
& = \bm\Sigma + (\bmu_1 - \bmu_0)^\T \Var(y_1^*)(\bmu_1 - \bmu_0)  + \bC^\T \Var(\bZ_1^*) \bC + (\bmu_1 - \bmu_0)^\T \Cov(y_1^*, \bZ_1^*) \bC \\ 
&\qquad + \bC^\T \Cov(\bZ_1^*, y_1^*) (\bmu_1 - \bmu_0) \\
& = \bm\Sigma + (\bmu_1 - \bmu_0)^\T\bSigma_{yy}(\bmu_1 - \bmu_0)  + \bC^\T \bSigma_{zz} \bC + (\bmu_1 - \bmu_0)^\T \bSigma_{yz} \bC + \bC^\T \bSigma_{zy} (\bmu_1 - \bmu_0). 
\end{array}
\label{eq:Sigma_xx}
\end{equation}
Recall that for any two random vectors $\bV_1$ and $\bV_2$, we have 
$$
\begin{array}{cl}
\Cov(\bV_1,\bV_2) & = E(\bV_1^T \bV_2) - E(\bV_1)^T E(\bV_2) \\
& = E[E(\bV_1^T \bV_2 \mid \bV_2)] - E[E(\bV_1 \mid \bV_2)]^T E(\bV_2) \\
& = E[E(\bV_1 \mid \bV_2)^T \bV_2] - E[E(\bV_1 \mid \bV_2)^T] E(\bV_2) \\
& = \Cov[E(\bV_1 \mid \bV_2),\bV_2],
\end{array}
$$
so we have
\begin{equation}\label{eq:Sigma_xz}
\begin{array}{cl}
& \bSigma_{xz} = \Cov [\bX^*_1,\bZ^*_1] = \Cov[E(\bX^*_1 \mid \bZ^*_1),\ \bZ^*_1] \\
= & \Cov [(\bmu_1 - \bmu_0) E(y^*_1 \mid \bZ^*_1) + \bmu_0 + \bZ^*_1 \bC, \ \bZ^*_1]
\\
= &  \Cov [(\bmu_1 - \bmu_0) E(y^*_1 \mid \bZ^*_1) , \ \bZ^*_1] + \Cov [\bmu_0 , \ \bZ^*_1] + \Cov [\bZ^*_1 \bC, \ \bZ^*_1] \\
= &  (\bmu_1 - \bmu_0)^\T \Cov[E(y^*_1 \mid \bZ^*_1), \bZ^*_1] + 0 + \bC^\T \Cov(\bZ^*_1, \bZ^*_1)
\\
= &  (\bmu_1 - \bmu_0)^\T \bSigma_{yz} + \bC^\T \bSigma_{zz}.
\end{array}
\end{equation}
Hence
$$
\begin{array}{cl}
& 
\bSigma_{xz}\bSigma_{zz}^{-1} \bSigma_{zx} \\
= & [(\bmu_1 - \bmu_0)^\T \bSigma_{yz} + \bC^\T \bSigma_{zz}] \bSigma_{zz}^{-1} [(\bmu_1 - \bmu_0)^\T \bSigma_{yz} + \bC^\T \bSigma_{zz}]^T \\
= & [(\bmu_1 - \bmu_0)^\T \bSigma_{yz} + \bC^\T \bSigma_{zz}] \bSigma_{zz}^{-1} [ \bSigma_{zy}(\bmu_1 - \bmu_0) +  \bSigma_{zz}\bC] \\
= & (\bmu_1 - \bmu_0)^\T \bSigma_{yz} \bSigma_{zz}^{-1} \bSigma_{zy}(\bmu_1 - \bmu_0) + (\bmu_1 - \bmu_0)^\T \bSigma_{yz}  \bC + \bC^\T  \bSigma_{zy}(\bmu_1 - \bmu_0)  + \bC^\T \bSigma_{zz} \bC. 
\end{array}
$$
Combine this expression with \eqref{eq:Sigma_xx}, we simplify the first factor in \eqref{eq:barb1} as
\begin{equation}
\label{eq:factor1}
\begin{array}{cl}
& \bSigma_{xx} - \bSigma_{xz}\bSigma_{zz}^{-1} \bSigma_{zx} \\
= & \bm\Sigma + (\bmu_1 - \bmu_0)^\T\bSigma_{yy}(\bmu_1 - \bmu_0) - (\bmu_1 - \bmu_0)^\T \bSigma_{yz} \bSigma_{zz}^{-1} \bSigma_{zy}(\bmu_1 - \bmu_0)\\
= & \bm\Sigma + (\bmu_1 - \bmu_0)^\T (\bSigma_{yy} - \bSigma_{yz} \bSigma_{zz}^{-1} \bSigma_{zy}) (\bmu_1 - \bmu_0).
\end{array}
\end{equation}
Furthermore, we have
\begin{equation}\label{eq:Sigma_xy}
\begin{array}{cl}
& \bSigma_{xy} = \Cov[\bX^*_1, y^*_1] \\
= & E[\Cov(\bX^*_1, y^*_1\mid \bZ^*_1)] + \Cov[E(\bX^*_1\mid \bZ^*_1), E(y^*_1\mid \bZ^*_1)] \\
= & E[\Cov((\bmu_1 - \bmu_0) y^*_1 + \bmu_0 + \bZ^*_1 \bC, \ y^*_1\mid \bZ^*_1)] \\
& \qquad + \Cov [E( (\bmu_1 - \bmu_0) y^*_1 \mid \bZ^*_1) + \bmu_0 + \bZ^*_1 \bC, \ E(y^*_1 \mid \bZ^*_1)]
\\
= & E[\Cov((\bmu_1 - \bmu_0) y^*_1, \ y^*_1\mid \bZ^*_1)] \\
& \qquad + \Cov [E( (\bmu_1 - \bmu_0) y^*_1 \mid \bZ^*_1), \ E(y^*_1 \mid \bZ^*_1)] + 0 + \Cov [\bZ^*_1 \bC, \ E(y^*_1 \mid \bZ^*_1)]
\\
= &  \Cov[(\bmu_1 - \bmu_0) y^*_1, \ y^*_1]  + \Cov [\bZ^*_1 \bC, \ E(y^*_1 \mid \bZ^*_1)] \\
= &  (\bmu_1 - \bmu_0)^\T \Cov(y^*_1, \ y^*_1)  + \bC^\T \Cov(\bZ^*_1, y^*_1)
\\
= &  (\bmu_1 - \bmu_0)^\T \bSigma_{yy} + \bC^\T \bSigma_{zy},
\end{array}
\end{equation}
where the second-to-last equality used \rev{the property $\Cov(\bV_1,\bV_2) = \Cov[E(\bV_1 \mid \bV_2),\bV_2] = \Cov[\bV_1, E(\bV_2 \mid \bV_1)]$ for any two random vector $\bV_1$ and $\bV_2$. This property holds because
$$
\begin{array}{cl}
 \Cov(\bV_1,\bV_2) & = E(\bV_1^T \bV_2) - E(\bV_1)^T E(\bV_2) \\
& = E[E(\bV_1^T \bV_2 \mid \bV_2)] - E[E(\bV_1 \mid \bV_2)]^T E(\bV_2) \\
& = E[E(\bV_1 \mid \bV_2)^T \bV_2] - E[E(\bV_1 \mid \bV_2)^T] E(\bV_2) \\
& = \Cov[E(\bV_1 \mid \bV_2),\bV_2].
\end{array}
$$ }

Combining the expression \eqref{eq:Sigma_xy} with \eqref{eq:Sigma_xz}, we simplify the second factor in \eqref{eq:barb1} as
\begin{equation}
\label{eq:factor2}
\begin{array}{cl}
& 
\bSigma_{xy} - \bSigma_{xz}\bSigma_{zz}^{-1}\bSigma_{zy} \\
= & (\bmu_1 - \bmu_0)^\T \bSigma_{yy} + \bC^\T \bSigma_{zy} - [(\bmu_1 - \bmu_0)^\T \bSigma_{yz} + \bC^\T \bSigma_{zz}] \bSigma_{zz}^{-1} \bSigma_{zy} \\
= & (\bmu_1 - \bmu_0)^\T \bSigma_{yy} + \bC^\T \bSigma_{zy} - (\bmu_1 - \bmu_0)^\T \bSigma_{yz} \bSigma_{zz}^{-1} \bSigma_{zy} - \bC^\T \bSigma_{zy}  \\
= & (\bmu_1 - \bmu_0)^\T \bSigma_{yy} - (\bmu_1 - \bmu_0)^\T \bSigma_{yz} \bSigma_{zz}^{-1} \bSigma_{zy}  \\
= & (\bmu_1 - \bmu_0)^\T (\bSigma_{yy} - \bSigma_{yz} \bSigma_{zz}^{-1} \bSigma_{zy}) . 
\end{array}
\end{equation}

Note that, equation \eqref{eq:barb1} can be rewritten as 
$
(\bSigma_{xx} - \bSigma_{xz}\bSigma_{zz}^{-1} \bSigma_{zx}) \bar\bb_1 = 
(\bSigma_{xy} - \bSigma_{xz}\bSigma_{zz}^{-1}\bSigma_{zy}).$
Substituting \eqref{eq:factor1} and \eqref{eq:factor2} into this expression, we have
$$
[\bm\Sigma + (\bmu_1 - \bmu_0)^\T (\bSigma_{yy} - \bSigma_{yz} \bSigma_{zz}^{-1} \bSigma_{zy}) (\bmu_1 - \bmu_0)]\bar\bb_1 = (\bmu_1 - \bmu_0)^\T (\bSigma_{yy} - \bSigma_{yz} \bSigma_{zz}^{-1} \bSigma_{zy}),
$$
which is equivalent to
$$
\begin{array}{cl}
\bm\Sigma \bar\bb_1 & = (\bmu_1 - \bmu_0)^\T (\bSigma_{yy} - \bSigma_{yz} \bSigma_{zz}^{-1} \bSigma_{zy}) - (\bmu_1 - \bmu_0)^\T (\bSigma_{yy} - \bSigma_{yz} \bSigma_{zz}^{-1} \bSigma_{zy}) (\bmu_1 - \bmu_0)\bar\bb_1 \\
& = (\bmu_1 - \bmu_0)^\T (\bSigma_{yy} - \bSigma_{yz} \bSigma_{zz}^{-1} \bSigma_{zy}) [1 - (\bmu_1 - \bmu_0)\bar\bb_1] \\
& = \bm\Sigma \bbeta_1 (\bSigma_{yy} - \bSigma_{yz} \bSigma_{zz}^{-1} \bSigma_{zy}) [1 - (\bmu_1 - \bmu_0)\bar\bb_1].
\end{array}
$$
where the last equality used definition of $\bbeta_1$ from \eqref{eq:beta}. Multiplying $\bm\Sigma^{-1}$ to both sides of above equation, we have
$$
\bar\bb_1 = \bbeta_1 (\bSigma_{yy} - \bSigma_{yz} \bSigma_{zz}^{-1} \bSigma_{zy}) [1 - (\bmu_1 - \bmu_0)\bar\bb_1].
$$
Hence, to finish the proof of this Lemma, i.e., to get $\bbeta_1= \bar\bb_1/\Var(\varepsilon_1)$, we only need to show  
\begin{equation}\label{eq:Var.eps}
\Var(\varepsilon_1) = \Var(y^*_1 - \bW_1^*\bar{\bb}) =(\bSigma_{yy} - \bSigma_{yz} \bSigma_{zz}^{-1} \bSigma_{zy}) [1 - (\bmu_1 - \bmu_0)\bar\bb_1].
\end{equation}

Recall the definition $\bar{\bm{b}} = \left[\Var(\bW_1^*)\right]^{-1} \Cov(\bW_1^*, y_1^*)$. Hence $\Var(\bW_1^*) \bar{\bm{b}} = \Cov(\bW_1^*, y_1^*)$, and we have
$$
\begin{aligned}
\Var(\varepsilon_1) & = \Var(y^*_1 -  \bW_1^*\bar{\bb}) \ \ = \Var(y^*_1) + \bar\bb^\T \Var(\bW_1^*) \bar\bb - 2 \Cov(y_1^*, \bW_1^*) \bar\bb \\
& = \Var(y^*_1) + \bar\bb^\T \Cov(\bW_1^*, y_1^*)  - 2 \Cov(y_1^*, \bW_1^*) \bar\bb \\
& = \Var(y^*_1) - \Cov(y_1^*, \bW_1^*) \bar\bb  \ \ = \Var(y^*_1) - [\Cov(y_1^*, \bX_1^*) \bar\bb_1 + \Cov(y_1^*, \bZ_1^*) \bar\bb_2]  \\ 
& = \bSigma_{yy} -  \bSigma_{yx} \bar\bb_1 - \bSigma_{yz} \bar\bb_2.
\end{aligned}
$$
Using \eqref{eq:barb2}, this becomes
$$
\begin{aligned}
\Var(\varepsilon_1) & = \bSigma_{yy} -  \bSigma_{yx} \bar\bb_1 - 
\bSigma_{yz}(\bSigma_{zz}^{-1}\bSigma_{zy} - \bSigma_{zz}^{-1} \bSigma_{zx} \bar\bb_1) \\
& = \bSigma_{yy} - \bSigma_{yz}\bSigma_{zz}^{-1}\bSigma_{zy}
- (\bSigma_{yx} - \bSigma_{yz} \bSigma_{zz}^{-1} \bSigma_{zx}) \bar\bb_1 \\
& = \bSigma_{yy} - \bSigma_{yz}\bSigma_{zz}^{-1}\bSigma_{zy}
- (\bSigma_{xy} - \bSigma_{xz} \bSigma_{zz}^{-1} \bSigma_{xy})^T \bar\bb_1,
\end{aligned}
$$
Using \eqref{eq:factor2}, this becomes
$$
\begin{aligned}
\Var(\varepsilon_1) & =  \bSigma_{yy} - \bSigma_{yz}\bSigma_{zz}^{-1}\bSigma_{zy}
- (\bSigma_{yy} - \bSigma_{yz}\bSigma_{zz}^{-1}\bSigma_{zy}) (\bmu_1 - \bmu_0) \bar\bb_1 \\
& = (\bSigma_{yy} - \bSigma_{yz}\bSigma_{zz}^{-1}\bSigma_{zy}) [1 - (\bmu_1 - \bmu_0) \bar\bb_1].
\end{aligned}
$$
This is indeed \eqref{eq:Var.eps}, thus the proof is finished.

\bibliographystyle{apalike}
\bibliography{refs}
\end{document}

\subsection{Proof of equation~\eqref{eq:log_relationship}}\label{sec:log_relationship}

\begin{equation}
\footnotesize
\begin{aligned}
& P(y_i^* = 1 \mid \bZstar_i, \bXstar_i) \\
= & \frac{P(y^*_i = 1, \bZstar_i, \bX^*_i)}{P(y^*_i = 1, \bZstar_i, \bX^*_i) + P(y^*_i = 0, \bZstar_i, \bX^*_i)} \\
= & \frac{ p_1(\bZstar_i)\exp\{-\frac{1}{2} (\bX_i^* - \bmu_1 - \bZstar_i \bC) \bm\Sigma^{-1} (\bX_i^* - \bmu_1- \bZstar_i \bC)^T\}}{p_1(\bZstar_i) \exp\{-\frac{1}{2} (\bX_i^* - \bmu_1 - \bZstar_i \bC) \bm\Sigma^{-1} (\bX_i^* - \bmu_1 - \bZstar_i \bC)^T\} + p_0(\bZstar_i)  \exp\{-\frac{1}{2} (\bX_i^* - \bmu_0 - \bZstar_i \bC) \bm\Sigma^{-1} (\bX_i^* - \bmu_0 - \bZstar_i \bC)^T\} } \\
= & \frac{1}{1 + \frac{p_0(\bZstar_i)}{p_1(\bZstar_i)}  \exp\{ \frac{\bmu_1 + \bmu_0 + 2 \bZstar_i \bC}{2} \bm\Sigma^{-1} (\bmu_1 - \bmu_0)^T  + \bX_i^* \ \bm\Sigma^{-1} (\bmu_0 - \bmu_1)^T\}} \\
= & \frac{1}{1 + \frac{1}{\exp(\gamma_0 + \bZstar_i\bm\gamma_1)}  \exp\{ \frac{\bmu_1 + \bmu_0 + 2 \bZstar_i \bC}{2} \bm\Sigma^{-1} (\bmu_1 - \bmu_0)^T  + \bX_i^* \ \bm\Sigma^{-1} (\bmu_0 - \bmu_1)^T\}} \\
= & \frac{ 1}{1 +   \exp\{ - \gamma_0 - \bZstar_i\bm\gamma_1  + \frac{\bmu_1 + \bmu_0 }{2} \bm\Sigma^{-1} (\bmu_1 - \bmu_0)^T + \bZstar_i \bC \bm\Sigma^{-1} (\bmu_1 - \bmu_0)^T - \bX_i^* \bm\Sigma^{-1} (\bmu_1 - \bmu_0)^T\}} \\
= & \frac{ 1}{1 +   \exp\{ - [\gamma_0 -\frac{\bmu_1 + \bmu_0 }{2} \bm\Sigma^{-1} (\bmu_1 - \bmu_0)^T]  - \bX_i^* [\bm\Sigma^{-1} (\bmu_1 - \bmu_0)^T] - \bZstar_i\bm [\gamma_1 - \bC \bm\Sigma^{-1} (\bmu_1 - \bmu_0)^T]
\}}.
\end{aligned}
\end{equation}
The last expression reduces to 
$\dfrac{1}{1 + \exp(-\beta_0 - \bXstar_i\bbeta_1  - \bZstar_i\bbeta_2)}$ as in \eqref{eq:log_relationship} when we denote
$$
\beta_0 = \gamma_0 -\frac{\bmu_1 + \bmu_0 }{2} \bm\Sigma^{-1} (\bmu_1 - \bmu_0)^T = \gamma_0 - \frac{1}{2} (\bmu_1 \bSigma^{-1}\bmu_1^\T - \bmu_0\bSigma^{-1}\bmu_0^\T ),
$$ 
$\bm\beta_1 = \bm\Sigma^{-1}(\bmu_1 - \bmu_0)^\T$ and $\bm\beta_2 = \bgamma_1 - \bC \bm\Sigma^{-1} (\bmu_1 - \bmu_0)^T = \bgamma_1 - \bC \bbeta_1$.

\subsection{Proof of Lemma~\ref{est.b1.cond.mixturemodel}}\label{sec:lemma1}
To simplify the notation, we let $\bSigma_{xz}$ denote the covariance between $\mathbf{X}_1^*$ and $\mathbf{Z}_1^*$, where $\mathbf{X}_1^*$ and $\bZ_1^*$ are the variables in the first row of the raw data, i.e $\bSigma_{xz} = E(\bX_1^{*\T} \bZ_1^*) - \left\{E(\bX_1^*\right\}^\T E(\bZ_1^*)$. Similar definitions hold for $\bSigma_{xy}$, $\bSigma_{zy}$, $\bSigma_{xx}$ and $\bSigma_{zz}$.

Hence, we can write 
$$
\bar{\bm{b}} = \begin{bmatrix}
\bSigma_{xx} & \bSigma_{xz} \\
\bSigma_{zx} & \bSigma_{zz} \\
\end{bmatrix}^{-1} \begin{bmatrix}
\bSigma_{xy}\\
\bSigma_{zy}
\end{bmatrix} = \begin{bmatrix}
\bar{\bm{b}}_1 \\
\bar{\bm{b}}_2 \\
\end{bmatrix}
$$
Using the Schur's matrix inversion formula, we have
\begin{align*}
& \begin{bmatrix}
\bSigma_{xx} & \bSigma_{xz} \\
\bSigma_{zx} & \bSigma_{zz} \\
\end{bmatrix}^{-1} \\
= & \begin{bmatrix}
(\bSigma_{xx} - \bSigma_{xz}\bSigma_{zz}^{-1} \bSigma_{zx})^{-1}& - \left( \bSigma_{xx} - \bSigma_{xz}\bSigma_{zz}^{-1} \bSigma_{zx}\right)^{-1}\bSigma_{xz}\bSigma_{zz}^{-1}\\
-\bSigma_{zz}^{-1}\bSigma_{zx}  \left( \bSigma_{xx} - \bSigma_{xz}\bSigma_{zz}^{-1} \bSigma_{zx}\right)^{-1} & \bSigma_{zz}^{-1} + \bSigma_{zz}^{-1} \bSigma_{zx} (\bSigma_{xx} - \bSigma_{xz}\bSigma_{zz}^{-1} \bSigma_{zx})^{-1} \bSigma_{xz} \bSigma_{zz}^{-1}
\end{bmatrix}
\end{align*}
so we have
\begin{align}
&\bar\bb_1 = 
(\bSigma_{xx} - \bSigma_{xz}\bSigma_{zz}^{-1} \bSigma_{zx})^{-1}(\bSigma_{xy} - \bSigma_{xz}\bSigma_{zz}^{-1}\bSigma_{zy}),
\label{eq:barb1}
\end{align} and 
\begin{equation}
\begin{array}{cl}
     \bar\bb_2 & =  -\bSigma_{zz}^{-1}\bSigma_{zx}  \left( \bSigma_{xx} - \bSigma_{xz}\bSigma_{zz}^{-1} \bSigma_{zx}\right)^{-1} \bSigma_{xy} + \bSigma_{zz}^{-1}\bSigma_{zy} \\
 & \qquad + \bSigma_{zz}^{-1} \bSigma_{zx} (\bSigma_{xx} - \bSigma_{xz}\bSigma_{zz}^{-1} \bSigma_{zx})^{-1} \bSigma_{xz} \bSigma_{zz}^{-1}\bSigma_{zy} \\
 & = \bSigma_{zz}^{-1}\bSigma_{zy} - \bSigma_{zz}^{-1} \bSigma_{zx} (\bSigma_{xx} - \bSigma_{xz}\bSigma_{zz}^{-1} \bSigma_{zx})^{-1}(\bSigma_{xy} - \bSigma_{xz}\bSigma_{zz}^{-1}\bSigma_{zy}) \\
 & = \bSigma_{zz}^{-1}\bSigma_{zy} - \bSigma_{zz}^{-1} \bSigma_{zx} \bar\bb_1.
\end{array}
\label{eq:barb2}
\end{equation}
Now we will express $\bar\bb_1$ in terms of parameters of the mixture model \eqref{conditionmixturemodel}-\eqref{eq:logisticZ}, through finding the expressions for the two factors in \eqref{eq:barb1}: $\bSigma_{xx} - \bSigma_{xz}\bSigma_{zz}^{-1} \bSigma_{zx}$ and $\bSigma_{xy} - \bSigma_{xz}\bSigma_{zz}^{-1}\bSigma_{zy}$. First, we have
\begin{equation}
\begin{array}{cl}
& \ \ \ \bSigma_{xx}  = \Var(\bX_1^*)  \\
& = E\left[\Var(\bX_1^*\mid y_1^*, \bZ_1^* ) \right] + \Var\left[E\left(\bX_1^*\mid y_1^*, \bZ_1^* \right) \right] \\
& = E\left[\bm\Sigma \right]  + \Var\left[(\bmu_1 - \bmu_0) y_1^* + \bmu_0 + \bZ_1^* \bC \right]\\
& = \bm\Sigma + \Var\left[(\bmu_1 - \bmu_0) y_1^* + \bZ_1^* \bC \right]\\
& = \bm\Sigma + \Var[(\bmu_1 - \bmu_0) y_1^*] + \Var[\bZ_1^* \bC ] 
+ \Cov[(\bmu_1 - \bmu_0) y_1^*, \bZ_1^* \bC] \\
& \qquad + \Cov[\bZ_1^* \bC, (\bmu_1 - \bmu_0) y_1^*]\\
& = \bm\Sigma + (\bmu_1 - \bmu_0)^\T \Var(y_1^*)(\bmu_1 - \bmu_0)  + \bC^\T \Var(\bZ_1^*) \bC + (\bmu_1 - \bmu_0)^\T \Cov(y_1^*, \bZ_1^*) \bC \\ 
&\qquad + \bC^\T \Cov(\bZ_1^*, y_1^*) (\bmu_1 - \bmu_0) \\
& = \bm\Sigma + (\bmu_1 - \bmu_0)^\T\bSigma_{yy}(\bmu_1 - \bmu_0)  + \bC^\T \bSigma_{zz} \bC + (\bmu_1 - \bmu_0)^\T \bSigma_{yz} \bC + \bC^\T \bSigma_{zy} (\bmu_1 - \bmu_0). 
\end{array}
\label{eq:Sigma_xx}
\end{equation}

For further simplifications, we will use the following property
\begin{equation}\label{eq:prop.A1}
\Cov(\bV_1,\bV_2) = \Cov[E(\bV_1 \mid \bV_2),\bV_2], \qquad \Cov(\bV_1,\bV_2) = \Cov[\bV_1, E(\bV_2 \mid \bV_1)].
\end{equation}
The property~\eqref{eq:prop.A1} holds because
$$
\begin{array}{cl}
 \Cov(\bV_1,\bV_2) & = E(\bV_1^T \bV_2) - E(\bV_1)^T E(\bV_2) \\
& = E[E(\bV_1^T \bV_2 \mid \bV_2)] - E[E(\bV_1 \mid \bV_2)]^T E(\bV_2) \\
& = E[E(\bV_1 \mid \bV_2)^T \bV_2] - E[E(\bV_1 \mid \bV_2)^T] E(\bV_2) \\
& = \Cov[E(\bV_1 \mid \bV_2),\bV_2].
\end{array}
$$

Using the property~\eqref{eq:prop.A1}, we have
\begin{equation}\label{eq:Sigma_xz}
\begin{array}{cl}
& \bSigma_{xz} = \Cov [\bX^*_1,\bZ^*_1] = \Cov[E(\bX^*_1 \mid \bZ^*_1),\ \bZ^*_1] \\
= & \Cov [(\bmu_1 - \bmu_0) E(y^*_1 \mid \bZ^*_1) + \bmu_0 + \bZ^*_1 \bC, \ \bZ^*_1]
\\
= &  \Cov [(\bmu_1 - \bmu_0) E(y^*_1 \mid \bZ^*_1) , \ \bZ^*_1] + \Cov [\bmu_0 , \ \bZ^*_1] + \Cov [\bZ^*_1 \bC, \ \bZ^*_1] \\
= &  (\bmu_1 - \bmu_0)^\T \Cov[E(y^*_1 \mid \bZ^*_1), \bZ^*_1] + 0 + \bC^\T \Cov(\bZ^*_1, \bZ^*_1)
\\
= &  (\bmu_1 - \bmu_0)^\T \bSigma_{yz} + \bC^\T \bSigma_{zz}.
\end{array}
\end{equation}
Hence
$$
\begin{array}{cl}
& 
\bSigma_{xz}\bSigma_{zz}^{-1} \bSigma_{zx} \\
= & [(\bmu_1 - \bmu_0)^\T \bSigma_{yz} + \bC^\T \bSigma_{zz}] \bSigma_{zz}^{-1} [(\bmu_1 - \bmu_0)^\T \bSigma_{yz} + \bC^\T \bSigma_{zz}]^T \\
= & [(\bmu_1 - \bmu_0)^\T \bSigma_{yz} + \bC^\T \bSigma_{zz}] \bSigma_{zz}^{-1} [ \bSigma_{zy}(\bmu_1 - \bmu_0) +  \bSigma_{zz}\bC] \\
= & (\bmu_1 - \bmu_0)^\T \bSigma_{yz} \bSigma_{zz}^{-1} \bSigma_{zy}(\bmu_1 - \bmu_0) + (\bmu_1 - \bmu_0)^\T \bSigma_{yz}  \bC + \bC^\T  \bSigma_{zy}(\bmu_1 - \bmu_0)  + \bC^\T \bSigma_{zz} \bC. 
\end{array}
$$
Combine this expression with \eqref{eq:Sigma_xx}, we simplify the first factor in \eqref{eq:barb1} as
\begin{equation}
\label{eq:factor1}
\begin{array}{cl}
& \bSigma_{xx} - \bSigma_{xz}\bSigma_{zz}^{-1} \bSigma_{zx} \\
= & \bm\Sigma + (\bmu_1 - \bmu_0)^\T\bSigma_{yy}(\bmu_1 - \bmu_0) - (\bmu_1 - \bmu_0)^\T \bSigma_{yz} \bSigma_{zz}^{-1} \bSigma_{zy}(\bmu_1 - \bmu_0)\\
= & \bm\Sigma + (\bmu_1 - \bmu_0)^\T (\bSigma_{yy} - \bSigma_{yz} \bSigma_{zz}^{-1} \bSigma_{zy}) (\bmu_1 - \bmu_0).
\end{array}
\end{equation}

Now
\begin{equation}\label{eq:Sigma_xy}
\begin{array}{cl}
& \bSigma_{xy} = \Cov[\bX^*_1, y^*_1] \\
= & E[\Cov(\bX^*_1, y^*_1\mid \bZ^*_1)] + \Cov[E(\bX^*_1\mid \bZ^*_1), E(y^*_1\mid \bZ^*_1)] \\
= & E[\Cov((\bmu_1 - \bmu_0) y^*_1 + \bmu_0 + \bZ^*_1 \bC, \ y^*_1\mid \bZ^*_1)] \\
& \qquad + \Cov [E( (\bmu_1 - \bmu_0) y^*_1 \mid \bZ^*_1) + \bmu_0 + \bZ^*_1 \bC, \ E(y^*_1 \mid \bZ^*_1)]
\\
= & E[\Cov((\bmu_1 - \bmu_0) y^*_1, \ y^*_1\mid \bZ^*_1)] \\
& \qquad + \Cov [E( (\bmu_1 - \bmu_0) y^*_1 \mid \bZ^*_1), \ E(y^*_1 \mid \bZ^*_1)] + 0 + \Cov [\bZ^*_1 \bC, \ E(y^*_1 \mid \bZ^*_1)]
\\
= &  \Cov[(\bmu_1 - \bmu_0) y^*_1, \ y^*_1]  + \Cov [\bZ^*_1 \bC, \ E(y^*_1 \mid \bZ^*_1)] \\
= &  (\bmu_1 - \bmu_0)^\T \Cov(y^*_1, \ y^*_1)  + \bC^\T \Cov(\bZ^*_1, y^*_1)
\\
= &  (\bmu_1 - \bmu_0)^\T \bSigma_{yy} + \bC^\T \bSigma_{zy},
\end{array}
\end{equation}
where the second to last equality used the property \eqref{eq:prop.A1}.
Combining this expression with \eqref{eq:Sigma_xz}, we simplify the second factor in \eqref{eq:barb1} as
\begin{equation}
\label{eq:factor2}
\begin{array}{cl}
& 
\bSigma_{xy} - \bSigma_{xz}\bSigma_{zz}^{-1}\bSigma_{zy} \\
= & (\bmu_1 - \bmu_0)^\T \bSigma_{yy} + \bC^\T \bSigma_{zy} - [(\bmu_1 - \bmu_0)^\T \bSigma_{yz} + \bC^\T \bSigma_{zz}] \bSigma_{zz}^{-1} \bSigma_{zy} \\
= & (\bmu_1 - \bmu_0)^\T \bSigma_{yy} + \bC^\T \bSigma_{zy} - (\bmu_1 - \bmu_0)^\T \bSigma_{yz} \bSigma_{zz}^{-1} \bSigma_{zy} - \bC^\T \bSigma_{zy}  \\
= & (\bmu_1 - \bmu_0)^\T \bSigma_{yy} - (\bmu_1 - \bmu_0)^\T \bSigma_{yz} \bSigma_{zz}^{-1} \bSigma_{zy}  \\
= & (\bmu_1 - \bmu_0)^\T (\bSigma_{yy} - \bSigma_{yz} \bSigma_{zz}^{-1} \bSigma_{zy}) . 
\end{array}
\end{equation}

Now, \eqref{eq:barb1} can be rewritten as 
$$
(\bSigma_{xx} - \bSigma_{xz}\bSigma_{zz}^{-1} \bSigma_{zx}) \bar\bb_1 = 
(\bSigma_{xy} - \bSigma_{xz}\bSigma_{zz}^{-1}\bSigma_{zy}).
$$
Substituting \eqref{eq:factor1} and \eqref{eq:factor2} into this expression, we have
$$
[\bm\Sigma + (\bmu_1 - \bmu_0)^\T (\bSigma_{yy} - \bSigma_{yz} \bSigma_{zz}^{-1} \bSigma_{zy}) (\bmu_1 - \bmu_0)]\bar\bb_1 = (\bmu_1 - \bmu_0)^\T (\bSigma_{yy} - \bSigma_{yz} \bSigma_{zz}^{-1} \bSigma_{zy}),
$$
which is equivalent to
$$
\begin{array}{cl}
\bm\Sigma \bar\bb_1 & = (\bmu_1 - \bmu_0)^\T (\bSigma_{yy} - \bSigma_{yz} \bSigma_{zz}^{-1} \bSigma_{zy}) - (\bmu_1 - \bmu_0)^\T (\bSigma_{yy} - \bSigma_{yz} \bSigma_{zz}^{-1} \bSigma_{zy}) (\bmu_1 - \bmu_0)\bar\bb_1 \\
& = (\bmu_1 - \bmu_0)^\T (\bSigma_{yy} - \bSigma_{yz} \bSigma_{zz}^{-1} \bSigma_{zy}) [1 - (\bmu_1 - \bmu_0)\bar\bb_1] \\
& = \bm\Sigma \bbeta_1 (\bSigma_{yy} - \bSigma_{yz} \bSigma_{zz}^{-1} \bSigma_{zy}) [1 - (\bmu_1 - \bmu_0)\bar\bb_1].
\end{array}
$$
where the last equality used definition of $\bbeta_1$ from \eqref{eq:beta}. 

Multiplying $\bm\Sigma^{-1}$ to both sides of above equation, we have
$$
\bar\bb_1 = \bbeta_1 (\bSigma_{yy} - \bSigma_{yz} \bSigma_{zz}^{-1} \bSigma_{zy}) [1 - (\bmu_1 - \bmu_0)\bar\bb_1].
$$
Hence, to finish the proof of this Lemma, i.e., to get $\bbeta_1= \bar\bb_1/\Var(\varepsilon_1)$, we only need to show that 
\begin{equation}\label{eq:Var.eps}
\Var(\varepsilon_1) = \Var(y^*_1 - \bW_1^*\bar{\bb}) =(\bSigma_{yy} - \bSigma_{yz} \bSigma_{zz}^{-1} \bSigma_{zy}) [1 - (\bmu_1 - \bmu_0)\bar\bb_1].
\end{equation}

Recall the definition $\bar{\bm{b}} = \left[\Var(\bW_1^*)\right]^{-1} \Cov(\bW_1^*, y_1^*)$. Hence $\Var(\bW_1^*) \bar{\bm{b}} = \Cov(\bW_1^*, y_1^*)$, and we have
$$
\begin{aligned}
\Var(\varepsilon_1) & = \Var(y^*_1 -  \bW_1^*\bar{\bb}) \ \ = \Var(y^*_1) + \bar\bb^\T \Var(\bW_1^*) \bar\bb - 2 \Cov(y_1^*, \bW_1^*) \bar\bb \\
& = \Var(y^*_1) + \bar\bb^\T \Cov(\bW_1^*, y_1^*)  - 2 \Cov(y_1^*, \bW_1^*) \bar\bb \\
& = \Var(y^*_1) - \Cov(y_1^*, \bW_1^*) \bar\bb  \ \ = \Var(y^*_1) - [\Cov(y_1^*, \bX_1^*) \bar\bb_1 + \Cov(y_1^*, \bZ_1^*) \bar\bb_2]  \\ 
& = \bSigma_{yy} -  \bSigma_{yx} \bar\bb_1 - \bSigma_{yz} \bar\bb_2.
\end{aligned}
$$
Using ~\eqref{eq:barb2}, this becomes
$$
\begin{aligned}
\Var(\varepsilon_1) & = \bSigma_{yy} -  \bSigma_{yx} \bar\bb_1 - 
\bSigma_{yz}(\bSigma_{zz}^{-1}\bSigma_{zy} - \bSigma_{zz}^{-1} \bSigma_{zx} \bar\bb_1) \\
& = \bSigma_{yy} - \bSigma_{yz}\bSigma_{zz}^{-1}\bSigma_{zy}
- (\bSigma_{yx} - \bSigma_{yz} \bSigma_{zz}^{-1} \bSigma_{zx}) \bar\bb_1 \\
& = \bSigma_{yy} - \bSigma_{yz}\bSigma_{zz}^{-1}\bSigma_{zy}
- (\bSigma_{xy} - \bSigma_{xz} \bSigma_{zz}^{-1} \bSigma_{xy})^T \bar\bb_1,
\end{aligned}
$$
Using \eqref{eq:factor2}, this becomes
$$
\begin{aligned}
\Var(\varepsilon_1) & =  \bSigma_{yy} - \bSigma_{yz}\bSigma_{zz}^{-1}\bSigma_{zy}
- (\bSigma_{yy} - \bSigma_{yz}\bSigma_{zz}^{-1}\bSigma_{zy}) (\bmu_1 - \bmu_0) \bar\bb_1 \\
& = (\bSigma_{yy} - \bSigma_{yz}\bSigma_{zz}^{-1}\bSigma_{zy}) [1 - (\bmu_1 - \bmu_0) \bar\bb_1].
\end{aligned}
$$

This is indeed \eqref{eq:Var.eps}, thus the proof is finished.

\subsection{Proof of Proposition \ref{prop:unbiasedscore}}\label{sec:proposition}
For the first estimating equation $\bm{m}_{i, \bm\theta}(\bm\theta, \phi)$, recall $\widetilde{\bW}_i = (1, \bW_i)$, $\bm\theta = (b_0, \bar{\bm{b}}^\T)^\T/\tau^2 = (b_0, \bar{\bm{b}}^\T)^\T \phi $, hence $\bm\theta/\phi =  (b_0, \bar{\bm{b}}^\T)^\T$. We have
\begin{equation} \label{eq:est.eq.1}
\begin{array}{cl}
E[m_{i, \bm\theta}(\bm\theta, \phi)] & = E\left\{\widetilde{\bW}_i^{\textsc{T}}y_i - \dfrac{1}{\phi} \left(\widetilde{\bW}_{i}^{\T} \widetilde{\bW}_i  - \sigma^2\bJ_{p+q}\right) \bm\theta \right\}  \\ & = E \left\{\begin{bmatrix} y_i \\\bW_i^\T y_i \end{bmatrix} - \begin{bmatrix} 1 & \bW_i \\
\bW_i^{\T} & \bW_i^{\T}\bW_i-\sigma^2 {\bI}_{p+q}\end{bmatrix} \begin{bmatrix} b_0 \\ \bar{\bm{b}} \end{bmatrix} \right\}  \\ & =  \begin{bmatrix}
E(y_i) - b_0 - E(\bW_i)\bar{\bm{b}}  \\
E(\bW_i^\T y_i) - E(\bW_i^{\T}) a - E(\bW_i^{\T} \bW_i - \sigma^2 \bI_{p+q}) \bar{\bm{b}} 
\end{bmatrix} .
\end{array}
\end{equation}

Because $y_i = y^*_i + v_i$ and $\bW_i = \bW^*_i + \bU_i$, where each element of  $v_i$ and $\bU_i$ has zero expectation and variance $\sigma^2$ and are independent of $y^*_i$ and $\bW^*_i$, we have 
\begin{equation}\label{eq:star.to.obs}
\begin{array}{cc}
E(y_i) = E(y^*_i), & \Var(y_i) = \Var(y^*_i) + \sigma^2, \\
E(\bW_i) = E(\bW_i^*), & \Var(\bW_i) = \Var(\bW_i^*) + \sigma^2 \bI_{p+q}.
\end{array}
\end{equation} 
As a result, the upper expression in ~\eqref{eq:est.eq.1} becomes:
\begin{equation} \label{eq:est.eq.1.1}
E(y_i) - b_0 - E(\bW_i) \bar{\bb} = E(y_i^*) - b_0 - E(\bW_i^*) \bar{\bb} = 0,
\end{equation}
where the last equality is due to the definition $b_0 = E(y_i^*) - E(\bW_i^*)\bar{\bb}$. That is, the upper expression in ~\eqref{eq:est.eq.1} is unbiased. 

Since $\bW_i = \bW^*_i + \bU_i$, and $\bW^*_i$ is independent of $\bU_i$, we have
$$
E(\bW_i^{\T} \bW_i) = E(\bW_i^{*\T} \bW^*_i) + 0 + E(\bU_i^{\T} \bU_i) = E(\bW_i^{*\T} \bW^*_i) + \sigma^2 \bI_{p+q}.
$$
Similarly, $y_i = y^*_i + v_i$ with $y^*_i$ is independent of $v_i$, implying
$$
E(y_i^2) = E(y_i^{*2}) + 0 + E(v_i^2) = E(y_i^{*2}) + \sigma^2.
$$
Thus
\begin{equation}\label{eq:star.square}
  E(y_i^2) -  \sigma^2 = E(y_i^{*2}), \qquad  E(\bW_i^{\T} \bW_i) - \sigma^2 \bI_{p+q} = E(\bW_i^{*\T} \bW^*_i).
\end{equation}

Using \eqref{eq:star.square} and $E(\bW_i) = E(\bW_i^*)$ from \eqref{eq:star.to.obs}, the bottom expression in~\eqref{eq:est.eq.1} becomes
$$
\begin{array}{cl}
& E(\bW_i^\T y_i) - E(\bW_i^\T) b_0 - [E(\bW_i^{\T} \bW_i) - \sigma^2\bI_{p+q}] \bar{\bm{b}} \\
= & E(\bW_i^\T y_i) - E[(\bW^*_i)^\T] b_0 - E[(\bW^*_i)^{\T} \bW^*_i] \bar{\bm{b}}. 
\end{array}
$$
Recall $y_i = y^*_i + v_i$, $\bW_i = \bW^*_i + \bU_i$. $v_i$ and $\bU_i$ are independent of each other with zero means, and independent from both $y^*_i$ and $\bW^*_i$, thus $E(\bW_i^\T y_i) = E[(\bW^*_i)^T(y^*_i)]$. Put this into the above expression, we have
$$
\begin{array}{cl}
& E(\bW_i^\T y_i) - E(\bW_i^\T) b_0 - [E(\bW_i^{\T} \bW_i) - \sigma^2\bI_{p+q}] \bar{\bm{b}} \\
= & E[(\bW^*_i)^T(y^*_i)] - E[(\bW^*_i)^\T] b_0 - E[(\bW^*_i)^{\T} \bW^*_i] \bar{\bm{b}} \\
= & E[(\bW^*_i)^T(y^*_i)] - E[(\bW^*_i)^\T] [E(y_i^*) - E(\bW_i^*)\bar{\bb}] - E[(\bW^*_i)^{\T} \bW^*_i] \bar{\bm{b}} \\
= & E[(\bW^*_i)^T(y^*_i)] - E[(\bW^*_i)^\T] E(y_i^*)  - \{E[(\bW^*_i)^{\T} \bW^*_i] - E[(\bW^*_i)^\T]E(\bW_i^*) \} \bar{\bm{b}} \\
= & \Cov(\bW_i^*, y_i^*) - \Var(\bW^*_i) \bar{\bb}. 
\end{array}
$$
This expression equals zero due the definition $\bar{\bb} = [\Var(\bW_i^*)]^{-1} \Cov(\bW_i^*, y_i^*) $. That is, the bottom expression in ~\eqref{eq:est.eq.1} is unbiased. Substituting this and \eqref{eq:est.eq.1.1} into \eqref{eq:est.eq.1}, we have proved the unbiasedness of the first estimating equation: $E[\bm{m}_{i, \bm\theta}(\bm\theta, \phi)] = \bm{0}$. 

For the second estimating equation $\bm{m}_{i, \phi}(\bm\theta, \phi)$, we have
$$
\begin{array}{cl}
2 E\{\bm{m}_{i, \bm\theta}(\bm\theta, \phi)\} & = E \left\{\dfrac{1}{\phi} - \left(y_i^2 - \sigma^2 \right) +  \dfrac{\bm\theta}{\phi}^\top \left(\widetilde{\bW}_i\widetilde{\bW}_i^\T - \sigma^2 \bJ_{p+q} \right) \dfrac{\bm\theta}{\phi}     \right\} \\
& = \dfrac{1}{\phi} - \left\{E(y_i^2) - \sigma^2\right\} + \begin{bmatrix} b_0 & \bar{\bb}^\T \end{bmatrix} \begin{bmatrix} 1 & E(\bW_i) \\
E(\bW_i^{\T}) & E(\bW_i^{\T}\bW_i - \sigma^2 \bI_{p+q}) \end{bmatrix} \begin{bmatrix} b_0 \\ \bar{\bm{b}} \end{bmatrix} 
\\
& = \dfrac{1}{\phi} - \left\{E(y_i^2) - \sigma^2\right\} + b_0^2 + 2aE(\bW_i )\bar{\bm{b}} +  \bar{\bm{b}}^\T [E(\bW_i^{\T}\bW_i) - \sigma^2 \bI_{p+q}] \bar{\bm{b}} 
\end{array}
$$
Using \eqref{eq:star.square} and $E(\bW_i) = E(\bW_i^*)$ from \eqref{eq:star.to.obs},  this becomes
$$
\begin{array}{cl}
2 E\{\bm{m}_{i, \bm\theta}(\bm\theta, \phi)\} 
& = \dfrac{1}{\phi} - \left\{E(y_i^2) - \sigma^2\right\} + b_0^2 + 2b_0E(\bW_i )\bar{\bm{b}} +  \bar{\bm{b}}^\T [E(\bW_i^{\T}\bW_i) - \sigma^2 \bI_{p+q}] \bar{\bm{b}} \\[1em]
& = \dfrac{1}{\phi} - E(y_i^{*2}) + b_0^2 + 2b_0E(\bW_i^* )\bar{\bm{b}} +  \bar{\bm{b}}^\T E(\bW_i^{*\T}\bW_i^*)  \bar{\bm{b}} 
\end{array}
$$
Recall that $1/\phi = \tau^2 = \Var(\varepsilon_i) = \Var(y_i^* - \widetilde{\bW}_i^*\bar\bb) = \Var(y_i^*) - \bar\bb^\T \Var(\bW_i^*)\bar{\bb}$,  the above expression becomes
$$
\begin{array}{cl}
2 E\{\bm{m}_{i, \bm\theta}(\bm\theta, \phi)\} 
& = \dfrac{1}{\phi} - E(y_i^{*2}) + b_0^2 + 2b_0E(\bW_i^* )\bar{\bm{b}} +  \bar{\bm{b}}^\T E(\bW_i^{*\T}\bW_i^*)  \bar{\bm{b}}  \\
& = \Var(y_i^*) - \bar\bb^\T \Var(\bW_i^*)\bar{\bb} - E(y_i^{*2}) + b_0^2 + 2b_0E(\bW_i^* )\bar{\bm{b}} +  \bar{\bm{b}}^\T E(\bW_i^{*\T}\bW_i^*)  \bar{\bm{b}} \\
& = [\Var(y_i^*) - E(y_i^{*2})] - \bar\bb^\T [\Var(\bW_i^*) - E(\bW_i^{*\T}\bW_i^*)] \bar{\bb}  + b_0^2 + 2b_0E(\bW_i^* )\bar{\bm{b}} 
\\
& = -\{E(y_i^*)\}^2  + \bar\bb^\T [E(\bW_i^{*\T}) E(\bW^*_i)] \bar{\bb}  + b_0^2 + 2b_0E(\bW_i^* )\bar{\bm{b}}
\\
& = -\{E(y_i^*)\}^2  + [E(\bW^*_i) \bar{\bb}]^2  + b_0[b_0 + 2E(\bW_i^* )\bar{\bm{b}}].
\end{array}
$$
Using the definition $b_0 = E(y_i^*) - E(\bW_i^*)\bar{\bb}$, we have
$$
\begin{array}{cl}
2 E\{\bm{m}_{i, \bm\theta}(\bm\theta, \phi)\} 
& = -\{E(y_i^*)\}^2  + [E(\bW^*_i) \bar{\bb}]^2   + b_0[b_0 + 2E(\bW_i^* )\bar{\bm{b}}] \\
& = -\{E(y_i^*)\}^2  + [E(\bW^*_i) \bar{\bb}]^2   + [E(y_i^*) - E(\bW_i^*)\bar{\bb}][E(y_i^*) + E(\bW_i^*)\bar{\bb}]\\
& = -\{E(y_i^*)\}^2  + [E(\bW^*_i) \bar{\bb}]^2   + \{E(y_i^*)\}^2 - [E(\bW^*_i) \bar{\bb}]^2 \\
& =0.
\end{array}
$$
That is, the second estimating equation is also unbiased. This finishes the proof.

\subsection{Proof of Theorem~\ref{thm: asymptotic.normality.correctedestimator}}\label{sec:theorem}

We prove Theorem~\ref{thm: asymptotic.normality.correctedestimator} using the classical theory of estimating equations. It suffices to establish the consistency of $(\hat{\bm\theta}^{(c)}, \hat{\phi}^{(c)})$, which is the solution of the estimating equation~\eqref{eq:rawestimatorsCONDITIONAL}: $\sum_{i=1}^{n}\bm{m}_{i, \bm\theta}(\bm\theta, \phi) =0$ and $\sum_{i=1}^{n}m_{i, \phi}(\bm\theta, \phi) =0$ with 
$$
\bm{m}_{i, \bm\theta}(\bm\theta, \phi) = \widetilde{\bW}_i^\T y_i - \dfrac{1}{\phi} (\widetilde{\bW}_i^\T\widetilde{\bW}_i - \sigma^2\bJ_{p+q}) \bm\theta
$$
and 
$$
m_{i, \phi}(\bm\theta, \phi) = \dfrac{1}{2\phi} - \dfrac{1}{2} \left(y_i^2 - \sigma^2 \right) +  \dfrac{1}{2\phi^2}\bm\theta^\top \left(\widetilde{\bW}_i^\T\widetilde{\bW}_i - \sigma^2\widetilde{\bI}_{p+q} \right)\bm\theta.
$$

To simplify the notations, we combine the equations in~\eqref{eq:rawestimatorsCONDITIONAL} as 
$\bm{m}_i(\bm\theta, \phi) =
[\bm{m}_{i, \bm\theta}(\bm\theta, \phi)^\T,  m_{i, \phi}(\bm\theta, \phi)]^\T.
$
We have shown above that the estimating equation is unbiased so that $E[\bm{m}_{i}(\bm\theta, \phi)] = \boldsymbol{0}$. 

Denote $\bm\Psi = (\bm\theta, \phi)$ and $\bS(\bm\Psi) = n^{-1} \sum_{i=1}^{n} \bm{m}_{i}(\bm\theta, \phi)$, then the estimator $\widehat{\bm\Psi}^{(c)} = (\hat{\bm\theta}^{(c)}, \hat\phi^{(c)})$ is the solution of $\bS(\bm\Psi) = \boldsymbol{0}$. That is, $\bS(\widehat{\bm\Psi}^{(c)}) = \bS(\widehat{\bm{\theta}}^{(c)}, \hat\phi^{(c)}) = \mathbf{0}$. 

By the Taylor's series expansion, we have
\begin{equation}\label{eq:taylor}
\mathbf{0} = \bS(\widehat{\bm\Psi}^{(c)}) = \bS({\bm\Psi}) + \nabla\bS(\bm\Psi) (\widehat{\bm\Psi}^{(c)} - \bm\Psi) + o_p(n^{-1/2}).
\end{equation}
Since $E[\bm{m}_{i}(\bm\theta, \phi)] = \boldsymbol{0}$, the quantity $\bS(\bm\Psi) = n^{-1} \sum_{i=1}^{n} \bm{m}_{i}(\bm\theta, \psi)$ is the average of $n$ iid random variables, so applying the central limit theorem yields
\begin{equation}\label{eq:clm}
\sqrt{n}\bS(\bm\Psi) \xrightarrow{d} N\left(\mathbf{0}, \bB \right),
\end{equation}
where $\bB = \Var[\bm{m}_1(\bm\theta, \phi)]$.
We make the following claim regarding the behavior of the first derivative $\nabla\bS(\bm\Psi)$:
\vspace{-1em}
\paragraph{Claim} Under the conditions stated in the theorem, we have $\nabla\bS(\bm\Psi) = n^{-1} \sum_{i=1}^{n} \nabla \bm{m}_i(\bm\theta, \phi) \xrightarrow{p} \bA $, where $\bA  = E_{\bm\theta, \phi}[\nabla m_1(\bm\theta, \phi)]$  is a negative definite matrix. 

If this claim is true, using \eqref{eq:taylor} and \eqref{eq:clm}, we would arrive at
$$
\sqrt{n} (\widehat{\bm\Psi}^{(c)} - \bm\Psi) = [-\nabla\bS(\bm\Psi)]^{-1}\sqrt{n}\bS(\bm\Psi) + o_p(1) 
\xrightarrow{d} N\left(\mathbf{0}, \bA^{-1} \bB \bA^{-1} \right)    
$$
as stated in the theorem.  

\subsubsection{Proof of the claim}

Direct calculation of the gradient shows that
\begin{equation}\label{eq:grad.mi}
\nabla \bm{m}_i(\bm\theta, \phi) = \begin{bmatrix}
-\phi^{-1}(\widetilde{\bW}_i^\T \widetilde{\bW}_i - \sigma^2 \bJ_{p+q}) & \phi^{-2} (\widetilde{\bW}_i^\T \widetilde{\bW}_i - \sigma^2 \bJ_{p+q}) \bm\theta \\[.5em]
\phi^{-2}\bm\theta^\T (\widetilde{\bW}_i^\T \widetilde{\bW}_i - \sigma^2 \bJ_{p+q}) & -1/(2\phi^{2}) -\phi^{-3} \bm\theta^\top (\widetilde{\bW}_i^\T \widetilde{\bW}_i - \sigma^2 \bJ_{p+q})\bm\theta
\end{bmatrix}
\end{equation}

Denote $\bQ_i = (\widetilde{\bW}_i^\T \widetilde{\bW}_i - \sigma^2 \bJ_{p+q})$, thus $\nabla\bS(\bm\Psi) = n^{-1} \sum_{i=1}^{n} \nabla \bm{m}_i(\bm\theta, \phi)$ becomes
\begin{equation}\label{eq:grad.Q}
\nabla\bS(\bm\Psi) =  \begin{bmatrix}
-\phi^{-1}(n^{-1} \sum_{i=1}^{n} \bQ_i) & \phi^{-2} (n^{-1} \sum_{i=1}^{n} \bQ_i) \bm\theta \\[.5em]
\phi^{-2}\bm\theta^\T (n^{-1} \sum_{i=1}^{n} \bQ_i) & -1/(2\phi^{2}) -\phi^{-3} \bm\theta^\top (n^{-1} \sum_{i=1}^{n} \bQ_i) \bm\theta
\end{bmatrix}.
\end{equation}

First, we prove that $n^{-1} \sum_{i=1}^{n} \bQ_i$ converges to $\bar{\bQ} = E(\bQ_1)$. 

Denote $\bQ_i^{(0)}=({\bW}_i^\T {\bW}_i - \sigma^2 {\bI}_{p+q})$ so that $\bQ_i = (\widetilde{\bW}_i^\T \widetilde{\bW}_i - \sigma^2 \bJ_{p+q}) = \begin{bmatrix}
1 &  \bW_i \\
\bW_i^\T &  \bQ_i^{(0)} \\
\end{bmatrix}$, we have
$$
n^{-1} \sum_{i=1}^{n} \bQ_i = \begin{bmatrix}
1 & n^{-1} \sum_{i=1}^{n} \bW_i \\
n^{-1} \sum_{i=1}^{n} \bW_i^\T &  n^{-1} \sum_{i=1}^{n} \bQ_i^{(0)} \\
\end{bmatrix}.
$$
For the off-diagonal sub-vector, the law of large number ensures that 
\begin{equation}\label{eq:converge.1}
   n^{-1} \sum_{i=1}^{n} \bW_i  \xrightarrow{p} E(\bW_1),
\end{equation}
as long as $\Var(\bW_1)$ is finite. The finiteness of $\Var(\bW_i)$ is ensured since $\Var(\bW_i) = \Var(\bW_i^*) + \sigma^2 \bI_{p+q}$ where $\Var(\bW_i^*)$ is finite by condition~\ref{cond:marginalmeanandvarianceZ}. 

For the bottom right block sub-matrix in $n^{-1} \sum_{i=1}^{n} \bQ_i^{(0)}$, we decompose $\bQ_i$ into four components:
\begin{align*}
\bQ_i^{(0)} & =  \left\{(\bW_i^* + \bU_i)^\T(\bW_i^* + \bU_i) - \sigma^2 \bI_{p+q} \right\} \\ & =  \bW_i^{*\T} \bW_i^* +  \bW_i^{*\T} \bU_i + \bU_i^{\T} \bW_i^* + \left(\bU_i^{\T} \bU_i - \sigma^2 \bI_{p+q} \right) \\ &= \bQ_i^{(1)} + \bQ_i^{(2)} +\bQ_i^{(3)} + \bQ_i^{(4)} .
\end{align*}
The law of large number ensures that, for $k=1,..,4$,  
\begin{equation}\label{eq:converge.k}
n^{-1} \sum_{i=1}^{n} \bQ_i^{(k)} \xrightarrow{p}  E(\bQ_1^{(k)}),
\end{equation}
if $n^{-1} \Var(\bQ_1^{(k)}) \to \mathbf{0}$ when $n \to \infty$. 

By condition~\ref{cond:marginalmeanandvarianceZ}, $\Var(\bQ_1^{(1)}) = Var(\bW_1^{*\T} \bW_1^*)$ is finite, thus $n^{-1} \Var(\bQ_1^{(1)}) \to \mathbf{0}$ when $n \to \infty$. 

Since $\bW_1^*$ and $\bU_1$ are independent and $\bU_1 \sim N(\mathbf{0}, \sigma^2 \bI_{p+q})$, then each element of $\bQ_1^{(2)}=\bW_i^{*\T} \bU_i$ and each element of $\bQ_1^{(3)}=\bU_i^{\T} \bW_i^*$ has variance of order $O(\sigma^2)$, and each element of $\bQ_1^{(4)}=(\bU_i^{\T} \bU_i - \sigma^2 \bI_{p+q})$ has variance of order $O(\sigma^4)$.
Hence, under the conditions for the Theorem, when both $\sigma^2$ and $n$ diverge to infinity but $\sigma^4/n \to 0$, $n^{-1} \Var(\bQ_1^{(k)}) \to \mathbf{0}$ for $k=2,3,4$. 

Putting together, \eqref{eq:converge.k} holds for all $k=1,2,3,4$. Therefore,
\begin{equation}\label{eq:converge.2}
   n^{-1} \sum_{i=1}^{n} \bQ_i^{(0)} \xrightarrow{p} 
   E(\bQ_1^{(1)}) + E(\bQ_1^{(2}) + E(\bQ_1^{(3)}) + E(\bQ_1^{(4)}) = E(\bQ_i^{(0)}).
\end{equation}

Combine \eqref{eq:converge.1} and \eqref{eq:converge.2}, we get 
$$
n^{-1} \sum_{i=1}^{n} \bQ_i \xrightarrow{p}  E(\bQ_1) = \bar{\bQ}. 
$$

Plug this into \eqref{eq:grad.Q}, we have
\begin{equation}
\nabla\bS(\bm\Psi) \xrightarrow{p}  \begin{bmatrix}
-\phi^{-1}\bar{\bQ} & \phi^{-2} \bar{\bQ} \bm\theta \\[.5em]
\phi^{-2}\bm\theta^\T \bar{\bQ} & -1/(2\phi^{2}) -\phi^{-3} \bm\theta^\top \bar{\bQ} \bm\theta
\end{bmatrix}.
\end{equation}

Using \eqref{eq:grad.mi}, it is easy to see that the above limit is
$$
\begin{bmatrix}
-\phi^{-1}\bar{\bQ} & \phi^{-2} \bar{\bQ} \bm\theta \\[.5em]
\phi^{-2}\bm\theta^\T \bar{\bQ} & -1/(2\phi^{2}) -\phi^{-3} \bm\theta^\top \bar{\bQ} \bm\theta
\end{bmatrix} = E[ \nabla \bm{m}_1(\bm\theta, \phi)] = \bA. 
$$

That is, we have proved that, when both $\sigma^2$ and $n$ diverge to infinity but $\sigma^4/n \to 0$,
$$\nabla\bS(\bm\Psi) \xrightarrow{p} \bA.$$

What remains is to prove that $\bA$  is indeed  negative definite. Equivalently, we show that $\bar{\bA} = - \bA$ is positive definite. 

Notice that 
$$
\bar{\bA} = \begin{bmatrix}
\phi^{-1}\bar{\bQ} & - \phi^{-2} \bar{\bQ} \bm\theta \\[.5em]
- \phi^{-2}\bm\theta^\T \bar{\bQ} & 1/(2\phi^{2}) + \phi^{-3} \bm\theta^\top \bar{\bQ} \bm\theta
\end{bmatrix}
$$
would be positive definite if $\bar{\bQ}$ is positive definite, due to the following arguments. 

If $\bar{\bQ}$ is positive definite, then the upper left block sub-matrix in $\bar{\bA}$, $\phi^{-1}\bar{\bQ}$, is also positive definite (since $\phi^{-1} = \tau^2$ is a positive scalar). Hence  all the  $k$-th leading principal minors of $\bar{\bA}$ are positive for $k=1,\ldots, p$. Applying the formula for the determinant of the block matrix, we have \begin{align*}
\det\left(\bar{\bA}\right) &= \det(\phi^{-1} \bar{\bQ}) \left\{(1/2)\phi^{-2} +\phi^{-3} \bm\theta^\top \bar{\bQ}\bm\theta - (-\phi^{-2}\bm\theta^\top \bar{\bQ} (\phi \bar{\bQ}^{-1})(-\phi^{-2}  \bar{\bQ} \bm\theta) \right\}
\\ &=\det(\phi^{-1} \bar{\bQ}) \left\{(1/2)\phi^{-2} - \phi^{-3} \bm\theta^\top  \bar{\bQ} \bm\theta + \phi^{-3} \bm\theta^\top  \bar{\bQ} \bm\theta \right\} 
\\& = \det(\phi^{-1}\bar{\bQ}) (1/2) \phi^{-2} 
\\& > 0,
\end{align*}
so the $(p+1)$ leading principal minor of $\bar{\bA}$ is also positive.  Hence, using the Sylvester's criterion, $\bar{\bA}$ is positive definite. 

Finally, we show that $\bar{\bQ}$ is positive definite. Recall
$$
\bar{\bQ} = E(\bQ_1) = \begin{bmatrix}
1 & E(\bW_1) \\
E(\bW_1^{\T}) & E(\bQ_1^{(0)})
\end{bmatrix}.
$$
Notice $E(\bW_1) = E(\bW_1^* + \bU_1) = E(\bW_1^*)$. Also,
\begin{align*}
E(\bQ_1^{(0)}) & = E(\bQ_1^{(1)}) + E(\bQ_1^{(2)}) +E(\bQ_1^{(3)}) + E(\bQ_1^{(4)}) \\
& = E(\bW_1^{*\T} \bW_1^*) + E(\bW_1^{*\T} \bU_1) +E(\bU_1^{\T} \bW_1^*) + E\left(\bU_i^{\T} \bU_i - \sigma^2 \bI_{p+q} \right) \\
& = E(\bW_1^{*\T} \bW_1^*) + \mathbf{0} + \mathbf{0} + \left(\sigma^2 \bI_{p+q}  - \sigma^2 \bI_{p+q} \right) \\
& = E(\bW_1^{*\T} \bW_1^*).
\end{align*}
Hence
$$
\bar{\bQ} = \begin{bmatrix}
1 & E(\bW_1^*) \\
E(\bW_1^{*\T}) & E(\bW_1^{*\T} \bW_1^*)
\end{bmatrix}
$$
which is indeed positive definite by condition 2.

\subsection*{Additional simulation results}
\begin{table}[ht]
\centering
\caption{Simulation results when the raw data were generated from the unconditional mixture normal distribution with $p_1 = 0.1$. Reported are $(10 \times)$ bias, $(100 \times)$ MSE for point estimators (MLE: naive conditional MLE, LS: naive least square, and cLS: corrected least square), and coverage probability for the confidence interval corresponding to each coefficient of the slope vector.}
\label{tab:sim_results_mixturemodel.extra1}
\resizebox{!}{.45\textheight}{\begin{tabular}{lll rrr rrr rrr}
\toprule[1.5pt]
$\sigma$ & $n$  & Coef. & \multicolumn{3}{c}{Bias} & \multicolumn{3}{c}{MSE} & \multicolumn{3}{c}{Coverage} \\
&  & & MLE & LS & cLS & MLE & LS & cLS & MLE & LS & cLS \\
\cmidrule(lr){4-6} \cmidrule(lr){7-9} \cmidrule(lr){10-12}
0 & $ 10^3$ & $\beta_1$ & -0.21 & -0.16 & -0.16 & 1.85 & 1.67 & 1.67 & 0.95 & 0.95 & 0.95\\
 &  & $\beta_2$ & 0.19 & 0.13 & 0.13 & 2.37 & 2.21 & 2.21 & 0.95 & 0.95 & 0.95\\
 &  & $\beta_3$ & -0.01 & -0.01 & -0.01 & 1.72 & 1.62 & 1.62 & 0.95 & 0.95 & 0.95\\
\addlinespace
 & $ 10^4$ & $\beta_1$ & 0.00 & 0.00 & 0.00 & 0.18 & 0.16 & 0.16 & 0.95 & 0.95 & 0.95\\
 &  & $\beta_2$ & 0.01 & 0.01 & 0.01 & 0.22 & 0.21 & 0.21 & 0.95 & 0.95 & 0.95\\
 &  & $\beta_3$ & -0.02 & -0.02 & -0.02 & 0.15 & 0.15 & 0.15 & 0.96 & 0.97 & 0.97\\
\addlinespace
 & $ 10^5$ & $\beta_1$ & 0.00 & 0.00 & 0.00 & 0.02 & 0.02 & 0.02 & 0.94 & 0.94 & 0.94\\
 &  & $\beta_2$ & 0.00 & 0.01 & 0.01 & 0.02 & 0.02 & 0.02 & 0.95 & 0.96 & 0.96\\
 &  & $\beta_3$ & 0.00 & 0.00 & 0.00 & 0.02 & 0.02 & 0.02 & 0.96 & 0.96 & 0.96\\
\addlinespace
 & $ 2\times 10^5$ & $\beta_1$ & 0.00 & 0.00 & 0.00 & 0.01 & 0.01 & 0.01 & 0.94 & 0.94 & 0.94\\
 &  & $\beta_2$ & 0.00 & 0.00 & 0.00 & 0.01 & 0.01 & 0.01 & 0.93 & 0.93 & 0.93\\
 &  & $\beta_3$ & 0.00 & 0.00 & 0.00 & 0.01 & 0.01 & 0.01 & 0.95 & 0.95 & 0.95\\
\addlinespace
0.3 & $ 10^3$ & $\beta_1$ & 1.27 & 5.91 & -0.21 & 6.68 & 35.64 & 5.30 & 0.63 & 0.00 & 0.95\\
 &  & $\beta_2$ & -1.42 & -5.98 & 0.28 & 7.82 & 36.65 & 6.61 & 0.64 & 0.00 & 0.96\\
 &  & $\beta_3$ & 0.35 & 0.17 & -0.13 & 3.35 & 0.70 & 4.19 & 0.81 & 0.94 & 0.96\\
\addlinespace
 & $ 10^4$ & $\beta_1$ & 1.48 & 5.91 & -0.03 & 2.61 & 34.96 & 0.47 & 0.14 & 0.00 & 0.95\\
 &  & $\beta_2$ & -1.70 & -6.01 & 0.02 & 3.31 & 36.21 & 0.54 & 0.11 & 0.00 & 0.97\\
 &  & $\beta_3$ & 0.46 & 0.22 & 0.01 & 0.48 & 0.11 & 0.35 & 0.70 & 0.89 & 0.96\\
\addlinespace
 & $ 10^5$ & $\beta_1$ & 1.52 & 5.91 & 0.00 & 2.35 & 34.99 & 0.05 & 0.00 & 0.00 & 0.95\\
 &  & $\beta_2$ & -1.72 & -6.01 & 0.00 & 3.02 & 36.17 & 0.06 & 0.00 & 0.00 & 0.95\\
 &  & $\beta_3$ & 0.44 & 0.21 & 0.00 & 0.23 & 0.05 & 0.04 & 0.12 & 0.26 & 0.94\\
\addlinespace
 & $ 2\times 10^5$ & $\beta_1$ & 1.51 & 5.91 & 0.00 & 2.31 & 34.97 & 0.02 & 0.00 & 0.00 & 0.96\\
 &  & $\beta_2$ & -1.72 & -6.01 & 0.00 & 3.00 & 36.18 & 0.03 & 0.00 & 0.00 & 0.95\\
 &  & $\beta_3$ & 0.44 & 0.21 & 0.00 & 0.21 & 0.05 & 0.02 & 0.01 & 0.05 & 0.95\\
\addlinespace
1 & $ 10^3$ & $\beta_1$ & 5.38 & 9.72 & 151.15 & 61.52 & 94.50 & 21448457.38 & 0.09 & 0.00 & 0.96\\
 &  & $\beta_2$ & -6.26 & -9.76 & -248.94 & 66.92 & 95.34 & 54472070.48 & 0.09 & 0.00 & 0.97\\
 &  & $\beta_3$ & 1.49 & 0.09 & -161.65 & 23.50 & 0.06 & 26627838.31 & 0.40 & 0.93 & 1.00\\
\addlinespace
 & $ 10^4$ & $\beta_1$ & 6.53 & 9.73 & -0.66 & 43.61 & 94.62 & 13.69 & 0.00 & 0.00 & 0.95\\
 &  & $\beta_2$ & -6.78 & -9.75 & 0.77 & 46.88 & 95.00 & 17.76 & 0.00 & 0.00 & 0.96\\
 &  & $\beta_3$ & 0.88 & 0.07 & -0.15 & 1.62 & 0.01 & 6.78 & 0.27 & 0.83 & 0.98\\
\addlinespace
 & $ 10^5$ & $\beta_1$ & 6.59 & 9.73 & -0.07 & 43.55 & 94.62 & 1.01 & 0.00 & 0.00 & 0.95\\
 &  & $\beta_2$ & -6.87 & -9.75 & 0.06 & 47.29 & 95.06 & 1.30 & 0.00 & 0.00 & 0.95\\
 &  & $\beta_3$ & 0.89 & 0.07 & -0.02 & 0.87 & 0.01 & 0.64 & 0.01 & 0.11 & 0.94\\
\addlinespace
 & $ 2\times 10^5$ & $\beta_1$ & 6.61 & 9.73 & -0.03 & 43.68 & 94.63 & 0.48 & 0.00 & 0.00 & 0.95\\
 &  & $\beta_2$ & -6.87 & -9.75 & 0.05 & 47.26 & 95.04 & 0.63 & 0.00 & 0.00 & 0.95\\
 &  & $\beta_3$ & 0.89 & 0.07 & -0.01 & 0.83 & 0.01 & 0.28 & 0.00 & 0.01 & 0.96\\
\addlinespace
3 & $ 10^3$ & $\beta_1$ & 9.72 & 10.00 & 28.34 & 110.88 & 99.95 & 142660.62 & 0.01 & 0.00 & 0.92\\
 &  & $\beta_2$ & -9.71 & -10.00 & -17.24 & 111.32 & 99.95 & 205185.92 & 0.02 & 0.00 & 0.92\\
 &  & $\beta_3$ & 0.48 & 0.00 & 7.53 & 29.39 & 0.00 & 151164.11 & 0.19 & 0.97 & 1.00\\
\addlinespace
 & $ 10^4$ & $\beta_1$ & 9.25 & 9.99 & 19.89 & 88.88 & 99.89 & 187187.10 & 0.00 & 0.00 & 0.93\\
 &  & $\beta_2$ & -9.32 & -9.99 & -10.12 & 89.33 & 99.89 & 190271.88 & 0.00 & 0.00 & 0.96\\
 &  & $\beta_3$ & 0.28 & 0.00 & -1.05 & 3.69 & 0.00 & 63354.14 & 0.15 & 0.94 & 0.99\\
\addlinespace
 & $ 10^5$ & $\beta_1$ & 9.47 & 9.99 & -94.81 & 89.77 & 99.90 & 11343398.03 & 0.00 & 0.00 & 0.92\\
 &  & $\beta_2$ & -9.46 & -9.99 & 114.14 & 89.66 & 99.90 & 15738168.07 & 0.00 & 0.00 & 0.95\\
 &  & $\beta_3$ & 0.24 & 0.00 & 78.27 & 0.19 & 0.00 & 6952301.39 & 0.11 & 0.88 & 1.00\\
\addlinespace
 & $ 2\times 10^5$ & $\beta_1$ & 9.47 & 9.99 & -8.17 & 89.68 & 99.90 & 4996.50 & 0.00 & 0.00 & 0.95\\
 &  & $\beta_2$ & -9.48 & -9.99 & 8.60 & 89.84 & 99.90 & 7689.45 & 0.00 & 0.00 & 0.93\\
 &  & $\beta_3$ & 0.26 & 0.00 & -0.56 & 0.12 & 0.00 & 778.56 & 0.09 & 0.82 & 1.00\\
 \bottomrule[1.5pt]
 \end{tabular}
 }
\end{table}

\begin{table}[ht]
\centering
\caption{Simulation results when the raw data were generated from the unconditional mixture normal distribution with $p_1 = 0.9$. Reported are $(10 \times)$ bias, $(100 \times)$ MSE for point estimators (MLE: naive conditional MLE, LS: naive least square, and cLS: corrected least square), and coverage probability for the confidence interval corresponding to each coefficient of the slope vector.}
\label{tab:sim_results_mixturemodel.extra2}
\resizebox{!}{.45\textheight}{\begin{tabular}{lll rrr rrr rrr}
\toprule[1.5pt]
$\sigma$ & $n$  & Coef. & \multicolumn{3}{c}{Bias} & \multicolumn{3}{c}{MSE} & \multicolumn{3}{c}{Coverage} \\
&  & & MLE & LS & cLS & MLE & LS & cLS & MLE & LS & cLS \\
\cmidrule(lr){4-6} \cmidrule(lr){7-9} \cmidrule(lr){10-12}
0 & $ 10^3$ & $\beta_1$ & -0.01 & 0.01 & 0.01 & 1.88 & 1.81 & 1.81 & 0.95 & 0.94 & 0.94\\
 &  & $\beta_2$ & 0.10 & 0.07 & 0.07 & 2.30 & 2.22 & 2.22 & 0.95 & 0.94 & 0.94\\
 &  & $\beta_3$ & -0.01 & -0.01 & -0.01 & 1.58 & 1.55 & 1.55 & 0.95 & 0.96 & 0.96\\
\addlinespace
 & $ 10^4$ & $\beta_1$ & -0.01 & -0.01 & -0.01 & 0.18 & 0.17 & 0.17 & 0.96 & 0.95 & 0.95\\
 &  & $\beta_2$ & 0.01 & 0.01 & 0.01 & 0.22 & 0.20 & 0.20 & 0.95 & 0.95 & 0.95\\
 &  & $\beta_3$ & -0.02 & -0.01 & -0.01 & 0.17 & 0.17 & 0.17 & 0.95 & 0.95 & 0.95\\
\addlinespace
 & $ 10^5$ & $\beta_1$ & 0.00 & 0.00 & 0.00 & 0.02 & 0.02 & 0.02 & 0.95 & 0.96 & 0.96\\
 &  & $\beta_2$ & 0.00 & -0.01 & -0.01 & 0.02 & 0.02 & 0.02 & 0.96 & 0.96 & 0.96\\
 &  & $\beta_3$ & 0.00 & 0.00 & 0.00 & 0.02 & 0.01 & 0.01 & 0.97 & 0.96 & 0.96\\
\addlinespace
 & $ 2\times 10^5$ & $\beta_1$ & 0.00 & -0.01 & -0.01 & 0.01 & 0.01 & 0.01 & 0.96 & 0.96 & 0.96\\
 &  & $\beta_2$ & 0.01 & 0.01 & 0.01 & 0.01 & 0.01 & 0.01 & 0.95 & 0.94 & 0.94\\
 &  & $\beta_3$ & 0.00 & 0.00 & 0.00 & 0.01 & 0.01 & 0.01 & 0.96 & 0.96 & 0.96\\
\addlinespace
0.3 & $ 10^3$ & $\beta_1$ & 1.24 & 5.89 & -0.22 & 6.47 & 35.44 & 5.34 & 0.63 & 0.00 & 0.95\\
 &  & $\beta_2$ & -1.45 & -5.98 & 0.24 & 7.05 & 36.57 & 5.98 & 0.64 & 0.00 & 0.96\\
 &  & $\beta_3$ & 0.45 & 0.21 & -0.01 & 3.31 & 0.67 & 3.81 & 0.81 & 0.95 & 0.96\\
\addlinespace
 & $ 10^4$ & $\beta_1$ & 1.51 & 5.92 & 0.02 & 2.71 & 35.17 & 0.49 & 0.13 & 0.00 & 0.94\\
 &  & $\beta_2$ & -1.70 & -6.02 & 0.00 & 3.34 & 36.27 & 0.56 & 0.11 & 0.00 & 0.97\\
 &  & $\beta_3$ & 0.43 & 0.21 & -0.02 & 0.47 & 0.11 & 0.37 & 0.71 & 0.88 & 0.97\\
\addlinespace
 & $ 10^5$ & $\beta_1$ & 1.52 & 5.91 & 0.00 & 2.34 & 34.99 & 0.05 & 0.00 & 0.00 & 0.95\\
 &  & $\beta_2$ & -1.71 & -6.01 & 0.01 & 2.98 & 36.14 & 0.06 & 0.00 & 0.00 & 0.95\\
 &  & $\beta_3$ & 0.44 & 0.21 & 0.00 & 0.22 & 0.05 & 0.04 & 0.12 & 0.27 & 0.94\\
\addlinespace
 & $ 2\times 10^5$ & $\beta_1$ & 1.52 & 5.91 & 0.00 & 2.32 & 34.97 & 0.02 & 0.00 & 0.00 & 0.95\\
 &  & $\beta_2$ & -1.73 & -6.01 & 0.00 & 3.01 & 36.18 & 0.03 & 0.00 & 0.00 & 0.96\\
 &  & $\beta_3$ & 0.44 & 0.21 & 0.00 & 0.21 & 0.05 & 0.02 & 0.01 & 0.05 & 0.94\\
\addlinespace
1 & $ 10^3$ & $\beta_1$ & 5.72 & 9.73 & -0.79 & 61.01 & 94.75 & 7701.20 & 0.10 & 0.00 & 0.95\\
 &  & $\beta_2$ & -5.96 & -9.75 & 2.66 & 68.01 & 95.13 & 18520.81 & 0.09 & 0.00 & 0.97\\
 &  & $\beta_3$ & 1.03 & 0.07 & -2.65 & 19.74 & 0.06 & 7757.13 & 0.38 & 0.93 & 1.00\\
\addlinespace
 & $ 10^4$ & $\beta_1$ & 6.56 & 9.73 & -0.50 & 44.03 & 94.65 & 12.72 & 0.00 & 0.00 & 0.95\\
 &  & $\beta_2$ & -6.83 & -9.75 & 0.54 & 47.60 & 95.07 & 15.83 & 0.00 & 0.00 & 0.96\\
 &  & $\beta_3$ & 0.88 & 0.07 & -0.10 & 1.58 & 0.01 & 6.44 & 0.26 & 0.83 & 0.98\\
\addlinespace
 & $ 10^5$ & $\beta_1$ & 6.59 & 9.73 & -0.08 & 43.54 & 94.61 & 0.99 & 0.00 & 0.00 & 0.95\\
 &  & $\beta_2$ & -6.86 & -9.75 & 0.09 & 47.15 & 95.03 & 1.34 & 0.00 & 0.00 & 0.96\\
 &  & $\beta_3$ & 0.89 & 0.07 & -0.03 & 0.86 & 0.01 & 0.55 & 0.00 & 0.10 & 0.96\\
\addlinespace
 & $ 2\times 10^5$ & $\beta_1$ & 6.59 & 9.73 & -0.06 & 43.49 & 94.62 & 0.43 & 0.00 & 0.00 & 0.97\\
 &  & $\beta_2$ & -6.87 & -9.75 & 0.05 & 47.18 & 95.04 & 0.60 & 0.00 & 0.00 & 0.95\\
 &  & $\beta_3$ & 0.90 & 0.07 & 0.00 & 0.84 & 0.01 & 0.28 & 0.00 & 0.00 & 0.95\\
\addlinespace
3 & $ 10^3$ & $\beta_1$ & 9.39 & 10.00 & 17.76 & 137.46 & 99.93 & 54132.92 & 0.02 & 0.00 & 0.91\\
 &  & $\beta_2$ & -9.10 & -10.00 & -19.11 & 233.16 & 99.96 & 631434.25 & 0.02 & 0.00 & 0.92\\
 &  & $\beta_3$ & 0.62 & 0.00 & 24.71 & 302.61 & 0.00 & 1310764.70 & 0.17 & 0.96 & 0.99\\
\addlinespace
 & $ 10^4$ & $\beta_1$ & 9.38 & 9.99 & 12.63 & 90.18 & 99.90 & 25558.47 & 0.00 & 0.00 & 0.92\\
 &  & $\beta_2$ & -9.37 & -9.99 & -34.07 & 89.86 & 99.90 & 987114.37 & 0.00 & 0.00 & 0.95\\
 &  & $\beta_3$ & 0.39 & 0.00 & 26.29 & 2.28 & 0.00 & 1118249.17 & 0.12 & 0.94 & 1.00\\
\addlinespace
 & $ 10^5$ & $\beta_1$ & 9.45 & 9.99 & -9.46 & 89.42 & 99.89 & 10222.59 & 0.00 & 0.00 & 0.93\\
 &  & $\beta_2$ & -9.47 & -9.99 & 10.35 & 89.72 & 99.90 & 16877.40 & 0.00 & 0.00 & 0.95\\
 &  & $\beta_3$ & 0.24 & 0.00 & -1.31 & 0.18 & 0.00 & 2072.33 & 0.12 & 0.89 & 1.00\\
\addlinespace
 & $ 2\times 10^5$ & $\beta_1$ & 9.47 & 9.99 & -3.12 & 89.70 & 99.90 & 779.23 & 0.00 & 0.00 & 0.93\\
 &  & $\beta_2$ & -9.49 & -10.00 & 2.94 & 90.12 & 99.90 & 819.94 & 0.00 & 0.00 & 0.93\\
 &  & $\beta_3$ & 0.23 & 0.00 & 0.07 & 0.12 & 0.00 & 67.16 & 0.08 & 0.84 & 1.00\\
 \bottomrule
 \end{tabular}
 }
\end{table}

\end{document}

\bigskip
\begin{center}
{\large\bf SUPPLEMENTARY MATERIAL}
\end{center}
The Supplementary Material contains the proofs for Proposition 3.1 and Theorem 3.1, as well as additional simulation results.
\appendix
\section{Appendix}

\subsection{Derivation of equation~\eqref{eq:log_relationship}}\label{sec:log_relationship}

\begin{equation}
\footnotesize
\begin{aligned}
& P(y_i^* = 1 \mid \bZstar_i, \bXstar_i) \\
= & \frac{P(y^*_i = 1, \bZstar_i, \bX^*_i)}{P(y^*_i = 1, \bZstar_i, \bX^*_i) + P(y^*_i = 0, \bZstar_i, \bX^*_i)} \\
= & \frac{ p_1(\bZstar_i)\exp\{-\frac{1}{2} (\bX_i^* - \bmu_1 - \bZstar_i \bC) \bm\Sigma^{-1} (\bX_i^* - \bmu_1- \bZstar_i \bC)^T\}}{p_1(\bZstar_i) \exp\{-\frac{1}{2} (\bX_i^* - \bmu_1 - \bZstar_i \bC) \bm\Sigma^{-1} (\bX_i^* - \bmu_1 - \bZstar_i \bC)^T\} + p_0(\bZstar_i)  \exp\{-\frac{1}{2} (\bX_i^* - \bmu_0 - \bZstar_i \bC) \bm\Sigma^{-1} (\bX_i^* - \bmu_0 - \bZstar_i \bC)^T\} } \\
= & \frac{1}{1 + \frac{p_0(\bZstar_i)}{p_1(\bZstar_i)}  \exp\{ \frac{\bmu_1 + \bmu_0 + 2 \bZstar_i \bC}{2} \bm\Sigma^{-1} (\bmu_1 - \bmu_0)^T  + \bX_i^* \ \bm\Sigma^{-1} (\bmu_0 - \bmu_1)^T\}} \\
= & \frac{1}{1 + \frac{1}{\exp(\gamma_0 + \bZstar_i\bm\gamma_1)}  \exp\{ \frac{\bmu_1 + \bmu_0 + 2 \bZstar_i \bC}{2} \bm\Sigma^{-1} (\bmu_1 - \bmu_0)^T  + \bX_i^* \ \bm\Sigma^{-1} (\bmu_0 - \bmu_1)^T\}} \\
= & \frac{ 1}{1 +   \exp\{ - \gamma_0 - \bZstar_i\bm\gamma_1  + \frac{\bmu_1 + \bmu_0 }{2} \bm\Sigma^{-1} (\bmu_1 - \bmu_0)^T + \bZstar_i \bC \bm\Sigma^{-1} (\bmu_1 - \bmu_0)^T - \bX_i^* \bm\Sigma^{-1} (\bmu_1 - \bmu_0)^T\}} \\
= & \frac{ 1}{1 +   \exp\{ - [\gamma_0 -\frac{\bmu_1 + \bmu_0 }{2} \bm\Sigma^{-1} (\bmu_1 - \bmu_0)^T]  - \bX_i^* [\bm\Sigma^{-1} (\bmu_1 - \bmu_0)^T] - \bZstar_i\bm [\bgamma_1 - \bC \bm\Sigma^{-1} (\bmu_1 - \bmu_0)^T]
\}}.
\end{aligned}
\end{equation}
The last expression reduces to 
$\dfrac{1}{1 + \exp(-\beta_0 - \bXstar_i\bbeta_1  - \bZstar_i\bbeta_2)}$ as in \eqref{eq:log_relationship} when we denote
$$
\beta_0 = \gamma_0 -\frac{\bmu_1 + \bmu_0 }{2} \bm\Sigma^{-1} (\bmu_1 - \bmu_0)^T = \gamma_0 - \frac{1}{2} (\bmu_1 \bSigma^{-1}\bmu_1^\T - \bmu_0\bSigma^{-1}\bmu_0^\T ),
$$ 
$\bm\beta_1 = \bm\Sigma^{-1}(\bmu_1 - \bmu_0)^\T$ and $\bm\beta_2 = \bgamma_1 - \bC \bm\Sigma^{-1} (\bmu_1 - \bmu_0)^T = \bgamma_1 - \bC \bbeta_1$.

\subsection{Proof of Lemma~\ref{est.b1.cond.mixturemodel}}\label{sec:lemma1}
To simplify the notation, we let $\bSigma_{xz}$ denote the covariance between $\mathbf{X}_1^*$ and $\mathbf{Z}_1^*$, where $\mathbf{X}_1^*$ and $\bZ_1^*$ are the variables in the first row of the raw data, i.e $\bSigma_{xz} = E(\bX_1^{*\T} \bZ_1^*) - \left\{E(\bX_1^*\right\}^\T E(\bZ_1^*)$. Similar definitions hold for $\bSigma_{xy}$, $\bSigma_{zy}$, $\bSigma_{xx}$ and $\bSigma_{zz}$.

Hence, we can write 
$$
\bar{\bm{b}} = \begin{bmatrix}
\bSigma_{xx} & \bSigma_{xz} \\
\bSigma_{zx} & \bSigma_{zz} \\
\end{bmatrix}^{-1} \begin{bmatrix}
\bSigma_{xy}\\
\bSigma_{zy}
\end{bmatrix} = \begin{bmatrix}
\bar{\bm{b}}_1 \\
\bar{\bm{b}}_2 \\
\end{bmatrix}
$$
Using the Schur's matrix inversion formula, we have
\begin{align*}
\footnotesize
& \begin{bmatrix}
\bSigma_{xx} & \bSigma_{xz} \\
\bSigma_{zx} & \bSigma_{zz} \\
\end{bmatrix}^{-1} 
\\ = & \begin{bmatrix}
(\bSigma_{xx} - \bSigma_{xz}\bSigma_{zz}^{-1} \bSigma_{zx})^{-1}& - \left( \bSigma_{xx} - \bSigma_{xz}\bSigma_{zz}^{-1} \bSigma_{zx}\right)^{-1}\bSigma_{xz}\bSigma_{zz}^{-1}\\
-\bSigma_{zz}^{-1}\bSigma_{zx}  \left( \bSigma_{xx} - \bSigma_{xz}\bSigma_{zz}^{-1} \bSigma_{zx}\right)^{-1} & \bSigma_{zz}^{-1} + \bSigma_{zz}^{-1} \bSigma_{zx} (\bSigma_{xx} - \bSigma_{xz}\bSigma_{zz}^{-1} \bSigma_{zx})^{-1} \bSigma_{xz} \bSigma_{zz}^{-1}
\end{bmatrix}
\end{align*}
so we have
\begin{align}
&\bar\bb_1 = 
(\bSigma_{xx} - \bSigma_{xz}\bSigma_{zz}^{-1} \bSigma_{zx})^{-1}(\bSigma_{xy} - \bSigma_{xz}\bSigma_{zz}^{-1}\bSigma_{zy}),
\label{eq:barb1}
\end{align} and 
\begin{equation}
\begin{array}{cl}
     \bar\bb_2 & =  -\bSigma_{zz}^{-1}\bSigma_{zx}  \left( \bSigma_{xx} - \bSigma_{xz}\bSigma_{zz}^{-1} \bSigma_{zx}\right)^{-1} \bSigma_{xy} + \bSigma_{zz}^{-1}\bSigma_{zy} \\
 & \qquad + \bSigma_{zz}^{-1} \bSigma_{zx} (\bSigma_{xx} - \bSigma_{xz}\bSigma_{zz}^{-1} \bSigma_{zx})^{-1} \bSigma_{xz} \bSigma_{zz}^{-1}\bSigma_{zy} \\
 & = \bSigma_{zz}^{-1}\bSigma_{zy} - \bSigma_{zz}^{-1} \bSigma_{zx} (\bSigma_{xx} - \bSigma_{xz}\bSigma_{zz}^{-1} \bSigma_{zx})^{-1}(\bSigma_{xy} - \bSigma_{xz}\bSigma_{zz}^{-1}\bSigma_{zy}) \\
 & = \bSigma_{zz}^{-1}\bSigma_{zy} - \bSigma_{zz}^{-1} \bSigma_{zx} \bar\bb_1.
\end{array}
\label{eq:barb2}
\end{equation}
Now we will express $\bar\bb_1$ in terms of parameters of the mixture model \eqref{conditionmixturemodel}-\eqref{eq:logisticZ}, through finding the expressions for the two factors in \eqref{eq:barb1}: $\bSigma_{xx} - \bSigma_{xz}\bSigma_{zz}^{-1} \bSigma_{zx}$ and $\bSigma_{xy} - \bSigma_{xz}\bSigma_{zz}^{-1}\bSigma_{zy}$. First, we have
\begin{equation}
\begin{array}{cl}
 \bSigma_{xx}  & = \Var(\bX_1^*)  \\
& = E\left[\Var(\bX_1^*\mid y_1^*, \bZ_1^* ) \right] + \Var\left[E\left(\bX_1^*\mid y_1^*, \bZ_1^* \right) \right] \\
& = E\left[\bm\Sigma \right]  + \Var\left[(\bmu_1 - \bmu_0) y_1^* + \bmu_0 + \bZ_1^* \bC \right]\\
& = \bm\Sigma + \Var\left[(\bmu_1 - \bmu_0) y_1^* + \bZ_1^* \bC \right]\\
& = \bm\Sigma + \Var[(\bmu_1 - \bmu_0) y_1^*] + \Var[\bZ_1^* \bC ] 
+ \Cov[(\bmu_1 - \bmu_0) y_1^*, \bZ_1^* \bC] + \Cov[\bZ_1^* \bC, (\bmu_1 - \bmu_0) y_1^*]\\
& = \bm\Sigma + (\bmu_1 - \bmu_0)^\T \Var(y_1^*)(\bmu_1 - \bmu_0)  + \bC^\T \Var(\bZ_1^*) \bC + (\bmu_1 - \bmu_0)^\T \Cov(y_1^*, \bZ_1^*) \bC \\ 
&\qquad + ~ \bC^\T \Cov(\bZ_1^*, y_1^*) (\bmu_1 - \bmu_0) \\
& = \bm\Sigma + (\bmu_1 - \bmu_0)^\T\bSigma_{yy}(\bmu_1 - \bmu_0)  + \bC^\T \bSigma_{zz} \bC + (\bmu_1 - \bmu_0)^\T \bSigma_{yz} \bC + \bC^\T \bSigma_{zy} (\bmu_1 - \bmu_0). 
\end{array}
\label{eq:Sigma_xx}
\end{equation}
Recall that for any two random vectors $\bV_1$ and $\bV_2$, we have 
\begin{equation}
\begin{array}{cl}
\Cov(\bV_1,\bV_2) & = E(\bV_1^T \bV_2) - E(\bV_1)^T E(\bV_2) \\
& = E[E(\bV_1^T \bV_2 \mid \bV_2)] - E[E(\bV_1 \mid \bV_2)]^T E(\bV_2) \\
& = E[E(\bV_1 \mid \bV_2)^T \bV_2] - E[E(\bV_1 \mid \bV_2)^T] E(\bV_2) \\
& = \Cov[E(\bV_1 \mid \bV_2),\bV_2],
\end{array}
\label{eq:prop.A1}
\end{equation}
so we obtain
\begin{equation}\label{eq:Sigma_xz}
\begin{array}{cl}
\bSigma_{xz} & =  \Cov [\bX^*_1,\bZ^*_1] = \Cov[E(\bX^*_1 \mid \bZ^*_1),\ \bZ^*_1] \\
& = \Cov [(\bmu_1 - \bmu_0) E(y^*_1 \mid \bZ^*_1) + \bmu_0 + \bZ^*_1 \bC, \ \bZ^*_1]
\\
& = \Cov [(\bmu_1 - \bmu_0) E(y^*_1 \mid \bZ^*_1) , \ \bZ^*_1] + \Cov [\bmu_0 , \ \bZ^*_1] + \Cov [\bZ^*_1 \bC, \ \bZ^*_1] \\
& =  (\bmu_1 - \bmu_0)^\T \Cov[E(y^*_1 \mid \bZ^*_1), \bZ^*_1] + 0 + \bC^\T \Cov(\bZ^*_1, \bZ^*_1)
\\
& =   (\bmu_1 - \bmu_0)^\T \bSigma_{yz} + \bC^\T \bSigma_{zz}.
\end{array}
\end{equation}
Hence
$$
\begin{array}{cl}
& 
\bSigma_{xz}\bSigma_{zz}^{-1} \bSigma_{zx} \\
&=  [(\bmu_1 - \bmu_0)^\T \bSigma_{yz} + \bC^\T \bSigma_{zz}] \bSigma_{zz}^{-1} [(\bmu_1 - \bmu_0)^\T \bSigma_{yz} + \bC^\T \bSigma_{zz}]^T \\
= & [(\bmu_1 - \bmu_0)^\T \bSigma_{yz} + \bC^\T \bSigma_{zz}] \bSigma_{zz}^{-1} [ \bSigma_{zy}(\bmu_1 - \bmu_0) +  \bSigma_{zz}\bC] \\
= & (\bmu_1 - \bmu_0)^\T \bSigma_{yz} \bSigma_{zz}^{-1} \bSigma_{zy}(\bmu_1 - \bmu_0) + (\bmu_1 - \bmu_0)^\T \bSigma_{yz}  \bC + \bC^\T  \bSigma_{zy}(\bmu_1 - \bmu_0)  + \bC^\T \bSigma_{zz} \bC. 
\end{array}
$$
Combine this expression with \eqref{eq:Sigma_xx}, we simplify the first factor in \eqref{eq:barb1} as
\begin{equation}
\label{eq:factor1}
\begin{array}{cl}
\bSigma_{xx} - \bSigma_{xz}\bSigma_{zz}^{-1} \bSigma_{zx}
& =  \bm\Sigma + (\bmu_1 - \bmu_0)^\T\bSigma_{yy}(\bmu_1 - \bmu_0) - (\bmu_1 - \bmu_0)^\T \bSigma_{yz} \bSigma_{zz}^{-1} \bSigma_{zy}(\bmu_1 - \bmu_0)\\
& =  \bm\Sigma + (\bmu_1 - \bmu_0)^\T (\bSigma_{yy} - \bSigma_{yz} \bSigma_{zz}^{-1} \bSigma_{zy}) (\bmu_1 - \bmu_0).
\end{array}
\end{equation}
Furthermore, we have
\begin{equation}\label{eq:Sigma_xy}
\begin{array}{cl}
\bSigma_{xy} & = \Cov[\bX^*_1, y^*_1] \\
 & =  E[\Cov(\bX^*_1, y^*_1\mid \bZ^*_1)] + \Cov[E(\bX^*_1\mid \bZ^*_1), E(y^*_1\mid \bZ^*_1)] \\
& = E[\Cov((\bmu_1 - \bmu_0) y^*_1 + \bmu_0 + \bZ^*_1 \bC, \ y^*_1\mid \bZ^*_1)] \\
& \qquad + ~\Cov [E( (\bmu_1 - \bmu_0) y^*_1 \mid \bZ^*_1) + \bmu_0 + \bZ^*_1 \bC, \ E(y^*_1 \mid \bZ^*_1)]
\\
& = E[\Cov((\bmu_1 - \bmu_0) y^*_1, \ y^*_1\mid \bZ^*_1)] \\
& \qquad + ~ \Cov [E( (\bmu_1 - \bmu_0) y^*_1 \mid \bZ^*_1), \ E(y^*_1 \mid \bZ^*_1)] + 0 + \Cov [\bZ^*_1 \bC, \ E(y^*_1 \mid \bZ^*_1)]
\\
& = \Cov[(\bmu_1 - \bmu_0) y^*_1, \ y^*_1]  + \Cov [\bZ^*_1 \bC, \ E(y^*_1 \mid \bZ^*_1)] \\
& =   (\bmu_1 - \bmu_0)^\T \Cov(y^*_1, \ y^*_1)  + \bC^\T \Cov(\bZ^*_1, y^*_1)
\\
& =   (\bmu_1 - \bmu_0)^\T \bSigma_{yy} + \bC^\T \bSigma_{zy},
\end{array}
\end{equation}
where the second to last equality used the property \eqref{eq:prop.A1}.
Combining this expression with \eqref{eq:Sigma_xz}, we simplify the second factor in \eqref{eq:barb1} as
\begin{equation}
\label{eq:factor2}
\begin{array}{cl}
& 
\bSigma_{xy} - \bSigma_{xz}\bSigma_{zz}^{-1}\bSigma_{zy} \\
= & (\bmu_1 - \bmu_0)^\T \bSigma_{yy} + \bC^\T \bSigma_{zy} - [(\bmu_1 - \bmu_0)^\T \bSigma_{yz} + \bC^\T \bSigma_{zz}] \bSigma_{zz}^{-1} \bSigma_{zy} \\
= & (\bmu_1 - \bmu_0)^\T \bSigma_{yy} + \bC^\T \bSigma_{zy} - (\bmu_1 - \bmu_0)^\T \bSigma_{yz} \bSigma_{zz}^{-1} \bSigma_{zy} - \bC^\T \bSigma_{zy}  \\
= & (\bmu_1 - \bmu_0)^\T \bSigma_{yy} - (\bmu_1 - \bmu_0)^\T \bSigma_{yz} \bSigma_{zz}^{-1} \bSigma_{zy}  \\
= & (\bmu_1 - \bmu_0)^\T (\bSigma_{yy} - \bSigma_{yz} \bSigma_{zz}^{-1} \bSigma_{zy}) . 
\end{array}
\end{equation}

Note that, equation \eqref{eq:barb1} can be rewritten as 
$
(\bSigma_{xx} - \bSigma_{xz}\bSigma_{zz}^{-1} \bSigma_{zx}) \bar\bb_1 = 
(\bSigma_{xy} - \bSigma_{xz}\bSigma_{zz}^{-1}\bSigma_{zy}).$
Substituting \eqref{eq:factor1} and \eqref{eq:factor2} into this expression, we have
$$
[\bm\Sigma + (\bmu_1 - \bmu_0)^\T (\bSigma_{yy} - \bSigma_{yz} \bSigma_{zz}^{-1} \bSigma_{zy}) (\bmu_1 - \bmu_0)]\bar\bb_1 = (\bmu_1 - \bmu_0)^\T (\bSigma_{yy} - \bSigma_{yz} \bSigma_{zz}^{-1} \bSigma_{zy}),
$$
which is equivalent to
$$
\begin{array}{cl}
\bm\Sigma \bar\bb_1 & = (\bmu_1 - \bmu_0)^\T (\bSigma_{yy} - \bSigma_{yz} \bSigma_{zz}^{-1} \bSigma_{zy}) - (\bmu_1 - \bmu_0)^\T (\bSigma_{yy} - \bSigma_{yz} \bSigma_{zz}^{-1} \bSigma_{zy}) (\bmu_1 - \bmu_0)\bar\bb_1 \\
& = (\bmu_1 - \bmu_0)^\T (\bSigma_{yy} - \bSigma_{yz} \bSigma_{zz}^{-1} \bSigma_{zy}) [1 - (\bmu_1 - \bmu_0)\bar\bb_1] \\
& = \bm\Sigma \bbeta_1 (\bSigma_{yy} - \bSigma_{yz} \bSigma_{zz}^{-1} \bSigma_{zy}) [1 - (\bmu_1 - \bmu_0)\bar\bb_1].
\end{array}
$$
where the last equality used definition of $\bbeta_1$ from \eqref{eq:beta}. Multiplying $\bm\Sigma^{-1}$ to both sides of above equation, we have
$$
\bar\bb_1 = \bbeta_1 (\bSigma_{yy} - \bSigma_{yz} \bSigma_{zz}^{-1} \bSigma_{zy}) [1 - (\bmu_1 - \bmu_0)\bar\bb_1].
$$
Hence, to finish the proof of this Lemma, i.e., to get $\bbeta_1= \bar\bb_1/\Var(\varepsilon_1)$, we only need to show  
\begin{equation}\label{eq:Var.eps}
\Var(\varepsilon_1) = \Var(y^*_1 - \bW_1^*\bar{\bb}) =(\bSigma_{yy} - \bSigma_{yz} \bSigma_{zz}^{-1} \bSigma_{zy}) [1 - (\bmu_1 - \bmu_0)\bar\bb_1].
\end{equation}
Recall the definition $\bar{\bm{b}} = \left[\Var(\bW_1^*)\right]^{-1} \Cov(\bW_1^*, y_1^*)$. Hence $\Var(\bW_1^*) \bar{\bm{b}} = \Cov(\bW_1^*, y_1^*)$, and we have
$$
\begin{aligned}
\Var(\varepsilon_1) & = \Var(y^*_1 -  \bW_1^*\bar{\bb}) \ \ = \Var(y^*_1) + \bar\bb^\T \Var(\bW_1^*) \bar\bb - 2 \Cov(y_1^*, \bW_1^*) \bar\bb \\
& = \Var(y^*_1) + \bar\bb^\T \Cov(\bW_1^*, y_1^*)  - 2 \Cov(y_1^*, \bW_1^*) \bar\bb \\
& = \Var(y^*_1) - \Cov(y_1^*, \bW_1^*) \bar\bb  \ \ = \Var(y^*_1) - [\Cov(y_1^*, \bX_1^*) \bar\bb_1 + \Cov(y_1^*, \bZ_1^*) \bar\bb_2]  \\ 
& = \bSigma_{yy} -  \bSigma_{yx} \bar\bb_1 - \bSigma_{yz} \bar\bb_2.
\end{aligned}
$$
Using ~\eqref{eq:barb2}, this becomes
$$
\begin{aligned}
\Var(\varepsilon_1) & = \bSigma_{yy} -  \bSigma_{yx} \bar\bb_1 - 
\bSigma_{yz}(\bSigma_{zz}^{-1}\bSigma_{zy} - \bSigma_{zz}^{-1} \bSigma_{zx} \bar\bb_1) \\
& = \bSigma_{yy} - \bSigma_{yz}\bSigma_{zz}^{-1}\bSigma_{zy}
- (\bSigma_{yx} - \bSigma_{yz} \bSigma_{zz}^{-1} \bSigma_{zx}) \bar\bb_1 \\
& = \bSigma_{yy} - \bSigma_{yz}\bSigma_{zz}^{-1}\bSigma_{zy}
- (\bSigma_{xy} - \bSigma_{xz} \bSigma_{zz}^{-1} \bSigma_{xy})^T \bar\bb_1,
\end{aligned}
$$
Using \eqref{eq:factor2}, this becomes
$$
\begin{aligned}
\Var(\varepsilon_1) & =  \bSigma_{yy} - \bSigma_{yz}\bSigma_{zz}^{-1}\bSigma_{zy}
- (\bSigma_{yy} - \bSigma_{yz}\bSigma_{zz}^{-1}\bSigma_{zy}) (\bmu_1 - \bmu_0) \bar\bb_1 \\
& = (\bSigma_{yy} - \bSigma_{yz}\bSigma_{zz}^{-1}\bSigma_{zy}) [1 - (\bmu_1 - \bmu_0) \bar\bb_1].
\end{aligned}
$$
This is indeed \eqref{eq:Var.eps}, thus the proof is finished.

\bibliographystyle{apalike}
\bibliography{arXiv}
\clearpage

\noindent\textbf{\Large SUPPLEMENTARY MATERIALS}

\renewcommand{\thesection}{S\arabic{section}}
\setcounter{section}{0}
\section{Proof of Proposition 3.1}
\label{sec:proposition}
For the first estimating equation $\bm{m}_{i, \bm\theta}(\bm\theta, \phi)$, recall $\widetilde{\bW}_i = (1, \bW_i)$, $\bm\theta = (b_0, \bar{\bm{b}}^\T)^\T/\tau^2 = (b_0, \bar{\bm{b}}^\T)^\T \phi $, hence $\bm\theta/\phi =  (b_0, \bar{\bm{b}}^\T)^\T$. We have
\begin{equation} \label{eq:est.eq.1}
\begin{array}{cl}
E[m_{i, \bm\theta}(\bm\theta, \phi)] & = E\left\{\widetilde{\bW}_i^{\textsc{T}}y_i - \dfrac{1}{\phi} \left(\widetilde{\bW}_{i}^{\T} \widetilde{\bW}_i  - \sigma^2\bJ_{p+q}\right) \bm\theta \right\}  \\ & = E \left\{\begin{bmatrix} y_i \\\bW_i^\T y_i \end{bmatrix} - \begin{bmatrix} 1 & \bW_i \\
\bW_i^{\T} & \bW_i^{\T}\bW_i-\sigma^2 {\bI}_{p+q}\end{bmatrix} \begin{bmatrix} b_0 \\ \bar{\bm{b}} \end{bmatrix} \right\}  \\ & =  \begin{bmatrix}
E(y_i) - b_0 - E(\bW_i)\bar{\bm{b}}  \\
E(\bW_i^\T y_i) - E(\bW_i^{\T}) a - E(\bW_i^{\T} \bW_i - \sigma^2 \bI_{p+q}) \bar{\bm{b}} 
\end{bmatrix} .
\end{array}
\end{equation}

Because $y_i = y^*_i + v_i$ and $\bW_i = \bW^*_i + \bU_i$, where each element of  $v_i$ and $\bU_i$ has zero expectation and variance $\sigma^2$ and are independent of $y^*_i$ and $\bW^*_i$, we have 
\begin{equation}\label{eq:star.to.obs}
\begin{array}{cc}
E(y_i) = E(y^*_i), & \Var(y_i) = \Var(y^*_i) + \sigma^2, \\
E(\bW_i) = E(\bW_i^*), & \Var(\bW_i) = \Var(\bW_i^*) + \sigma^2 \bI_{p+q}.
\end{array}
\end{equation} 
As a result, the upper expression in ~\eqref{eq:est.eq.1} becomes:
\begin{equation} \label{eq:est.eq.1.1}
E(y_i) - b_0 - E(\bW_i) \bar{\bb} = E(y_i^*) - b_0 - E(\bW_i^*) \bar{\bb} = 0,
\end{equation}
where the last equality is due to the definition $b_0 = E(y_i^*) - E(\bW_i^*)\bar{\bb}$. That is, the upper expression in ~\eqref{eq:est.eq.1} is unbiased. 

Since $\bW_i = \bW^*_i + \bU_i$, and $\bW^*_i$ is independent of $\bU_i$, we have
$$
E(\bW_i^{\T} \bW_i) = E(\bW_i^{*\T} \bW^*_i) + 0 + E(\bU_i^{\T} \bU_i) = E(\bW_i^{*\T} \bW^*_i) + \sigma^2 \bI_{p+q}.
$$
Similarly, $y_i = y^*_i + v_i$ with $y^*_i$ is independent of $v_i$, implying
$$
E(y_i^2) = E(y_i^{*2}) + 0 + E(v_i^2) = E(y_i^{*2}) + \sigma^2.
$$
Thus
\begin{equation}\label{eq:star.square}
  E(y_i^2) -  \sigma^2 = E(y_i^{*2}), \qquad  E(\bW_i^{\T} \bW_i) - \sigma^2 \bI_{p+q} = E(\bW_i^{*\T} \bW^*_i).
\end{equation}

Using \eqref{eq:star.square} and $E(\bW_i) = E(\bW_i^*)$ from \eqref{eq:star.to.obs}, the bottom expression in~\eqref{eq:est.eq.1} becomes
$$
\begin{array}{cl}
& E(\bW_i^\T y_i) - E(\bW_i^\T) b_0 - [E(\bW_i^{\T} \bW_i) - \sigma^2\bI_{p+q}] \bar{\bm{b}} \\
= & E(\bW_i^\T y_i) - E[(\bW^*_i)^\T] b_0 - E[(\bW^*_i)^{\T} \bW^*_i] \bar{\bm{b}}. 
\end{array}
$$
Recall $y_i = y^*_i + v_i$, $\bW_i = \bW^*_i + \bU_i$. $v_i$ and $\bU_i$ are independent of each other with zero means, and independent from both $y^*_i$ and $\bW^*_i$, thus $E(\bW_i^\T y_i) = E[(\bW^*_i)^T(y^*_i)]$. Put this into the above expression, we have
$$
\begin{array}{cl}
& E(\bW_i^\T y_i) - E(\bW_i^\T) b_0 - [E(\bW_i^{\T} \bW_i) - \sigma^2\bI_{p+q}] \bar{\bm{b}} \\
= & E[(\bW^*_i)^T(y^*_i)] - E[(\bW^*_i)^\T] b_0 - E[(\bW^*_i)^{\T} \bW^*_i] \bar{\bm{b}} \\
= & E[(\bW^*_i)^T(y^*_i)] - E[(\bW^*_i)^\T] [E(y_i^*) - E(\bW_i^*)\bar{\bb}] - E[(\bW^*_i)^{\T} \bW^*_i] \bar{\bm{b}} \\
= & E[(\bW^*_i)^T(y^*_i)] - E[(\bW^*_i)^\T] E(y_i^*)  - \{E[(\bW^*_i)^{\T} \bW^*_i] - E[(\bW^*_i)^\T]E(\bW_i^*) \} \bar{\bm{b}} \\
= & \Cov(\bW_i^*, y_i^*) - \Var(\bW^*_i) \bar{\bb}. 
\end{array}
$$
This expression equals zero due the definition $\bar{\bb} = [\Var(\bW_i^*)]^{-1} \Cov(\bW_i^*, y_i^*) $. That is, the bottom expression in ~\eqref{eq:est.eq.1} is unbiased. Substituting this and \eqref{eq:est.eq.1.1} into \eqref{eq:est.eq.1}, we have proved the unbiasedness of the first estimating equation: $E[\bm{m}_{i, \bm\theta}(\bm\theta, \phi)] = \bm{0}$. 

For the second estimating equation $\bm{m}_{i, \phi}(\bm\theta, \phi)$, we have
$$
\begin{array}{cl}
2 E\{\bm{m}_{i, \bm\theta}(\bm\theta, \phi)\} & = E \left\{\dfrac{1}{\phi} - \left(y_i^2 - \sigma^2 \right) +  \dfrac{\bm\theta}{\phi}^\top \left(\widetilde{\bW}_i\widetilde{\bW}_i^\T - \sigma^2 \bJ_{p+q} \right) \dfrac{\bm\theta}{\phi}     \right\} \\
& = \dfrac{1}{\phi} - \left\{E(y_i^2) - \sigma^2\right\} + \begin{bmatrix} b_0 & \bar{\bb}^\T \end{bmatrix} \begin{bmatrix} 1 & E(\bW_i) \\
E(\bW_i^{\T}) & E(\bW_i^{\T}\bW_i - \sigma^2 \bI_{p+q}) \end{bmatrix} \begin{bmatrix} b_0 \\ \bar{\bm{b}} \end{bmatrix} 
\\
& = \dfrac{1}{\phi} - \left\{E(y_i^2) - \sigma^2\right\} + b_0^2 + 2aE(\bW_i )\bar{\bm{b}} +  \bar{\bm{b}}^\T [E(\bW_i^{\T}\bW_i) - \sigma^2 \bI_{p+q}] \bar{\bm{b}} 
\end{array}
$$
Using \eqref{eq:star.square} and $E(\bW_i) = E(\bW_i^*)$ from \eqref{eq:star.to.obs},  this becomes
$$
\begin{array}{cl}
2 E\{\bm{m}_{i, \bm\theta}(\bm\theta, \phi)\} 
& = \dfrac{1}{\phi} - \left\{E(y_i^2) - \sigma^2\right\} + b_0^2 + 2b_0E(\bW_i )\bar{\bm{b}} +  \bar{\bm{b}}^\T [E(\bW_i^{\T}\bW_i) - \sigma^2 \bI_{p+q}] \bar{\bm{b}} \\[1em]
& = \dfrac{1}{\phi} - E(y_i^{*2}) + b_0^2 + 2b_0E(\bW_i^* )\bar{\bm{b}} +  \bar{\bm{b}}^\T E(\bW_i^{*\T}\bW_i^*)  \bar{\bm{b}} 
\end{array}
$$
Recall that $1/\phi = \tau^2 = \Var(\varepsilon_i) = \Var(y_i^* - \widetilde{\bW}_i^*\bar\bb) = \Var(y_i^*) - \bar\bb^\T \Var(\bW_i^*)\bar{\bb}$,  the above expression becomes
$$
\begin{array}{cl}
2 E\{\bm{m}_{i, \bm\theta}(\bm\theta, \phi)\} 
& = \dfrac{1}{\phi} - E(y_i^{*2}) + b_0^2 + 2b_0E(\bW_i^* )\bar{\bm{b}} +  \bar{\bm{b}}^\T E(\bW_i^{*\T}\bW_i^*)  \bar{\bm{b}}  \\
& = \Var(y_i^*) - \bar\bb^\T \Var(\bW_i^*)\bar{\bb} - E(y_i^{*2}) + b_0^2 + 2b_0E(\bW_i^* )\bar{\bm{b}} +  \bar{\bm{b}}^\T E(\bW_i^{*\T}\bW_i^*)  \bar{\bm{b}} \\
& = [\Var(y_i^*) - E(y_i^{*2})] - \bar\bb^\T [\Var(\bW_i^*) - E(\bW_i^{*\T}\bW_i^*)] \bar{\bb}  + b_0^2 + 2b_0E(\bW_i^* )\bar{\bm{b}} 
\\
& = -\{E(y_i^*)\}^2  + \bar\bb^\T [E(\bW_i^{*\T}) E(\bW^*_i)] \bar{\bb}  + b_0^2 + 2b_0E(\bW_i^* )\bar{\bm{b}}
\\
& = -\{E(y_i^*)\}^2  + [E(\bW^*_i) \bar{\bb}]^2  + b_0[b_0 + 2E(\bW_i^* )\bar{\bm{b}}].
\end{array}
$$
Using the definition $b_0 = E(y_i^*) - E(\bW_i^*)\bar{\bb}$, we have
$$
\begin{array}{cl}
2 E\{\bm{m}_{i, \bm\theta}(\bm\theta, \phi)\} 
& = -\{E(y_i^*)\}^2  + [E(\bW^*_i) \bar{\bb}]^2   + b_0[b_0 + 2E(\bW_i^* )\bar{\bm{b}}] \\
& = -\{E(y_i^*)\}^2  + [E(\bW^*_i) \bar{\bb}]^2   + [E(y_i^*) - E(\bW_i^*)\bar{\bb}][E(y_i^*) + E(\bW_i^*)\bar{\bb}]\\
& = -\{E(y_i^*)\}^2  + [E(\bW^*_i) \bar{\bb}]^2   + \{E(y_i^*)\}^2 - [E(\bW^*_i) \bar{\bb}]^2 \\
& =0.
\end{array}
$$
That is, the second estimating equation is also unbiased. This finishes the proof.

\section{Proof of Theorem 3.1}\label{sec:theorem}

We prove this theorem using the classical theory of estimating equations. It suffices to establish the consistency of $(\hat{\bm\theta}^{(c)}, \hat{\phi}^{(c)})$, which is the solution of the estimating equation $\sum_{i=1}^{n}\bm{m}_{i, \bm\theta}(\bm\theta, \phi) =0$ and $\sum_{i=1}^{n}m_{i, \phi}(\bm\theta, \phi) =0$ with 
$$
\bm{m}_{i, \bm\theta}(\bm\theta, \phi) = \widetilde{\bW}_i^\T y_i - \dfrac{1}{\phi} (\widetilde{\bW}_i^\T\widetilde{\bW}_i - \sigma^2\bJ_{p+q}) \bm\theta
$$
and 
$$
m_{i, \phi}(\bm\theta, \phi) = \dfrac{1}{2\phi} - \dfrac{1}{2} \left(y_i^2 - \sigma^2 \right) +  \dfrac{1}{2\phi^2}\bm\theta^\top \left(\widetilde{\bW}_i^\T\widetilde{\bW}_i - \sigma^2\widetilde{\bI}_{p+q} \right)\bm\theta.
$$

To simplify the notations, we combine the above equations as 
$\bm{m}_i(\bm\theta, \phi) =
[\bm{m}_{i, \bm\theta}(\bm\theta, \phi)^\T,  m_{i, \phi}(\bm\theta, \phi)]^\T.
$
We have shown above that the estimating equation is unbiased so that $E[\bm{m}_{i}(\bm\theta, \phi)] = \boldsymbol{0}$. 

Denote $\bm\Psi = (\bm\theta, \phi)$ and $\bS(\bm\Psi) = n^{-1} \sum_{i=1}^{n} \bm{m}_{i}(\bm\theta, \phi)$, then the estimator $\widehat{\bm\Psi}^{(c)} = (\hat{\bm\theta}^{(c)}, \hat\phi^{(c)})$ is the solution of $\bS(\bm\Psi) = \boldsymbol{0}$. That is, $\bS(\widehat{\bm\Psi}^{(c)}) = \bS(\widehat{\bm{\theta}}^{(c)}, \hat\phi^{(c)}) = \mathbf{0}$. 

By the Taylor's series expansion, we have
\begin{equation}\label{eq:taylor}
\mathbf{0} = \bS(\widehat{\bm\Psi}^{(c)}) = \bS({\bm\Psi}) + \nabla\bS(\bm\Psi) (\widehat{\bm\Psi}^{(c)} - \bm\Psi) + o_p(n^{-1/2}).
\end{equation}
Since $E[\bm{m}_{i}(\bm\theta, \phi)] = \boldsymbol{0}$, the quantity $\bS(\bm\Psi) = n^{-1} \sum_{i=1}^{n} \bm{m}_{i}(\bm\theta, \psi)$ is the average of $n$ iid random variables, so applying the central limit theorem yields
\begin{equation}\label{eq:clm}
\sqrt{n}\bS(\bm\Psi) \xrightarrow{d} N\left(\mathbf{0}, \bB \right),
\end{equation}
where $\bB = \Var[\bm{m}_1(\bm\theta, \phi)]$.
We make the following claim regarding the behavior of the first derivative $\nabla\bS(\bm\Psi)$:
\vspace{-1em}
\paragraph{Claim} Under the conditions stated in the theorem, we have $\nabla\bS(\bm\Psi) = n^{-1} \sum_{i=1}^{n} \nabla \bm{m}_i(\bm\theta, \phi) \xrightarrow{p} \bA $, where $\bA  = E_{\bm\theta, \phi}[\nabla m_1(\bm\theta, \phi)]$  is a negative definite matrix. 

If this claim is true, using \eqref{eq:taylor} and \eqref{eq:clm}, we would arrive at
$$
\sqrt{n} (\widehat{\bm\Psi}^{(c)} - \bm\Psi) = [-\nabla\bS(\bm\Psi)]^{-1}\sqrt{n}\bS(\bm\Psi) + o_p(1) 
\xrightarrow{d} N\left(\mathbf{0}, \bA^{-1} \bB \bA^{-1} \right)    
$$
as stated in the theorem.  

\subsubsection{Proof of the claim}

Direct calculation of the gradient shows that
\begin{equation}\label{eq:grad.mi}
\nabla \bm{m}_i(\bm\theta, \phi) = \begin{bmatrix}
-\phi^{-1}(\widetilde{\bW}_i^\T \widetilde{\bW}_i - \sigma^2 \bJ_{p+q}) & \phi^{-2} (\widetilde{\bW}_i^\T \widetilde{\bW}_i - \sigma^2 \bJ_{p+q}) \bm\theta \\[.5em]
\phi^{-2}\bm\theta^\T (\widetilde{\bW}_i^\T \widetilde{\bW}_i - \sigma^2 \bJ_{p+q}) & -1/(2\phi^{2}) -\phi^{-3} \bm\theta^\top (\widetilde{\bW}_i^\T \widetilde{\bW}_i - \sigma^2 \bJ_{p+q})\bm\theta
\end{bmatrix}
\end{equation}

Denote $\bQ_i = (\widetilde{\bW}_i^\T \widetilde{\bW}_i - \sigma^2 \bJ_{p+q})$, thus $\nabla\bS(\bm\Psi) = n^{-1} \sum_{i=1}^{n} \nabla \bm{m}_i(\bm\theta, \phi)$ becomes
\begin{equation}\label{eq:grad.Q}
\nabla\bS(\bm\Psi) =  \begin{bmatrix}
-\phi^{-1}(n^{-1} \sum_{i=1}^{n} \bQ_i) & \phi^{-2} (n^{-1} \sum_{i=1}^{n} \bQ_i) \bm\theta \\[.5em]
\phi^{-2}\bm\theta^\T (n^{-1} \sum_{i=1}^{n} \bQ_i) & -1/(2\phi^{2}) -\phi^{-3} \bm\theta^\top (n^{-1} \sum_{i=1}^{n} \bQ_i) \bm\theta
\end{bmatrix}.
\end{equation}

First, we prove that $n^{-1} \sum_{i=1}^{n} \bQ_i$ converges to $\bar{\bQ} = E(\bQ_1)$. 

Denote $\bQ_i^{(0)}=({\bW}_i^\T {\bW}_i - \sigma^2 {\bI}_{p+q})$ so that $\bQ_i = (\widetilde{\bW}_i^\T \widetilde{\bW}_i - \sigma^2 \bJ_{p+q}) = \begin{bmatrix}
1 &  \bW_i \\
\bW_i^\T &  \bQ_i^{(0)} \\
\end{bmatrix}$, we have
$$
n^{-1} \sum_{i=1}^{n} \bQ_i = \begin{bmatrix}
1 & n^{-1} \sum_{i=1}^{n} \bW_i \\
n^{-1} \sum_{i=1}^{n} \bW_i^\T &  n^{-1} \sum_{i=1}^{n} \bQ_i^{(0)} \\
\end{bmatrix}.
$$
For the off-diagonal sub-vector, the law of large number ensures that 
\begin{equation}\label{eq:converge.1}
   n^{-1} \sum_{i=1}^{n} \bW_i  \xrightarrow{p} E(\bW_1),
\end{equation}
as long as $\Var(\bW_1)$ is finite. The finiteness of $\Var(\bW_i)$ is ensured since $\Var(\bW_i) = \Var(\bW_i^*) + \sigma^2 \bI_{p+q}$ where $\Var(\bW_i^*)$ is finite by condition C2

For the bottom right block sub-matrix in $n^{-1} \sum_{i=1}^{n} \bQ_i^{(0)}$, we decompose $\bQ_i$ into four components:
\begin{align*}
\bQ_i^{(0)} & =  \left\{(\bW_i^* + \bU_i)^\T(\bW_i^* + \bU_i) - \sigma^2 \bI_{p+q} \right\} \\ & =  \bW_i^{*\T} \bW_i^* +  \bW_i^{*\T} \bU_i + \bU_i^{\T} \bW_i^* + \left(\bU_i^{\T} \bU_i - \sigma^2 \bI_{p+q} \right) \\ &= \bQ_i^{(1)} + \bQ_i^{(2)} +\bQ_i^{(3)} + \bQ_i^{(4)} .
\end{align*}
The law of large number ensures that, for $k=1,..,4$,  
\begin{equation}\label{eq:converge.k}
n^{-1} \sum_{i=1}^{n} \bQ_i^{(k)} \xrightarrow{p}  E(\bQ_1^{(k)}),
\end{equation}
if $n^{-1} \Var(\bQ_1^{(k)}) \to \mathbf{0}$ when $n \to \infty$. 

By condition C2, $\Var(\bQ_1^{(1)}) = \Var(\bW_1^{*\T} \bW_1^*)$ is finite, thus $n^{-1} \Var(\bQ_1^{(1)}) \to \mathbf{0}$ when $n \to \infty$. 

Since $\bW_1^*$ and $\bU_1$ are independent and $\bU_1 \sim N(\mathbf{0}, \sigma^2 \bI_{p+q})$, then each element of $\bQ_1^{(2)}=\bW_i^{*\T} \bU_i$ and each element of $\bQ_1^{(3)}=\bU_i^{\T} \bW_i^*$ has variance of order $O(\sigma^2)$, and each element of $\bQ_1^{(4)}=(\bU_i^{\T} \bU_i - \sigma^2 \bI_{p+q})$ has variance of order $O(\sigma^4)$.
Hence, under the conditions for the Theorem, when both $\sigma^2$ and $n$ diverge to infinity but $\sigma^4/n \to 0$, $n^{-1} \Var(\bQ_1^{(k)}) \to \mathbf{0}$ for $k=2,3,4$. 

Putting together, \eqref{eq:converge.k} holds for all $k=1,2,3,4$. Therefore,
\begin{equation}\label{eq:converge.2}
   n^{-1} \sum_{i=1}^{n} \bQ_i^{(0)} \xrightarrow{p} 
   E(\bQ_1^{(1)}) + E(\bQ_1^{(2}) + E(\bQ_1^{(3)}) + E(\bQ_1^{(4)}) = E(\bQ_i^{(0)}).
\end{equation}

Combine \eqref{eq:converge.1} and \eqref{eq:converge.2}, we get 
$$
n^{-1} \sum_{i=1}^{n} \bQ_i \xrightarrow{p}  E(\bQ_1) = \bar{\bQ}. 
$$

Plug this into \eqref{eq:grad.Q}, we have
\begin{equation}
\nabla\bS(\bm\Psi) \xrightarrow{p}  \begin{bmatrix}
-\phi^{-1}\bar{\bQ} & \phi^{-2} \bar{\bQ} \bm\theta \\[.5em]
\phi^{-2}\bm\theta^\T \bar{\bQ} & -1/(2\phi^{2}) -\phi^{-3} \bm\theta^\top \bar{\bQ} \bm\theta
\end{bmatrix}.
\end{equation}

Using \eqref{eq:grad.mi}, it is easy to see that the above limit is
$$
\begin{bmatrix}
-\phi^{-1}\bar{\bQ} & \phi^{-2} \bar{\bQ} \bm\theta \\[.5em]
\phi^{-2}\bm\theta^\T \bar{\bQ} & -1/(2\phi^{2}) -\phi^{-3} \bm\theta^\top \bar{\bQ} \bm\theta
\end{bmatrix} = E[ \nabla \bm{m}_1(\bm\theta, \phi)] = \bA. 
$$

That is, we have proved that, when both $\sigma^2$ and $n$ diverge to infinity but $\sigma^4/n \to 0$,
$$\nabla\bS(\bm\Psi) \xrightarrow{p} \bA.$$

What remains is to prove that $\bA$  is indeed  negative definite. Equivalently, we show that $\bar{\bA} = - \bA$ is positive definite. 

Notice that 
$$
\bar{\bA} = \begin{bmatrix}
\phi^{-1}\bar{\bQ} & - \phi^{-2} \bar{\bQ} \bm\theta \\[.5em]
- \phi^{-2}\bm\theta^\T \bar{\bQ} & 1/(2\phi^{2}) + \phi^{-3} \bm\theta^\top \bar{\bQ} \bm\theta
\end{bmatrix}
$$
would be positive definite if $\bar{\bQ}$ is positive definite, due to the following arguments. 

If $\bar{\bQ}$ is positive definite, then the upper left block sub-matrix in $\bar{\bA}$, $\phi^{-1}\bar{\bQ}$, is also positive definite (since $\phi^{-1} = \tau^2$ is a positive scalar). Hence  all the  $k$-th leading principal minors of $\bar{\bA}$ are positive for $k=1,\ldots, p$. Applying the formula for the determinant of the block matrix, we have \begin{align*}
\det\left(\bar{\bA}\right) &= \det(\phi^{-1} \bar{\bQ}) \left\{(1/2)\phi^{-2} +\phi^{-3} \bm\theta^\top \bar{\bQ}\bm\theta - (-\phi^{-2}\bm\theta^\top \bar{\bQ} (\phi \bar{\bQ}^{-1})(-\phi^{-2}  \bar{\bQ} \bm\theta) \right\}
\\ &=\det(\phi^{-1} \bar{\bQ}) \left\{(1/2)\phi^{-2} - \phi^{-3} \bm\theta^\top  \bar{\bQ} \bm\theta + \phi^{-3} \bm\theta^\top  \bar{\bQ} \bm\theta \right\} 
\\& = \det(\phi^{-1}\bar{\bQ}) (1/2) \phi^{-2} 
\\& > 0,
\end{align*}
so the $(p+1)$ leading principal minor of $\bar{\bA}$ is also positive.  Hence, using the Sylvester's criterion, $\bar{\bA}$ is positive definite. 

Finally, we show that $\bar{\bQ}$ is positive definite. Recall
$$
\bar{\bQ} = E(\bQ_1) = \begin{bmatrix}
1 & E(\bW_1) \\
E(\bW_1^{\T}) & E(\bQ_1^{(0)})
\end{bmatrix}.
$$
Notice $E(\bW_1) = E(\bW_1^* + \bU_1) = E(\bW_1^*)$. Also,
\begin{align*}
E(\bQ_1^{(0)}) & = E(\bQ_1^{(1)}) + E(\bQ_1^{(2)}) +E(\bQ_1^{(3)}) + E(\bQ_1^{(4)}) \\
& = E(\bW_1^{*\T} \bW_1^*) + E(\bW_1^{*\T} \bU_1) +E(\bU_1^{\T} \bW_1^*) + E\left(\bU_i^{\T} \bU_i - \sigma^2 \bI_{p+q} \right) \\
& = E(\bW_1^{*\T} \bW_1^*) + \mathbf{0} + \mathbf{0} + \left(\sigma^2 \bI_{p+q}  - \sigma^2 \bI_{p+q} \right) \\
& = E(\bW_1^{*\T} \bW_1^*).
\end{align*}
Hence
$$
\bar{\bQ} = \begin{bmatrix}
1 & E(\bW_1^*) \\
E(\bW_1^{*\T}) & E(\bW_1^{*\T} \bW_1^*)
\end{bmatrix}
$$
which is indeed positive definite by condition 2.

\section{Additional simulation results}
\begin{table}[ht]
\centering
\caption{Simulation results when the raw data were generated from the unconditional mixture normal distribution with $p_1 = 0.1$. Reported are $(10 \times)$ bias, $(100 \times)$ MSE for point estimators (MLE: naive conditional MLE, LS: naive least square, and cLS: corrected least square), and coverage probability for the confidence interval corresponding to each slope coefficient.}
\label{tab:sim_results_mixturemodel.extra1}
\resizebox{!}{.4\textheight}{\begin{tabular}{lll rrr rrr rrr}
\toprule[1.5pt]
$\sigma$ & $n$  & Coef. & \multicolumn{3}{c}{Bias} & \multicolumn{3}{c}{MSE} & \multicolumn{3}{c}{Coverage} \\
&  & & MLE & LS & cLS & MLE & LS & cLS & MLE & LS & cLS \\
\cmidrule(lr){4-6} \cmidrule(lr){7-9} \cmidrule(lr){10-12}
0 & $ 10^3$ & $\beta_1$ & -0.21 & -0.16 & -0.16 & 1.85 & 1.67 & 1.67 & 0.95 & 0.95 & 0.95\\
 &  & $\beta_2$ & 0.19 & 0.13 & 0.13 & 2.37 & 2.21 & 2.21 & 0.95 & 0.95 & 0.95\\
 &  & $\beta_3$ & -0.01 & -0.01 & -0.01 & 1.72 & 1.62 & 1.62 & 0.95 & 0.95 & 0.95\\
\addlinespace
 & $ 10^4$ & $\beta_1$ & 0.00 & 0.00 & 0.00 & 0.18 & 0.16 & 0.16 & 0.95 & 0.95 & 0.95\\
 &  & $\beta_2$ & 0.01 & 0.01 & 0.01 & 0.22 & 0.21 & 0.21 & 0.95 & 0.95 & 0.95\\
 &  & $\beta_3$ & -0.02 & -0.02 & -0.02 & 0.15 & 0.15 & 0.15 & 0.96 & 0.97 & 0.97\\
\addlinespace
 %& $ 10^5$ & $\beta_1$ & 0.00 & 0.00 & 0.00 & 0.02 & 0.02 & 0.02 & 0.94 & 0.94 & 0.94\\
 %&  & $\beta_2$ & 0.00 & 0.01 & 0.01 & 0.02 & 0.02 & 0.02 & 0.95 & 0.96 & 0.96\\
 %&  & $\beta_3$ & 0.00 & 0.00 & 0.00 & 0.02 & 0.02 & 0.02 & 0.96 & 0.96 & 0.96\\
%\addlinespace
 & $ 2\times 10^5$ & $\beta_1$ & 0.00 & 0.00 & 0.00 & 0.01 & 0.01 & 0.01 & 0.94 & 0.94 & 0.94\\
 &  & $\beta_2$ & 0.00 & 0.00 & 0.00 & 0.01 & 0.01 & 0.01 & 0.93 & 0.93 & 0.93\\
 &  & $\beta_3$ & 0.00 & 0.00 & 0.00 & 0.01 & 0.01 & 0.01 & 0.95 & 0.95 & 0.95\\
\addlinespace
0.3 & $ 10^3$ & $\beta_1$ & 1.27 & 5.91 & -0.21 & 6.68 & 35.64 & 5.30 & 0.63 & 0.00 & 0.95\\
 &  & $\beta_2$ & -1.42 & -5.98 & 0.28 & 7.82 & 36.65 & 6.61 & 0.64 & 0.00 & 0.96\\
 &  & $\beta_3$ & 0.35 & 0.17 & -0.13 & 3.35 & 0.70 & 4.19 & 0.81 & 0.94 & 0.96\\
\addlinespace
 & $ 10^4$ & $\beta_1$ & 1.48 & 5.91 & -0.03 & 2.61 & 34.96 & 0.47 & 0.14 & 0.00 & 0.95\\
 &  & $\beta_2$ & -1.70 & -6.01 & 0.02 & 3.31 & 36.21 & 0.54 & 0.11 & 0.00 & 0.97\\
 &  & $\beta_3$ & 0.46 & 0.22 & 0.01 & 0.48 & 0.11 & 0.35 & 0.70 & 0.89 & 0.96\\
\addlinespace
 %& $ 10^5$ & $\beta_1$ & 1.52 & 5.91 & 0.00 & 2.35 & 34.99 & 0.05 & 0.00 & 0.00 & 0.95\\
 %&  & $\beta_2$ & -1.72 & -6.01 & 0.00 & 3.02 & 36.17 & 0.06 & 0.00 & 0.00 & 0.95\\
 %&  & $\beta_3$ & 0.44 & 0.21 & 0.00 & 0.23 & 0.05 & 0.04 & 0.12 & 0.26 & 0.94\\
%\addlinespace
 & $ 2\times 10^5$ & $\beta_1$ & 1.51 & 5.91 & 0.00 & 2.31 & 34.97 & 0.02 & 0.00 & 0.00 & 0.96\\
 &  & $\beta_2$ & -1.72 & -6.01 & 0.00 & 3.00 & 36.18 & 0.03 & 0.00 & 0.00 & 0.95\\
 &  & $\beta_3$ & 0.44 & 0.21 & 0.00 & 0.21 & 0.05 & 0.02 & 0.01 & 0.05 & 0.95\\
\addlinespace
1 & $ 10^3$ & $\beta_1$ & 5.38 & 9.72 & 151.15 & 61.52 & 94.50 & 21448457.38 & 0.09 & 0.00 & 0.96\\
 &  & $\beta_2$ & -6.26 & -9.76 & -248.94 & 66.92 & 95.34 & 54472070.48 & 0.09 & 0.00 & 0.97\\
 &  & $\beta_3$ & 1.49 & 0.09 & -161.65 & 23.50 & 0.06 & 26627838.31 & 0.40 & 0.93 & 1.00\\
\addlinespace
 & $ 10^4$ & $\beta_1$ & 6.53 & 9.73 & -0.66 & 43.61 & 94.62 & 13.69 & 0.00 & 0.00 & 0.95\\
 &  & $\beta_2$ & -6.78 & -9.75 & 0.77 & 46.88 & 95.00 & 17.76 & 0.00 & 0.00 & 0.96\\
 &  & $\beta_3$ & 0.88 & 0.07 & -0.15 & 1.62 & 0.01 & 6.78 & 0.27 & 0.83 & 0.98\\
\addlinespace
 %& $ 10^5$ & $\beta_1$ & 6.59 & 9.73 & -0.07 & 43.55 & 94.62 & 1.01 & 0.00 & 0.00 & 0.95\\
 %&  & $\beta_2$ & -6.87 & -9.75 & 0.06 & 47.29 & 95.06 & 1.30 & 0.00 & 0.00 & 0.95\\
 
 %&  & $\beta_3$ & 0.89 & 0.07 & -0.02 & 0.87 & 0.01 & 0.64 & 0.01 & 0.11 & 0.94\\
%\addlinespace
 & $ 2\times 10^5$ & $\beta_1$ & 6.61 & 9.73 & -0.03 & 43.68 & 94.63 & 0.48 & 0.00 & 0.00 & 0.95\\
 &  & $\beta_2$ & -6.87 & -9.75 & 0.05 & 47.26 & 95.04 & 0.63 & 0.00 & 0.00 & 0.95\\
 &  & $\beta_3$ & 0.89 & 0.07 & -0.01 & 0.83 & 0.01 & 0.28 & 0.00 & 0.01 & 0.96\\
\addlinespace
3 & $ 10^3$ & $\beta_1$ & 9.72 & 10.00 & 28.34 & 110.88 & 99.95 & 142660.62 & 0.01 & 0.00 & 0.92\\
 &  & $\beta_2$ & -9.71 & -10.00 & -17.24 & 111.32 & 99.95 & 205185.92 & 0.02 & 0.00 & 0.92\\
 &  & $\beta_3$ & 0.48 & 0.00 & 7.53 & 29.39 & 0.00 & 151164.11 & 0.19 & 0.97 & 1.00\\
\addlinespace
 & $ 10^4$ & $\beta_1$ & 9.25 & 9.99 & 19.89 & 88.88 & 99.89 & 187187.10 & 0.00 & 0.00 & 0.93\\
 &  & $\beta_2$ & -9.32 & -9.99 & -10.12 & 89.33 & 99.89 & 190271.88 & 0.00 & 0.00 & 0.96\\
 &  & $\beta_3$ & 0.28 & 0.00 & -1.05 & 3.69 & 0.00 & 63354.14 & 0.15 & 0.94 & 0.99\\
\addlinespace
 %& $ 10^5$ & $\beta_1$ & 9.47 & 9.99 & -94.81 & 89.77 & 99.90 & 11343398.03 & 0.00 & 0.00 & 0.92\\
 %&  & $\beta_2$ & -9.46 & -9.99 & 114.14 & 89.66 & 99.90 & 15738168.07 & 0.00 & 0.00 & 0.95\\
 %&  & $\beta_3$ & 0.24 & 0.00 & 78.27 & 0.19 & 0.00 & 6952301.39 & 0.11 & 0.88 & 1.00\\
%\addlinespace
 & $ 2\times 10^5$ & $\beta_1$ & 9.47 & 9.99 & -8.17 & 89.68 & 99.90 & 4996.50 & 0.00 & 0.00 & 0.95\\
 &  & $\beta_2$ & -9.48 & -9.99 & 8.60 & 89.84 & 99.90 & 7689.45 & 0.00 & 0.00 & 0.93\\
 &  & $\beta_3$ & 0.26 & 0.00 & -0.56 & 0.12 & 0.00 & 778.56 & 0.09 & 0.82 & 1.00\\
 \bottomrule[1.5pt]
 \end{tabular}
 }
\end{table}

\begin{table}[ht]
\centering
\caption{Simulation results when the raw data were generated from the unconditional mixture normal distribution with $p_1 = 0.9$. Reported are $(10 \times)$ bias, $(100 \times)$ MSE for point estimators (MLE: naive conditional MLE, LS: naive least square, and cLS: corrected least square), and coverage probability for the confidence interval corresponding to each slope coefficient.}
\label{tab:sim_results_mixturemodel.extra2}
\resizebox{!}{.4\textheight}{\begin{tabular}{lll rrr rrr rrr}
\toprule[1.5pt]
$\sigma$ & $n$  & Coef. & \multicolumn{3}{c}{Bias} & \multicolumn{3}{c}{MSE} & \multicolumn{3}{c}{Coverage} \\
&  & & MLE & LS & cLS & MLE & LS & cLS & MLE & LS & cLS \\
\cmidrule(lr){4-6} \cmidrule(lr){7-9} \cmidrule(lr){10-12}
0 & $ 10^3$ & $\beta_1$ & -0.01 & 0.01 & 0.01 & 1.88 & 1.81 & 1.81 & 0.95 & 0.94 & 0.94\\
 &  & $\beta_2$ & 0.10 & 0.07 & 0.07 & 2.30 & 2.22 & 2.22 & 0.95 & 0.94 & 0.94\\
 &  & $\beta_3$ & -0.01 & -0.01 & -0.01 & 1.58 & 1.55 & 1.55 & 0.95 & 0.96 & 0.96\\
\addlinespace
 & $ 10^4$ & $\beta_1$ & -0.01 & -0.01 & -0.01 & 0.18 & 0.17 & 0.17 & 0.96 & 0.95 & 0.95\\
 &  & $\beta_2$ & 0.01 & 0.01 & 0.01 & 0.22 & 0.20 & 0.20 & 0.95 & 0.95 & 0.95\\
 &  & $\beta_3$ & -0.02 & -0.01 & -0.01 & 0.17 & 0.17 & 0.17 & 0.95 & 0.95 & 0.95\\
\addlinespace
%& $ 10^5$ & $\beta_1$ & 0.00 & 0.00 & 0.00 & 0.02 & 0.02 & 0.02 & 0.95 & 0.96 & 0.96\\
 %&  & $\beta_2$ & 0.00 & -0.01 & -0.01 & 0.02 & 0.02 & 0.02 & 0.96 & 0.96 & 0.96\\
 %&  & $\beta_3$ & 0.00 & 0.00 & 0.00 & 0.02 & 0.01 & 0.01 & 0.97 & 0.96 & 0.96\\
\addlinespace
 & $ 2\times 10^5$ & $\beta_1$ & 0.00 & -0.01 & -0.01 & 0.01 & 0.01 & 0.01 & 0.96 & 0.96 & 0.96\\
 &  & $\beta_2$ & 0.01 & 0.01 & 0.01 & 0.01 & 0.01 & 0.01 & 0.95 & 0.94 & 0.94\\
 &  & $\beta_3$ & 0.00 & 0.00 & 0.00 & 0.01 & 0.01 & 0.01 & 0.96 & 0.96 & 0.96\\
\addlinespace
0.3 & $ 10^3$ & $\beta_1$ & 1.24 & 5.89 & -0.22 & 6.47 & 35.44 & 5.34 & 0.63 & 0.00 & 0.95\\
 &  & $\beta_2$ & -1.45 & -5.98 & 0.24 & 7.05 & 36.57 & 5.98 & 0.64 & 0.00 & 0.96\\
 &  & $\beta_3$ & 0.45 & 0.21 & -0.01 & 3.31 & 0.67 & 3.81 & 0.81 & 0.95 & 0.96\\
\addlinespace
 & $ 10^4$ & $\beta_1$ & 1.51 & 5.92 & 0.02 & 2.71 & 35.17 & 0.49 & 0.13 & 0.00 & 0.94\\
 &  & $\beta_2$ & -1.70 & -6.02 & 0.00 & 3.34 & 36.27 & 0.56 & 0.11 & 0.00 & 0.97\\
 &  & $\beta_3$ & 0.43 & 0.21 & -0.02 & 0.47 & 0.11 & 0.37 & 0.71 & 0.88 & 0.97\\
\addlinespace
 %& $ 10^5$ & $\beta_1$ & 1.52 & 5.91 & 0.00 & 2.34 & 34.99 & 0.05 & 0.00 & 0.00 & 0.95\\
 %&  & $\beta_2$ & -1.71 & -6.01 & 0.01 & 2.98 & 36.14 & 0.06 & 0.00 & 0.00 & 0.95\\
 %&  & $\beta_3$ & 0.44 & 0.21 & 0.00 & 0.22 & 0.05 & 0.04 & 0.12 & 0.27 & 0.94\\
\addlinespace
 & $ 2\times 10^5$ & $\beta_1$ & 1.52 & 5.91 & 0.00 & 2.32 & 34.97 & 0.02 & 0.00 & 0.00 & 0.95\\
 &  & $\beta_2$ & -1.73 & -6.01 & 0.00 & 3.01 & 36.18 & 0.03 & 0.00 & 0.00 & 0.96\\
 &  & $\beta_3$ & 0.44 & 0.21 & 0.00 & 0.21 & 0.05 & 0.02 & 0.01 & 0.05 & 0.94\\
\addlinespace
1 & $ 10^3$ & $\beta_1$ & 5.72 & 9.73 & -0.79 & 61.01 & 94.75 & 7701.20 & 0.10 & 0.00 & 0.95\\
 &  & $\beta_2$ & -5.96 & -9.75 & 2.66 & 68.01 & 95.13 & 18520.81 & 0.09 & 0.00 & 0.97\\
 &  & $\beta_3$ & 1.03 & 0.07 & -2.65 & 19.74 & 0.06 & 7757.13 & 0.38 & 0.93 & 1.00\\
\addlinespace
 & $ 10^4$ & $\beta_1$ & 6.56 & 9.73 & -0.50 & 44.03 & 94.65 & 12.72 & 0.00 & 0.00 & 0.95\\
 &  & $\beta_2$ & -6.83 & -9.75 & 0.54 & 47.60 & 95.07 & 15.83 & 0.00 & 0.00 & 0.96\\
 &  & $\beta_3$ & 0.88 & 0.07 & -0.10 & 1.58 & 0.01 & 6.44 & 0.26 & 0.83 & 0.98\\
\addlinespace
% & $ 10^5$ & $\beta_1$ & 6.59 & 9.73 & -0.08 & 43.54 & 94.61 & 0.99 & 0.00 & 0.00 & 0.95\\
% &  & $\beta_2$ & -6.86 & -9.75 & 0.09 & 47.15 & 95.03 & 1.34 & 0.00 & 0.00 & 0.96\\
% &  & $\beta_3$ & 0.89 & 0.07 & -0.03 & 0.86 & 0.01 & 0.55 & 0.00 & 0.10 & 0.96\\
\addlinespace
 & $ 2\times 10^5$ & $\beta_1$ & 6.59 & 9.73 & -0.06 & 43.49 & 94.62 & 0.43 & 0.00 & 0.00 & 0.97\\
 &  & $\beta_2$ & -6.87 & -9.75 & 0.05 & 47.18 & 95.04 & 0.60 & 0.00 & 0.00 & 0.95\\
 &  & $\beta_3$ & 0.90 & 0.07 & 0.00 & 0.84 & 0.01 & 0.28 & 0.00 & 0.00 & 0.95\\
\addlinespace
3 & $ 10^3$ & $\beta_1$ & 9.39 & 10.00 & 17.76 & 137.46 & 99.93 & 54132.92 & 0.02 & 0.00 & 0.91\\
 &  & $\beta_2$ & -9.10 & -10.00 & -19.11 & 233.16 & 99.96 & 631434.25 & 0.02 & 0.00 & 0.92\\
 &  & $\beta_3$ & 0.62 & 0.00 & 24.71 & 302.61 & 0.00 & 1310764.70 & 0.17 & 0.96 & 0.99\\
\addlinespace
 & $ 10^4$ & $\beta_1$ & 9.38 & 9.99 & 12.63 & 90.18 & 99.90 & 25558.47 & 0.00 & 0.00 & 0.92\\
 &  & $\beta_2$ & -9.37 & -9.99 & -34.07 & 89.86 & 99.90 & 987114.37 & 0.00 & 0.00 & 0.95\\
 &  & $\beta_3$ & 0.39 & 0.00 & 26.29 & 2.28 & 0.00 & 1118249.17 & 0.12 & 0.94 & 1.00\\
\addlinespace
% & $ 10^5$ & $\beta_1$ & 9.45 & 9.99 & -9.46 & 89.42 & 99.89 & 10222.59 & 0.00 & 0.00 & 0.93\\
 %&  & $\beta_2$ & -9.47 & -9.99 & 10.35 & 89.72 & 99.90 & 16877.40 & 0.00 & 0.00 & 0.95\\
 %&  & $\beta_3$ & 0.24 & 0.00 & -1.31 & 0.18 & 0.00 & 2072.33 & 0.12 & 0.89 & 1.00\\
\addlinespace
 & $ 2\times 10^5$ & $\beta_1$ & 9.47 & 9.99 & -3.12 & 89.70 & 99.90 & 779.23 & 0.00 & 0.00 & 0.93\\
 &  & $\beta_2$ & -9.49 & -10.00 & 2.94 & 90.12 & 99.90 & 819.94 & 0.00 & 0.00 & 0.93\\
 &  & $\beta_3$ & 0.23 & 0.00 & 0.07 & 0.12 & 0.00 & 67.16 & 0.08 & 0.84 & 1.00\\
 \bottomrule
 \end{tabular}
 }
\end{table}

\end{document}